# Copying along Arithmetic Progressions Cannot Simulate Interleaving with Polylogarithmic Loss

Matthias G. Mayer
Computational Rational Agents Laboratory (CORAL)
matthias.georg.mayer@gmail.com

Abstract. Two classical ways to compress a word are straight-line programs (SLPs), which build it by concatenating smaller pieces, and the Lempel–Ziv 77 (LZ77) factorisation, which copies earlier parts of it; their sizes are measures of repetitiveness. LZ77 approximates the smallest SLP efficiently: it has no more factors than a smallest SLP has size, and yields an SLP at most a logarithmic factor larger in near-linear time. Motivated by online prediction, Kosoy introduced zipline programs, which are analogous to SLPs but interleave their pieces round-robin instead of concatenating them. Her generalisation of LZ77, the arithmetic repetition system, describes a word by copies from one arithmetic progression of positions to a later one. Do arithmetic repetition systems approximate the smallest zipline program as well? Kosoy proved that the least size $\mathrm{ARC}(x)$ of an arithmetic repetition system for a word $x$ of length $n$ is at most a factor $O(n^{1/3})$ larger than the least size $\mathrm{ZLC}(x)$ of a zipline program for it, and asked whether this factor is tight. Her polynomial-time next-letter predictor makes $O(\mathrm{ARC}(x)\log n)$ mistakes, so a polylogarithmic factor would bound its mistakes by $\mathrm{ZLC}(x)\,\mathrm{polylog}(n)$, which no known polynomial-time predictor achieves. We show that the optimal exponent is 1/4, not 1/3: $\mathrm{ARC}(x) \leq 2n^{1/4}\,\mathrm{ZLC}(x)$ for every word $x$, and at every large enough length $n$ some binary word has $\mathrm{ARC}(x) \geq n^{1/4-o(1)}\,\mathrm{ZLC}(x)$.

## 1. Introduction

Two classical ways to describe a repetitive word are a *straight-line program* (SLP), which builds it by concatenating smaller pieces, and the *Lempel–Ziv 77 (LZ77) factorisation* [1], which copies earlier parts of it. The study of *measures of repetitiveness* compares the sizes of smallest descriptions of different kinds for the same word [2]. For these two kinds, copying simulates concatenation without loss: the LZ77 factorisation of a word has at most as many factors as a smallest SLP for the word in Chomsky normal form has rules [3, Thm. 1]. Conversely, concatenation simulates copying at a logarithmic loss: an SLP at most $O(\log n)$ times larger than the factorisation can be computed from it in near-linear time [3, Thm. 2], so LZ77 approximates the smallest SLP efficiently. More generally, the measures based on substring complexity, string attractors, the LZ77 factorisation, SLPs and the runs of the Burrows–Wheeler transform are within polylogarithmic factors of each other [2, 4, 5].

We ask whether copying along arithmetic progressions simulates interleaving. Kosoy [6] introduced *zipline programs*, which are analogous to SLPs with concatenation replaced by round-robin interleaving: a rooted acyclic graph in which each vertex has a list of children, vertices or letters, repetitions allowed, and the word it stands for, its *value*, zips theirs, reading them round-robin and stopping at the first missing symbol. For a word $x \in \Sigma^n$, $\Sigma$ a finite alphabet, we call the least number of edges of a zipline program whose value has $x$ as a prefix the *zipline complexity* $\mathrm{ZLC}(x)$ (Definition 3.4). ZLC lies outside the polylogarithmic cluster above, in either direction (Remark E.8). Her generalisation of LZ77 is the *arithmetic repetition system*[1]: a set of literals $x[i] = a$ and of equations $x[\sigma + \tau t] = x[\sigma' + \tau' t]$ for all $t < \ell$ (Definition 3.2), such that every position of $x$ is given by a literal or copied by an equation from an earlier position. We write $\mathrm{ARC}(x)$ for the least number of literals and equations of such a system. It is a strong form of copying: $\mathrm{ARC}(x) \leq \mathrm{LZC}(x)$ [6, Prop. 4.3], where $\mathrm{LZC}(x)$ is the number of factors of the LZ77 factorisation of $x$ with self-reference allowed, and it can be exponentially smaller than ZLC: some binary words have $\mathrm{ARC}(x) = O(\log n)$ but $\mathrm{ZLC}(x) = \Omega(n/\log n)$ (Proposition E.1). Kosoy proved $\mathrm{ARC}(x) = O\big(n^{1/3}\,\mathrm{ZLC}(x)\big)$ [6, Prop. 4.7] and asked "whether the factor $O(\sqrt[3]{n})$ in the above is tight". The answer matters for online prediction: her polynomial-time predictor for ARC makes $O(\mathrm{ARC}(x)\log n)$ mistakes, so a polylogarithmic factor would give it a guarantee for ZLC that no known polynomial-time predictor achieves.

We answer the question: the optimal exponent of $n$ in this factor is $1/4$, not $1/3$. Every word has $\mathrm{ARC}(x) \leq 2n^{1/4}\,\mathrm{ZLC}(x)$, and at every large enough length $n$ some binary word has $\mathrm{ARC}(x) \geq n^{1/4-o(1)}\,\mathrm{ZLC}(x)$. So even this strong form of copying cannot simulate interleaving at a polylogarithmic loss.

**Where the measures come from.** Kosoy's series [6, 7, 8] seeks sequence predictors that are computationally efficient, have strong provable mistake bounds and make a vanishing fraction of mistakes on rich natural classes of deterministic sequences, a combination with few known examples [6, §1]. Its predictors work against stringological complexity measures, candidate tractable analogues of Kolmogorov complexity [7, §1]. She calls a predictor *statistically efficient* for a measure $C$ if it makes at most $(C(x)+1)\,\mathrm{polylog}(n)$ mistakes on every $x \in \Sigma^n$ [6, Def. 2.3], *computationally feasible* if its state has size polynomial in $n$ and it predicts and updates in time polynomial in that size [6, Def. 2.4], and *computationally efficient* if moreover its state has size polynomial in $C(x)$ and $\log n$ [6, Def. 2.5].

Her simplest measures, the *left-to-right* and the *right-to-left $k$-automaticity*, are the least number of states of an automaton that computes each symbol from the base-$k$ digits of its position (padded with leading zeros to one length), read from the most or from the least significant digit [6, §1, Def. 3.1]; both have statistically and computationally efficient predictors [7], [6, Thm. 3.4]. They depend on the arbitrary choice of base $k$ [6, §1]. To drop the dependence on the base, Kosoy passes to SLPs on the left-to-right side and defines a statistically and computationally efficient predictor [7, Thm. 4.6, eq. (20)]. On the right-to-left side she passes to zipline programs [6, §1, §4], but whether ZLC has a statistically efficient, computationally feasible predictor is open [6, §4.2].

A restriction makes copying enough. A zipline program is *synchronous* if, for each vertex, all paths from the root to it have the same product of the numbers of children along the path; $\mathrm{sZLC}(x)$ is the least number of edges of a synchronous program for $x$ (Definition 3.4). Kosoy proved $\mathrm{ARC}(x) \leq \mathrm{sZLC}(x)$ [6, Prop. 4.9], so her ARC-MUWU, statistically efficient and computationally feasible with $O(\mathrm{ARC}(x)\log n)$ mistakes [6, Thm. 4.11], is also statistically efficient for sZLC. The restriction is real: by our lower bound (Theorem 1.2), $\mathrm{sZLC}(x)$ can exceed $\mathrm{ZLC}(x)$ by a factor $n^{1/4-o(1)}$ (Corollary C.22). sZLC still covers right-to-left automaticity in every base [6, Prop. 4.10]; for computationally efficient prediction, Kosoy restricts it further [8].

If $\mathrm{ARC}(x) \leq \mathrm{ZLC}(x)\,\mathrm{polylog}(n)$ held for every word, ARC-MUWU would be a statistically efficient, computationally feasible predictor for ZLC. With Kosoy's bound it is statistically efficient for $\sqrt[3]{n}\cdot\mathrm{ZLC}(x)$ [6, §4.2], and with ours for $n^{1/4}\,\mathrm{ZLC}(x)$ (Theorem 1.1);

[1] A closer analogue of LZ77 for interleaving would allow only equations whose progressions run to the end of the word.

by our lower bound, no bound $\mathrm{ARC}(x) \le \mathrm{ZLC}(x)\,\mathrm{polylog}(n)$ holds, so its guarantee for ARC does not imply one for ZLC through this route.

## 1.1. The Optimal Exponent

Kosoy's bound implies an inequality $\mathrm{ARC}(x) \le Cn^c\,\mathrm{ZLC}(x)$ for all $n \ge 1$ and $x \in \Sigma^n$, with $c = 1/3$ and $C = 2\sqrt[3]{2} + 2$; we determine the optimal exponent $c$.

**Theorem 1.1** (Upper bound; proof in Section 4)**.** *For every finite alphabet $\Sigma$ and every $x \in \Sigma^n$ with $n \ge 1$,*

$$\mathrm{ARC}(x) \le 2n^{1/4}\,\mathrm{ZLC}(x).$$

**Theorem 1.2** (Lower bound; sketch in Section 5, proof in Appendix C.4)**.** *There are constants $C > 0$ and $n_0$ such that for every integer $n \ge n_0$ some binary word $x$ of length $n$ satisfies*

$$\mathrm{ARC}(x) \ge n^{1/4} \exp\Bigl(-C\sqrt{\log n \log\log n}\Bigr)\,\mathrm{ZLC}(x).$$

*In particular, for every $a < 1/4$ and every large enough $n$, some binary word $x$ of length $n$ has $\mathrm{ARC}(x) \ge n^a\,\mathrm{ZLC}(x)$.*

Its words have $\mathrm{ZLC}(x) = n^{1/2+o(1)}$ (Remark C.21); other regimes of $\mathrm{ZLC}(x)$ are an interesting further question (Section 6).

**Corollary 1.3** (Optimal exponent; proof in Appendix A.1)**.** *Let $\Sigma$ be a finite alphabet with at least two letters. The least exponent $c$ such that for some constant $C$, $\mathrm{ARC}(x) \le Cn^c\,\mathrm{ZLC}(x)$ for all $n \ge 1$ and $x \in \Sigma^n$ is $c = 1/4$.*

## 1.2. Techniques: Exit Edges and Reused Random Bits

Both bounds rest on Lemma 3.8: of two progressions with offsets below their steps, the one of larger step is *lower* (places an index at a smaller position) only at small indices, below $\sqrt{n}$ inside a word of length $n$.

*Upper bound.* Each root path places a prefix of the value of its end along a progression, an *occurrence*. The paths of the least occurrences (least step, then offset) form a spanning tree of the program; every other occurrence leaves it at its *exit edge* and is lowest within the word only below $\sqrt{n}$. The indices where an occurrence is lowest form an interval, its *lowest interval*; there is nothing lower to copy from, so the interval is passed down and split among the children of its vertex; each part is copied from lower occurrences of that child by at most two equations (from the occurrences lower on the longest initial and longest final segment of the part), or is a literal if the child is a letter, and where the child is lowest in turn, we go down further. So at its vertex a lowest interval costs up to two equations per child, however short it is, and with up to $O\bigl(n^{1/3}\bigr)$ lowest intervals per vertex this gives the order of Kosoy's bound. But short intervals can crowd the small indices. There we instead handle each index below about $n^{1/4}$ times the number of children on its own: it goes to a single child, where one literal giving its letter or one equation copying it from an earlier position suffices, so these early indices cost $O\bigl(n^{1/4}\bigr)$ per edge, however many intervals lie there. What remains of the intervals lies late, where few occurrences are lowest, and costs only $O\bigl(n^{1/4}\bigr)$ per edge on average (Section 4).

*Lower bound.* We build a program for a word $x$ with many lowest intervals of occurrences that are not least (the costly ones in the upper bound) and fill the intervals with random bits that cannot be copied in bulk. Some of its vertices get $H$ occurrences each, which take turns being lowest. Edges of these vertices lead to random words from a small pool; a bit is *new* if no earlier position shows it through the same edge. An equation that copied many new bits would have to match random bits of many edges (we place the occurrences at random offsets so that an equation cannot concentrate on a few edges), so with high probability every possible equation copies only $O(\log^2 n)$ new bits of a fixed constant fraction of the edges. Hence an arithmetic repetition system for $x$ needs about $H/O(\log^2 n)$ equations and literals per edge of our program, while $\mathrm{ZLC}(x)$ is at most its number of edges. Therefore $\mathrm{ARC}(x) \geq H\,\mathrm{ZLC}(x)/O(\log^2 n)$. Our construction needs words of length $n = H^{4+o(1)}$, so $\mathrm{ARC}(x) \geq n^{1/4-o(1)}\,\mathrm{ZLC}(x)$.

## 1.3. Formal Verification

Every numbered definition, statement and proof step of this paper's own development is formalised and proved in Lean 4 [9] with Mathlib [10], using only the standard axioms of Lean. The exceptions are the comparisons in Remark E.8 that rest on cited results. Of the cited results, only Kosoy's ARC $\leq$ sZLC, Lemma 3.7 and two Chebyshev bounds in Mathlib are formalised; formal deductions that use another cited comparison take it as a hypothesis. The README of the Lean repository [11] gives the declarations of each numbered block and proof step, and the differences in form between the formal statements and the text, such as explicit constants in place of $O$-notation.

## 1.4. Use of AI Tools

This paper was written with extensive use of AI systems: Anthropic's Claude Opus, Sonnet and Haiku 5.5 and OpenAI's Codex (gpt-6-astra). In a research and writing workflow designed by the author, AI agents found the constructions and proofs, drafted the text, wrote the Lean formalisation, and reviewed each other's work across the two model families. The technical details are machine-checked in Lean, with the exceptions stated in Section 1.3. The author chose the questions and research directions, selected results and proofs, commented extensively on the drafts, read the main text in full and edited it, reviewed the appendix proofs, and checked the main Lean definitions and statements. The author takes responsibility for the content of the paper; Appendix F gives the details.

# 2. Related Work

**Kosoy's results.** For a narrower class of programs, Kosoy gives a statistically and computationally efficient predictor for the least width (the largest number of vertices at one depth) of a layered program (a synchronous program whose vertices at each depth share one arity from a fixed finite set) [8, Def. 5, Prop. 6, Def. 7, Thm. 17].

**Interleaving and automatic sequences.** By Grabmayer et al., for each fixed $k \geq 2$, a stream is $k$-automatic iff it is defined by a zip-$k$ specification, a system of stream equations built with prefixing and the $k$-ary zip [12, Thm. 52]; zips of different arities define exactly the mix-automatic sequences [12, Thm. 64], [13].

**Measures of repetitiveness.** Polynomial gaps like ours, with known exponents, occur between compressors and the smallest grammar [14, 15], between smallest L-systems and the measure based on substring complexity [16], and, for two-dimensional strings [17, 18], between that measure and smallest string attractors [19].

An arithmetic repetition system is akin to a macro scheme [20], with copies along arithmetic progressions; arithmetical complexity counts the distinct words of each length along them [21], [22, §1].

## 3. Preliminaries

We write $[m] := \{0, ..., m-1\}$; $\mathbb{N}$ contains 0, and logarithms are to base 2. Words $u \in \Sigma^*$ over the finite non-empty alphabet $\Sigma$ are indexed from 0: $u = u[0]u[1]...u[|u|-1]$, of length $|u|$; $u[: j] := u[0]...u[j-1]$, and $v \sqsubseteq u$ means that $v$ is a prefix of $u$.

An equation of an arithmetic repetition system asserts that two readings of the word along arithmetic progressions, each stopped at the end of the word, are equal.

**Definition 3.1** (Arithmetic progressions, step, length). A (finite) *arithmetic progression* $\alpha$ with integer start, *step* and *length* $\sigma \geq 0$, $\tau \geq 1$, $|\alpha| := \ell \geq 1$ is the map $\alpha : [\ell] \to \mathbb{N}$, $\alpha(t) = \sigma + \tau t$; it *contains* the positions of its image, and where only these matter we identify it with its image. For $u \in \Sigma^*$, $u[\alpha] := u[\alpha(0)]u[\alpha(1)]...u[\alpha(k-1)]$, where $k := |\{t < \ell : \alpha(t) < |u|\}|$.

**Definition 3.2** (Covering, ARS, size, ARC [6, Defs. 4.1, 4.2]). An ordered pair $(\alpha, \alpha')$ of arithmetic progressions *covers* a position $i$ if $\alpha(t) < \alpha'(t) = i$ for some $t < \min(|\alpha|, |\alpha'|)$. A pair $(L, \Phi)$ of a set $L \subseteq \Sigma \times [n]$ of *literals* and a finite set $\Phi$ of such pairs, its *equations*, is called an *arithmetic repetition system* (ARS) for $x \in \Sigma^n$ if

(a) $x[i] = a$ for every $(a, i) \in L$;

(b) $|\alpha| = |\alpha'|$ and $x[\alpha] = x[\alpha']$ for every $(\alpha, \alpha') \in \Phi$;

(c) for every $i < n$, either $(x[i], i) \in L$, or some $(\alpha, \alpha') \in \Phi$ covers $i$.

We call $|L| + |\Phi|$ its *size*, and the least size for $x$ the *arithmetic repetition complexity* $\mathrm{ARC}(x)$.

The zip reads round-robin and stops at the first missing symbol: $\mathrm{zip}(abc, AB, 1234) = aA1bB2c$.

**Definition 3.3** (Zip [6, eqs. (7), (8)]). For words $u_0, ..., u_{d-1} \in \Sigma^*$, $d \geq 1$, let $\ell := \min_i |u_i|$ and let $j$ be the least index with $|u_j| = \ell$. We define $\mathrm{zip}(u_0, ..., u_{d-1})$ as the word of length $d\ell + j$ whose symbol at position $i < d\ell + j$ is $u_{i \bmod d}[\lfloor i/d \rfloor]$.

**Definition 3.4** (ZLP, edges, values, size, ZLC, sZLC [6, Defs. 4.5, 4.6, 4.8]). A triple $P = (Q, q_0, \delta)$ of a finite set $Q$ of *vertices*, a *root* $q_0 \in Q$ and a map $\delta : Q \to (Q \sqcup \Sigma)^*$ is called a *zipline program* (ZLP) if every vertex $q$ has *arity* $d_q := |\delta(q)| \geq 2$, the multigraph with an edge $q \to \delta(q)[j]$ for each $j < d_q$ is acyclic, and $q_0$ is the only vertex that occurs in no list $\delta(q)$. We call the pair $(q, j)$ with $q \in Q$ and $j < d_q$ an *edge* of $P$, or edge $j$ of $q$, with *index* $j$; it *leads to* $\delta(q)[j]$, which we call the *child* on it. We define the *values* $\mathrm{val}(a) := a$ for $a \in \Sigma$, $\mathrm{val}(q) := \mathrm{zip}\big(\mathrm{val}(\delta(q)[0]), ..., \mathrm{val}\big(\delta(q)\big[d_q - 1\big]\big)\big)$ and $\mathrm{val}(P) := \mathrm{val}(q_0)$. The *size* $|P| := \sum_q d_q$ is the number of edges. A ZLP is called *synchronous* if there is a map $\varphi : Q \to \mathbb{N}$ with $\varphi(q_0) = 1$ and $\varphi(q') = d_q \varphi(q)$ for all vertices $q, q' \in Q$ such that $q'$ occurs in $\delta(q)$. We define the *zipline complexity* $\mathrm{ZLC}(x)$ of a word $x$ as the least size of a ZLP $P$ with $x \sqsubseteq \mathrm{val}(P)$, and $\mathrm{sZLC}(x)$ likewise over synchronous $P$.

**Definition 3.5** (Paths, product, root paths, height, depth). Let $z$ be a vertex. We call a sequence $\pi = (j_1, ..., j_k)$ $(k \geq 0)$ of edge indices a *path* from $z$ if $w_0 := z$ and $w_i := \delta(w_{i-1})[j_i]$ are vertices for all $i \leq k$; its edges are $(w_{i-1}, j_i)$. Its *product* is $p(\pi) := d_{w_0} \cdots d_{w_{k-1}}$ (1 for $k = 0$), and we call $w_k$ its *end*. We call a path from $q_0$ a *root path*. A path $\pi$ from $z$, followed by an edge of its end that leads to a letter, is called a *path from $z$ to a letter*; its edges are those of $\pi$ and this edge. The *height* of $z$ is the largest number of edges of such a path, and the *depth* of $P$ is the height of $q_0$. Every vertex is the end of a root path.

Let $P = (Q, q_0, \delta)$ be a ZLP. Each root path to a vertex $q$ determines an infinite arithmetic progression $\rho : \mathbb{N} \to \mathbb{N}$, an *occurrence* of $q$ (Figure 1), whose step is the product of the arities on the path: whenever $\rho(t) < |\operatorname{val}(P)|$, position $\rho(t)$ of $\operatorname{val}(P)$ holds $\operatorname{val}(q)[t]$.

**Definition 3.6** (Occurrences, step, offset, child occurrences, descendants [6, App. E.3])**.** The sets $\operatorname{occ}_q$ of *occurrences* of the vertices $q \in Q$, maps $\mathbb{N} \to \mathbb{N}$, are defined recursively along the acyclic graph of $P$: $\operatorname{occ}_{q_0} := \{t \mapsto t\}$, and for a vertex $q' \neq q_0$,

$$\operatorname{occ}_{q'} := \left\{t \mapsto \rho\big(d_q t + j\big) : q \in Q, \rho \in \operatorname{occ}_q, j < d_q, \delta(q)[j] = q'\right\}.$$

Each occurrence of $q$ is built along a root path with end $q$, unique by Lemma 3.7 (b) below; we call it the *root path* of the occurrence. By induction along the root path, an occurrence of $q$ has the form $\rho(t) = st + i$, where $s$ is the product of its root path and $0 \leq i < s$; we call $s$ its *step* and $i$ its *offset*. The *child occurrence* of $\rho \in \operatorname{occ}_q$ through an edge $j$ to a vertex $q'$ is $\rho_j : t \mapsto \rho\big(d_q t + j\big)$, which belongs to $\operatorname{occ}_{q'}$. An occurrence $\rho'$ is a *descendant* of $\rho$ if its root path extends that of $\rho$, allowing equality.

**Lemma 3.7** ([6, App. E.3]; proof in Appendix A.2)**.**

*(a) If $x \sqsubseteq \operatorname{val}(P)$, $\rho \in \operatorname{occ}_q$ and $\rho(t) < |x|$, then $t < |\operatorname{val}(q)|$ and $x[\rho(t)] = \operatorname{val}(q)[t]$.*

*(b) Distinct root paths to the same vertex give occurrences with disjoint images.*

*Proof idea.* (a) By induction along the root path. (b) Two distinct root paths to $q$ part at a vertex $w$, through edges $j \neq j'$. The occurrence $\eta$ of $w$ that they share is injective, and the two continuations have images in $\eta(j + d_w\mathbb{N})$ and $\eta(j' + d_w\mathbb{N})$, which are disjoint. ◻

In Figure 1, the vertex $q$ has the three occurrences $2t+1$, $4t$ and $4t+2$, of two different steps. Both bounds rest on this comparison of two such progressions (Section 4, Section 5):

**Lemma 3.8** (Which progression is lower)**.** *Let $s > s^*$, $i \geq 0$ and $0 \leq i^* < s^*$ be integers, and let $t \in \mathbb{N}$ be an index at which the progression $st + i$ is* lower *than $s^*t + i^*$, that is, $st + i < s^*t + i^*$. Then*

*(a) $(s - s^*)t < i^* - i < s^*$;*

*(b) $t < s^*/(s - s^*)$;*

*(c) $t < \sqrt{n/(s - s^*)}$ for every $n$ with $st + i < n$.*

*Proof.* Rearranging gives $(s - s^*)t < i^* - i$, and $i^* - i \leq i^* < s^*$; this is (a), and (b) follows. By (a), $(s - s^*)t^2 \leq s^*t \leq st + i < n$, which gives (c). ◻

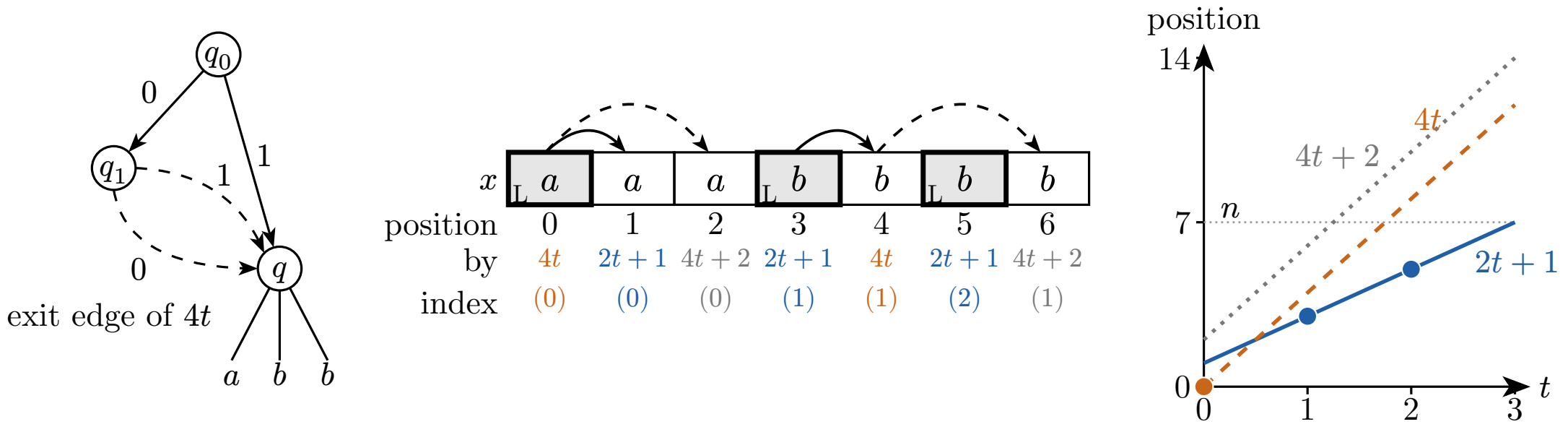

Figure 1: A small program (Remark A.3) with value $x = aaabbbb$.
*Left:* the program; between vertices, the solid edges form the tree $\mathcal{T}$ of least paths (Lemma 4.6).
*Middle:* under each position of $x$, the occurrence $st + i$ of $q$ that writes it and the index $t$; shaded (L): literals; arcs: the three equations of the conversion with one piece per lowest interval, the dashed pair one equation (Remark A.4).
*Right:* the occurrences $4t$ (dashed), $2t + 1$ (solid, the least) and $4t + 2$ (dotted) of $q$, with lowest intervals (Definition 4.2) $\{0\}$, $\{1, 2\}$ and $\varnothing$; dots: the lowest positions 0, 3, 5 at the indices 0, 1, 2.

# 4. The Upper Bound

Throughout this section, $P = (Q, q_0, \delta)$ is a ZLP with $x \sqsubseteq \mathrm{val}(P)$, $n := |x| \geq 1$, $m := |P|$, $d_0 := d_{q_0}$ and $N := n/d_0$; we build an ARS for $x$ of size $O\big(n^{1/4}m\big)$. By the cost of a part of the construction we mean the number of literals and equations it adds to the ARS.

**Copying.** A ZLP can be small because one vertex can be reached from the root along many paths, each placing a prefix of its value in $x$ along an occurrence $\rho(t) = st + i$ (Lemma 3.7). As $\rho - \eta$ is affine for two occurrences $\rho$, $\eta$ of one vertex, the indices with $\eta(t) < \rho(t) < n$ form an initial or a final segment of $\{t \in \mathbb{N} : \rho(t) < n\}$, and one equation $x[\eta(t)] = x[\rho(t)]$ copies $\rho$ from $\eta$ on all of it.

**The recursion.** At an index $t$, the occurrences $\rho$ of a vertex map $t$ to different positions (Lemma 3.7 (b)); we call the one with the smallest $\rho(t)$ *lowest* at $t$, and *lowest within the word* if moreover $\rho(t) < n$. At such an index no lower occurrence of the vertex can be copied from, but, writing $t = dt' + j$ with $j < d$ for the arity $d$ of the vertex, the child through edge $j$ writes the same position at its index $t'$; we descend until a letter (a literal) or a vertex with a lower occurrence (an equation).

**Least occurrences.** As $\rho - \eta$ is affine, the indices where an occurrence $\rho$ is lowest within the word form an interval, its *lowest interval.* Treating it as one *piece*, at a cost of up to a literal or two equations for each of the $d$ edges of a vertex (Lemma 4.4), with $O\big(n^{1/3}\big)$ non-empty lowest intervals per vertex, gives $O\big(n^{1/3}m\big)$, the order of Kosoy's bound [6, Prop. 4.7, Lemma E.2]. To do better we need to know which occurrences are lowest where. We organise the construction around the *least* occurrence of each vertex: smallest step, then smallest offset (Definition 4.5); it eventually lies below all other occurrences. Their root paths, the *least paths*, form a spanning tree of $P$: every prefix of a least path is itself a least path (Lemma 4.6).

**Exit edges.** The root path of every non-least occurrence leaves the tree at a first edge, its *exit edge* $e$ (Definition 4.7). We build the ARS vertex by vertex, but we add up its cost by exit edge: a vertex can have too many occurrences to count directly, whereas those with exit edge $e$ share their root path up to $e$, so they all continue one occurrence $\zeta_e$ of the vertex that $e$ leads to, the *occurrence through* $e$, each along its own path $\pi$ from there. As $\zeta_e$ is not least (Lemma A.1 (a)), the least occurrence of its vertex has a smaller step or lies below it everywhere, so if $\zeta_e$ is lowest within the word at an index $u$, then $u < \sqrt{n}$ (Lemma 3.8 (c)). If the occurrence along $\pi$ is lowest within the word at $t$, then $\zeta_e$ writes the same position at an index $u \geq p(\pi)t$ and is lowest within the word there, with $p(\pi)$ the product of the arities on $\pi$ (Lemma 4.6 (a)), so $p(\pi)t < \sqrt{n}$.

**The threshold.** The occurrences with exit edge $e$ can be many, and a lowest interval as one piece costs up to $2d$ at a vertex of arity $d$, however short it is. But only one occurrence is lowest at each index, and, as $p(\pi)t < \sqrt{n}$, few are lowest at large indices. So we fix an integer $\mu \geq 1$ and split each lowest interval of a vertex of arity $d$ at index $\mu d$. Below it we pay per index: each index $dt' + j$ is its own piece, which descends only through edge $j$ and costs at most 1; these are at most $\mu d$ indices, so $\mu$ per edge of the vertex, however many occurrences share them. From $\mu d$ on we pay per occurrence: the rest of the interval, if any, is one *late piece* of cost up to $2d$. For the least occurrences, one per vertex, this is at most 2 per edge. Among the occurrences with exit edge $e$, the one along $\pi$, ending at a vertex of arity $d$, has a late piece only if $p(\pi)\mu d < \sqrt{n}$, so few have one, and we show that together they cost less than $4\sqrt{n}/\mu$ (Section 4.3). With at most $m$ exit edges, this gives size $O\big((\mu + \sqrt{n}/\mu)m\big)$, least at $\mu \approx n^{1/4}$ (Theorem 4.1).

As non-root steps are multiples of $d_0$, in fact $u < \sqrt{N}$ (Lemma 4.8), and in terms of $N$ we prove:

**Theorem 4.1** (proof in Appendix A.5)**.** $\mathrm{ARC}(x) \le 2^{5/4} N^{1/4}\, m$.

*Proof of Theorem 1.1.* Take a ZLP $P$ with $x \sqsubseteq \mathrm{val}(P)$ and $|P| = \mathrm{ZLC}(x)$. As $d_0 \ge 2$, Theorem 4.1 gives $\mathrm{ARC}(x) \le 2^{5/4}(n/2)^{1/4}\,\mathrm{ZLC}(x) = 2n^{1/4}\,\mathrm{ZLC}(x)$. □

## 4.1. Lowest Intervals and the Conversion

**Definition 4.2** (First positions, lowest intervals, reach, $n_q$)**.** For a vertex $q$ and $t \in \mathbb{N}$ put $F_q(t) := \min_{\rho \in \mathrm{occ}_q} \rho(t)$. If $F_q(t) < n$, we call it the *first position* of $(q, t)$. An occurrence $\rho \in \mathrm{occ}_q$ is called *lowest* at $t$ if $\rho(t) = F_q(t)$. The *lowest interval* of $\rho \in \mathrm{occ}_q$ is $I_\rho := \{t \in \mathbb{N} : \rho(t) = F_q(t) < n\}$, and $\rho$ is called *lowest within the word* at the indices $t \in I_\rho$; its *reach* is $R_\rho := \max(I_\rho \cup \{0\})$. Put $n_q := |\{t : F_q(t) < n\}|$.

Each $I_\rho$ is an interval (possibly empty), as for another occurrence $\eta$ the indices with $\rho(t) < \eta(t)$ form an initial or a final segment, and the non-empty ones partition $[n_q]$ (Appendix A.3).

**Definition 4.3** (Meeting an edge, comparison sets)**.** Let $\rho \in \mathrm{occ}_q$, let $J \subseteq I_\rho$ be an interval and $(q, j)$ an edge; $J$ *meets* edge $j$ if $d_q t + j \in J$ for some $t \in \mathbb{N}$. If $(q, j)$ leads to a vertex $q'$, put $U := \{t : d_q t + j \in J\}$ and $\chi(t) := \rho(d_q t + j)$, in $\mathrm{occ}_{q'}$; for $\eta \in \mathrm{occ}_{q'}$, the *comparison set* of $\eta$ is $U_\eta := \{t \in U : \eta(t) < \chi(t)\}$.

**Lemma 4.4** (Conversion; proof in Appendix A.3)**.** *Partition every lowest interval into intervals, its* pieces*. There is an ARS for $x$ consisting of the following, for each piece $J$ of the lowest interval of an occurrence $\rho$ of $q$ and each edge $j$ of $q$ that $J$ meets:*

- *if $j$ leads to a letter $a$ and $j \in J$: the literal $(a, F_q(j))$ (as $\mathrm{val}(q)[j] = a$);*
- *if $j$ leads to a vertex, with $U$ and the comparison sets of Definition 4.3 for $\rho$, $J$ and $j$:*
  - *no equation if all comparison sets are empty;*
  - *at most one if all non-empty comparison sets are initial segments of $U$ (a set equal to $U$ counts as initial), in particular if $U$ has one element;*
  - *at most two in any case.*

*Proof idea.* Occurrences differ by affine maps, so the comparison sets are initial or final segments of $U$, and those of each kind are nested: for the $\eta$ with the longest set of each kind, the equation $x[\chi(t)] = x[\eta(t)]$, $t \in U_\eta$, copies the positions $\chi(t)$ in $x$ from the earlier positions $\eta(t)$, and so copies all that any of them can; so two suffice. If there are only initial segments, the longest alone suffices, and if all sets are empty, there is nothing to copy. By the recursion at the start of this section, these equations and the literals cover $x$. □

## 4.2. Least Occurrences and Exit Edges

**Definition 4.5** (Precedence, least occurrences and paths, tree edges, $m_{\mathcal{T}}$, $m_\Sigma$)**.** An occurrence of $q$ of step $s$ and offset $i$ *precedes* one of step $s'$ and offset $i'$ if $s < s'$, or $s = s'$ and $i < i'$. An occurrence is determined by its pair $(s, i)$, so the finite set $\mathrm{occ}_q$ has a least element $\rho_q^*$; we call it the *least occurrence* of $q$, and its root path the *least path* of $q$. We call the last edge of the least path of a vertex $q \ne q_0$ a *tree edge*. Let $\mathcal{T}$ be the set of tree edges, $m_{\mathcal{T}} := |\mathcal{T}|$ and $m_\Sigma$ the number of edges to letters.

**Lemma 4.6** (Least paths; proof in Appendix A.4)**.**

(a) *Let* $\rho \in \mathrm{occ}_q$*, let* $\zeta \in \mathrm{occ}_z$ *be the occurrence along a prefix of its root path, ending at* $z$*, and let* $\pi$ *be the rest of the path (from* $z$ *to* $q$*). Then there is an integer* $b$ *with* $0 \le b < p(\pi)$ *such that* $\rho(t) = \zeta(p(\pi)t + b)$ *for all* $t$*, and* $\zeta$ *is lowest at* $p(\pi)t + b$ *whenever* $\rho$ *is lowest at* $t$*.*

(b) *Every prefix of a least path is a least path. Hence* $\mathcal{T}$ *is a spanning tree of* $P$ *(Figure 1), so* $m_{\mathcal{T}} = |Q| - 1$*, and a root path all of whose edges lie in* $\mathcal{T}$ *is the least path of its end.*

**Definition 4.7** (Exit edges, $\zeta_e$)**.** Let $\rho \in \mathrm{occ}_q$. If $\rho \ne \rho_q^*$, its *exit edge* is the first edge of its root path outside $\mathcal{T}$ (it exists by Lemma 4.6 (b)). For an edge $e = (w, j) \notin \mathcal{T}$ that leads to a vertex, the *occurrence through* $e$ is the child occurrence $\zeta_e$ of $\rho_w^*$ through $j$, and its *reach* is $R_e := R_{\zeta_e}$.

For every integer $k \ge 1$, the reaches of $k$ distinct occurrences of one vertex that are lowest within the word somewhere but not least sum to less than $\sqrt{kN}$:

**Lemma 4.8** (Reaches)**.** *Let* $z \ne q_0$*, let* $\xi_1, ..., \xi_k$ $(k \ge 1)$ *be distinct members of* $\{\rho \in \mathrm{occ}_z : I_\rho \ne \varnothing, \rho \ne \rho_z^*\}$*, and let* $r := R_{\xi_1} + ... + R_{\xi_k}$*. Then* $r(r+2) < kN$*.*

*Proof.* All steps of occurrences of $z$ are multiples of $d_0$, as $z \ne q_0$; let $d_0\sigma$ be the least, the step of $\rho_z^*$, and write $\rho_z^*$ as $t \mapsto S_0 t + \iota_0$. Of two occurrences with equal steps one lies below the other everywhere (Lemma 3.7 (b)), so the $\xi_i$, each lowest somewhere and other than $\rho_z^*$, have distinct steps above $S_0$; ordered by step, write them as $t \mapsto S_l t + \iota_l$ $(1 \le l \le k)$. The reach $R_l$ of the $l$-th is the last index at which it is lowest within the word, so there it lies strictly below its predecessor ($\rho_z^*$ for $l = 1$; Lemma 3.7 (b)). As consecutive steps differ by at least $d_0$, Lemma 3.8 (a) gives

$$S_l R_l + \iota_l < S_{l-1} R_l + \iota_{l-1}, \quad \text{so} \quad d_0 R_l \le (S_l - S_{l-1}) R_l < \iota_{l-1} - \iota_l.$$

Summing over $l = 1, ..., k$, the offsets telescope:

$$d_0 r = d_0(R_1 + ... + R_k) < \iota_0 - \iota_k \le \iota_0 < d_0\sigma,$$

so $r \le \sigma - 1$, as reaches are integers. Each $\xi_i$ also has step at least $d_0(\sigma + 1)$ and lies below $n$ at its reach, so $(\sigma + 1)R_l < N$. Summing, $r(r+2) \le r(\sigma + 1) < kN$. □

## 4.3. Early and Late Pieces

**Definition 4.9** (Early and late pieces, cost, $C_{\mathrm{late}(\rho)}$)**.** Let $\mu \ge 1$ be an integer. The *early* pieces of $\rho \in \mathrm{occ}_q$ are the singletons $\{t\}$ with $t \in I_\rho$ and $t < \mu d_q$; the other indices of $I_\rho$, if any, form its *late* piece. The *cost* of a piece $J$ is the number of literals and equations that Lemma 4.4 gives for $J$. For an occurrence $\rho$, $C_{\mathrm{late}(\rho)}$ is the total cost of the late pieces of the descendants of $\rho$.

By Lemma 4.4, $\mathrm{ARC}(x)$ is at most the total cost of the pieces.

**Lemma 4.10** (Early pieces; proof in Appendix A.5)**.** *The early pieces cost at most* $\mu(m - m_{\mathcal{T}})$*.*

*Proof idea.* A vertex $q$ has at most $\min(n_q, \mu d_q)$ early singletons, each of cost at most 1. If, for each $q \ne q_0$, $\min(n_q, \mu)$ early singletons of vertices above $q$ cost nothing, distinct for distinct pairs $(q, t)$, then the total is at most $\mu d_0 + \sum_{q \ne q_0} \mu(d_q - 1) = \mu(m - m_{\mathcal{T}})$, as $\min(n_q, \mu d_q) - \min(n_q, \mu) \le \mu(d_q - 1)$ and each $q \ne q_0$ has one tree edge. For $t < \min(n_q, \mu)$, let $(w, j)$ be the last edge of the root path of the occurrence $\rho$ of $q$ lowest at $t$: the singleton $\{d_w t + j\}$ of the occurrence of $w$ before that edge is early (Lemma 4.6 (a)) and meets only edge $j$, where all comparison sets are empty, as $\rho$ is lowest at $t$, so it costs nothing (Lemma 4.4). These singletons are distinct, as $w$ and $d_w t + j$ determine $j$ and $t$, and so $q$. □

**Lemma 4.11** (Late pieces of least occurrences). *The late pieces of the least occurrences cost at most* $2(m - m_\Sigma)$.

*Proof.* The late piece of the least occurrence of a vertex $q$ carries no literal, as an edge index $j < d_q \le \mu d_q$ lies in no late piece, and costs at most two equations per edge to a vertex (Lemma 4.4). □

**Lemma 4.12** (Late pieces of descendants; proof in Appendix A.5). *Let $\rho$ be an occurrence of a vertex $q$. Then $C_{\mathrm{late}(\rho)} = 0$ if $R_\rho < \mu d_q$, and $C_{\mathrm{late}(\rho)} \le 4\big(R_\rho/\mu - 1\big)$ otherwise.*

*Proof idea.* By induction on the height of $q$: the descendants of $\rho$ other than $\rho$ split among its child occurrences $\rho_j$, whose reaches shrink by the arity, $d_q R_{\rho_j} \le R_\rho$ (Lemma 4.6 (a)). Below the threshold, $R_\rho < \mu d_q$, $\rho$ has no late piece and no child reaches its own threshold. At or above it, the late piece of $\rho$ costs at most $2d_q$ (Lemma 4.4), and each of the at most $d_q$ children costs 0 or at most $4\big(R_{\rho_j}/\mu - 1\big) \le 4\big(R_\rho/(d_q\mu) - 1\big)$ by induction and this inequality; so

$$C_{\mathrm{late}(\rho)} \le 2d_q + d_q \cdot 4\left(\frac{R_\rho}{d_q\mu} - 1\right) \le 4\left(\frac{R_\rho}{\mu} - 1\right) \quad \big(d_q \ge 2\big).$$

□

**Lemma 4.13** (Adding up; proof in Appendix A.5). *Let $\mu \ge 1$ be an integer and $\lambda \ge \mu + 2$ real. If $N \le \mu^2\lambda(\lambda - \mu)/4$, then* $\mathrm{ARC}(x) \le \lambda m$.

*Proof idea.* Split the pieces into the early ones, the late ones of the least occurrences and the other late ones (Lemma A.1). Let $\mathcal{E}$ be the set of exit edges. By Lemma 4.10 and Lemma 4.11,

$$\mathrm{ARC}(x) \le \mu(m - m_{\mathcal{T}}) + 2(m - m_\Sigma) + \sum_{e\in\mathcal{E}} C_{\mathrm{late}(\zeta_e)} \le (\mu + 2)m - \mu m_{\mathcal{T}} + \sum_{e\in\mathcal{E}} C_{\mathrm{late}(\zeta_e)}.$$

So it suffices that $\sum_{e\in\mathcal{E}} C_{\mathrm{late}(\zeta_e)} \le (\lambda - \mu - 2)m + \mu m_{\mathcal{T}}$. By Lemma 4.12, $C_{\mathrm{late}(\zeta_e)} = 0$ if $R_e < 2\mu$ and $C_{\mathrm{late}(\zeta_e)} \le 4(R_e/\mu - 1)$ otherwise. The reaches of $k \ge 1$ edges with $R_e \ge 2\mu$ into one vertex sum to some $r$ with $r(r+2) < kN$ (Lemma 4.8), so by the AM-GM inequality

$$\frac{4r}{\mu} < \frac{4\sqrt{kN}}{\mu} = 2\sqrt{(\lambda-\mu)k \cdot \frac{4N}{\mu^2(\lambda-\mu)}} \le (\lambda-\mu)k + \frac{4N}{\mu^2(\lambda-\mu)} \le (\lambda-\mu)k + \lambda,$$

so these edges add at most $4r/\mu - 4k \le (\lambda - \mu - 4)k + \lambda \le (\lambda - \mu - 2)(k+1) + \mu$, as $k \ge 1$. Summing over the vertices, each with its tree edge, gives $\sum_{e\in\mathcal{E}} C_{\mathrm{late}(\zeta_e)} \le (\lambda - \mu - 2)m + \mu m_{\mathcal{T}}$, as $\lambda \ge \mu + 2$ for the remaining edges. □

*Proof idea of Theorem 4.1.* With $\lambda := 2\mu \ge 4$, Lemma 4.13 needs only $2N \le \mu^4$, so $\mu \approx (2N)^{1/4}$ gives $\mathrm{ARC}(x) \le \lambda m \approx 2(2N)^{1/4} m$. For the exact bound we round $\mu$ up and keep $\lambda = 2(2N)^{1/4}$; the condition $\lambda \ge \mu + 2$ then holds for $N > 24$. For $N \le 24$, where a suitable integer $\mu$ need not exist, a short separate count gives the same bound. □

## 5. The Lower Bound

We build a program for a word $x$ with many lowest intervals of occurrences that are not least (the costly ones in the upper bound) and fill the intervals with random bits that cannot be copied in bulk. To fill them, edges we call *slots* lead to words chosen at random from a small *pool* of random words, and each position of $x$ shows a bit of its slot's word or 0. Each slot shows $H+1$ *new* bits: bits of its word that no earlier position shows through it.

**No bulk copying.** An equation copies along two arithmetic progressions; cut where its new bits first involve $u$ distinct slots, it is fixed by the start and the step of each of its two progressions, so at most $n^4$ candidates matter. One that copies new bits of $u$ distinct slots holds only with probability about $2^{-u}$, so, with $u$ of order $\log n$, with high probability none holds. In our construction, with high probability every progression through $u^2$ new bits of the slots we select (a constant fraction) shows new bits of $u$ distinct slots; so each equation copies fewer than $u^2 = O(\log^2 n)$ of these, and $\mathrm{ARC}(x) \geq H\,\mathrm{ZLC}(x)/O(\log^2 n)$.

## 5.1. Slot Words and the Spread-Set Bound

We first show, for a class of random words, that new bits cannot be copied in bulk.

**Definition 5.1** (Slot words). Let $D \geq 1$ be an integer and $\Omega$ a finite set, whose elements we call *slots*. The *pool* is a family of $D$ infinite words $Y_a = Y_{a,0}Y_{a,1}\cdots$ ($a \in [D]$) whose *bits* $Y_{a,r}$ are independent and uniform; each slot $\omega$ gets a uniform *pointer* $\beta(\omega) \in [D]$, independently of each other and of the bits. A *slot word* $x \in \{0,1\}^n$ with *pool size* $D$ is given by a fixed partial map $\psi$ from $[n]$ to $\Omega \times \mathbb{N}$, its *layout*:

$$x[i] := \begin{cases} Y_{\beta(\omega),r} & \text{if } \psi(i) = (\omega, r), \\ 0 & \text{if } \psi(i) \text{ is undefined.} \end{cases}$$

For $A \subseteq [n]$, the slots of which $A$ shows some bit first form

$$\Omega_{\mathrm{new}}(A) := \{\omega \in \Omega : \exists i \in A, r \in \mathbb{N} : \psi(i) = (\omega, r) \text{ and } \forall j < i : \psi(j) \neq (\omega, r)\}.$$

**Lemma 5.2** (Copying new bits; proof in Appendix B). *Let $u \geq 2$ be an integer, $x$ a slot word with pool size $u^2$, and $j_0 < i_0, \ldots, j_{u-1} < i_{u-1}$ positions of $x$ with $|\Omega_{\mathrm{new}}(\{i_0, \ldots, i_{u-1}\})| = u$. Then* $\Pr[\forall l \in [u] : x[j_l] = x[i_l]] \leq \exp(6)2^{-u}$.

*Proof idea.* Without pointer collisions, the $d_l := x[i_l] \oplus x[j_l]$ are independent uniform bits, since each contains a pool bit first shown at $i_l$, so all vanish with probability $2^{-u}$. Collisions among $u^2$ pool words cost the factor $\exp(6)$. □

**Definition 5.3** (Spread sets). For a slot word and an integer $u \geq 2$, a set $S$ of its positions is *spread* if $|\Omega_{\mathrm{new}}(A)| \geq u$ for every arithmetic progression $A \subseteq [n]$ with $|A \cap S| \geq u^2$.

**Definition 5.4** (Forward equations). Let $x \in \Sigma^n$. A *forward equation* in $x$ is a pair $(\alpha, \alpha')$ of arithmetic progressions with $|\alpha| = |\alpha'|$, $x[\alpha] = x[\alpha']$ and $\alpha(t) < \alpha'(t) < n$ for all $t < |\alpha|$, as an equation of an ARS for $x$ may be (Definition 3.2); its *targets* are the positions $\alpha'(t)$.

**Lemma 5.5** (Target count). *Let $x \in \Sigma^n$, $S \subseteq [n]$ and $h \geq 1$ an integer. If every forward equation in $x$ has at most $h$ targets in $S$, then* $\mathrm{ARC}(x) \geq |S|\,/h$.

*Proof.* Let $(L, \Phi)$ be an ARS for $x$ of size $\mathrm{ARC}(x)$. By Lemma B.1, the positions that an equation of $\Phi$ covers are the targets of a forward equation in $x$, so at most $h$ of them lie in $S$. As every position is given by $L$ or covered (Definition 3.2 (c)), $|S| \leq |L| + h\,|\Phi| \leq h\,\mathrm{ARC}(x)$. □

**Proposition 5.6** (Spread-set bound). *Let $u \geq 2$ be an integer, $x$ a slot word of length $n$ with pool size $u^2$, and $S$ a spread set of its positions. With probability at least* $1 - \exp(6)n^4 2^{-u}$, *every forward equation in $x$ has fewer than $u^2$ targets in $S$.*

*Proof.* For $p \in S$ and $\tau' \in [n] \setminus \{0\}$ such that $|\{s : p + \tau' s \in S\}| \geq u^2$, let $t_*$ be the least $t$ with $|\{s \leq t : p + \tau' s \in S\}| = u^2$. For $\sigma \in [n]$ and $\tau \in [n] \setminus \{0\}$, the pair of the progressions $\alpha(t) \coloneqq \sigma + \tau t$ and $\alpha'(t) \coloneqq p + \tau' t$ $(t \le t_*)$ is a *candidate* if $\forall t \le t_* : \alpha(t) < \alpha'(t)$; it *holds* if $\forall t \le t_* : x[\alpha(t)] = x[\alpha'(t)]$. There are at most $|S|\ n^3 \le n^4$ candidates, and they depend on the layout and $S$ only.

1. *With probability at least* $1 - \exp(6)n^4 2^{-u}$*, no candidate holds.*
   Fix a candidate. By the choice of $t_*$, its target set $A \coloneqq \alpha'([t_* + 1]) \subseteq [n]$ is an arithmetic progression with $|A \cap S| = u^2$, so $|\Omega_{\text{new}}(A)| \ge u$ as $S$ is spread. Let $\omega_0, ..., \omega_{u-1}$ be the $u$ least slots of $\Omega_{\text{new}}(A)$ (in a fixed order of $\Omega$) and $t_l \le t_*$ the least $t$ such that $\alpha'(t)$ shows a bit of $\omega_l$ first. Then $i_l \coloneqq \alpha'(t_l)$ shows a bit of $\omega_l$ first, and as each position shows at most one slot, $|\Omega_{\text{new}}(\{i_0, ..., i_{u-1}\})| = u$. With $j_l \coloneqq \alpha(t_l) < i_l$, Lemma 5.2 gives $\Pr[\text{the candidate holds}] \le \Pr[\forall l \in [u] : x[j_l] = x[i_l]] \le \exp(6)2^{-u}$, and a union bound over the at most $n^4$ candidates gives the claim.
2. *If no candidate holds, every forward equation in $x$ has fewer than $u^2$ targets in $S$.*
   Suppose a forward equation $(\alpha, \alpha')$ in $x$ has $u^2$ targets in $S$. Let $t_0$ be the least index with $p \coloneqq \alpha'(t_0) \in S$, and $\sigma \coloneqq \alpha(t_0)$. The steps $\tau, \tau'$ of $\alpha, \alpha'$ are below $n$, as each has two positions below $n$. So $t \mapsto (\alpha(t_0 + t), \alpha'(t_0 + t))$ for $t \le t_*$, with $t_*$ as above for $p$ and $\tau'$, is a candidate that holds. □

## 5.2. The Relay Words

Our construction, the *relay word* $x_k$ (Definition C.8), is a random binary word of length $n = 2^{\Theta(k^2 \log k)}$; it has a large spread set:

**Theorem 5.7** (Large spread sets; proof in Appendix C.4)**.** *For every $k \ge 2$, the relay word $x_k$ with suitable runner offsets (random ones work with high probability) is a slot word of length $n$ with pool size $u^2$ and a set $\Omega$ of slots, such that*

*(a)* $5 \log n \le u = O(\log n)$ *and* $\log n = \Theta(k^2 \log k)$*;*
*(b)* $x_k$ *has a spread set* $S$ *with* $|S| \ge n^{1/4 - O(1/k)}\ |\Omega|$*;*
*(c)* $\text{ZLC}(x_k) \le 2\ |\Omega|$ *for every outcome of the pointers and the bits.*

**Corollary 5.8** (Relay words; explicit constants in Theorem C.20)**.** *For every $k \ge 2$, with probability $1 - O(1/n)$, the relay word $x_k$ satisfies* $\text{ARC}(x_k) \ge n^{1/4 - O(1/k)}\, \text{ZLC}(x_k)$*.*

*Proof.* Theorem 5.7, Proposition 5.6 and Lemma 5.5 $(h = u^2 - 1)$ give $\text{ARC}(x_k) \ge |S|\ /u^2 \ge n^{1/4 - O(1/k)}\, \text{ZLC}(x_k)/(2u^2)$ with probability $1 - \exp(6)n^4 2^{-u} \ge 1 - \exp(6)/n$, as $2^u \ge n^5$, and $2u^2 = n^{O(1/k)}$. □

*Proof idea of Theorem 1.2.* Corollary 5.8 gives words with $\text{ARC}(x) \ge n^{1/4 - O(1/k)}\, \text{ZLC}(x)$ for every $k$, so $\text{ARC}(x) \ge n^{1/4 - o(1)}\, \text{ZLC}(x)$ for infinitely many $n$; prefixes of longer variants of $x_k$ (same $k$) cover the lengths in between at a loss of the same order (Appendix C.4). The explicit loss in Theorem 1.2 comes from the best $k$ for a given $n$, the largest $k$ with $|x_k| \le n$: as $\log n = \Theta(k^2 \log k)$, $k = \Theta\Big(\sqrt{\log n / \log \log n}\Big)$, so $n^{O(1/k)} = \exp(O(\sqrt{\log n \log \log n}))$. □

By Theorem 1.1, for $\lambda > 0$ no word with $\text{ARC}(x) \ge \lambda\, \text{ZLC}(x)$ is shorter than $\lambda^4/16$: the bounds meet up to $n^{o(1)}$.

*Proof idea of Theorem 5.7.* **The construction.** To build the intervals, we give vertices we call *columns* exactly $H$ occurrences each, one through each of $H$ special vertices we call *runners*. We divide the indices of a column's value into consecutive *rounds* of equal length, and most indices of round $h$ lie in the lowest interval of the occurrence through runner $h$; all but one of these $H$ occurrences are not least. To fill these intervals, each edge of a column, a *slot*, leads to a word chosen at random from a small *pool* of random words. The column zips these words: round $h$ of its value consists of bit $h$ of each slot's word. Each index of the column's value appears first in $x$ where its occurrence is lowest, so bit $h$ of a slot appears first through runner $h$ (for most slots). Therefore most of each lowest interval shows new bits. Above the runners, a tree gives each runner its own step and, through random edge choices, a random offset. Below them, *shift gadgets* lead to every column: small subprograms that zip columns and zero words in a cyclically shifted order, which multiplies the step by a number we choose. The program has at most two edges per slot, which gives (c).

**Spread.** We take for $S$ the new bits of the *inner* slots (Definition C.10), which lie away from the ends of their column and of the list of columns; they are a constant fraction of all slots. If a progression $A$ has $|A \cap S| \geq u^2$ but $|\Omega_{\text{new}}(A)| < u$, then by pigeonhole it contains new bits of one inner slot through at least $u+1$ different runners. For a large step, the random offsets of the runners make this unlikely (Lemma C.18). For a small step, $A$ also runs through the whole round of one of these bits; there the columns zip the new bits of many slots at even spacing from it, so $A$, with its small step, contains one in every few of them: new bits of at least $u$ slots (Lemma C.19).

**The length.** An occurrence with step $s$ puts index $t$ of a column at a position at least $st$, and a column has $H+1$ rounds of $E = H^{o(1)}$ indices, so our words have length about $sH$ for the largest step $s$. Where the lowest occurrence of a column passes from one runner to the next (a *handover*), the occurrence with the larger step must stay lower than the one with the smaller step up to a late index. That needs a head start (the index times the difference of the steps), and a head start is less than the smaller step. The steps through the $H$ runners are distinct, and the shift gadgets multiply them by $H^{1+o(1)}$, so the steps through a middle and the last runner differ by $H^{2+o(1)}$. A middle runner must stay lower than the last one up to index about $EH/2$, which needs a head start of $H^{3+o(1)}$, so we take steps of $H^{3+o(1)}$. Hence $n = H^{4+o(1)}$, and each inner slot puts $H = n^{1/4-o(1)}$ new bits into $S$, which gives (b). □

## 6. Open Problems

(a) What is the order of $\max\{\text{ARC}(x)/\text{ZLC}(x) : x \in \{0,1\}^n\}$: $\Theta(n^{1/4})$, or smaller by an unbounded factor $n^{o(1)}$, such as $\exp(\Theta(\sqrt{\log n \log\log n}))$, our loss for the relay words?

(b) More generally, determine $\max\{\text{ARC}(x) : x \in \{0,1\}^n, \text{ZLC}(x) = m\}$; the relay words have $m = n^{1/2+o(1)}$ (Remark C.21), and (a) asks only for the largest ratio to $m$.

(c) Is $\text{ARC}(x) \leq C\,\text{ZLC}(x)^c$ for all $x$, with $C, c$ depending only on $\Sigma$ (then ARC-MUWU makes $O(\text{ZLC}(x)^c \log n)$ mistakes, $n \geq 2$), or is ARC superpolynomial in ZLC?

(d) Determine the optimal exponent when ZLC is restricted to programs of each fixed depth (Definition 3.5).

(e) The relay words are random. Can words $x$ with $\text{ARC}(x) \geq n^{1/4-o(1)}\,\text{ZLC}(x)$ be computed deterministically in time polynomial in their length $n$?

# A. Proofs for Sections 1–4

We prove Corollary 1.3, the deferred lemmas of Section 3 and Section 4, and Theorem 4.1.

An ARS determines $x$: by induction on $i$, $x[i]$ is given by a literal or equals an earlier symbol of $x$ (Definition 3.2 (b), (c)). The $n$ literals $(x[i], i)$ form an ARS, so $\mathrm{ARC}(x) \le n$.

## A.1. The Alphabet

**Corollary 1.3** (Optimal exponent, restated)**.** *Let $\Sigma$ be a finite alphabet with at least two letters. The least exponent $c$ such that for some constant $C$, $\mathrm{ARC}(x) \le Cn^c \,\mathrm{ZLC}(x)$ for all $n \ge 1$ and $x \in \Sigma^n$ is $c = 1/4$.*

*Proof.* By Theorem 1.1, the inequality holds with $c = 1/4$ and $C = 2$. Let $c < 1/4$ and $C > 0$, and fix $a$ with $c < a < 1/4$. Without loss of generality $0, 1 \in \Sigma$; then the binary words of Theorem 1.2 are words over $\Sigma$, with the same ARC (an ARS for $x$ over either alphabet has only literals $(x[i], i)$, so it is one over the other) and at most the same ZLC (a ZLP over $\{0, 1\}$ is one over $\Sigma$). By Theorem 1.2, for every large enough $n$ some binary word $x$ of length $n$ has $\mathrm{ARC}(x) \ge n^a \,\mathrm{ZLC}(x)$, hence also over $\Sigma$; take such an $n$ with $n^{a-c} > C$. As $\mathrm{ZLC}(x) \ge 1$,

$$\mathrm{ARC}(x) \ge n^a \,\mathrm{ZLC}(x) = n^{a-c} \cdot n^c \,\mathrm{ZLC}(x) > Cn^c \,\mathrm{ZLC}(x),$$

so $x \in \Sigma^n$ violates the inequality. □

## A.2. Occurrences

**Lemma 3.7** ([6, App. E.3], restated)**.**

*(a) If $x \sqsubseteq \mathrm{val}(P)$, $\rho \in \mathrm{occ}_q$ and $\rho(t) < |x|$, then $t < |\mathrm{val}(q)|$ and $x[\rho(t)] = \mathrm{val}(q)[t]$.*

*(b) Distinct root paths to the same vertex give occurrences with disjoint images.*

*Proof.* (a) Let $n := |x|$. By induction along the root path. We show, for every vertex $q$ and every $\rho \in \mathrm{occ}_q$,

$$\forall t \in \mathbb{N}: \quad \rho(t) < n \Rightarrow t < |\mathrm{val}(q)| \text{ and } x[\rho(t)] = \mathrm{val}(q)[t].$$

For the root, $\rho(t) = t$, so $t < n \le |\mathrm{val}(P)| = |\mathrm{val}(q_0)|$ and $x[t] = \mathrm{val}(P)[t]$, as $x \sqsubseteq \mathrm{val}(P)$. Let $\rho' \in \mathrm{occ}_{q'}$ be the child occurrence of $\rho \in \mathrm{occ}_q$ through edge $j$ of $q$, which leads to $q'$, so that $\rho'(t) = \rho(t')$ with $t' := d_q t + j$ (Definition 3.6), and suppose $\rho'(t) < n$. By induction, $t' < |\mathrm{val}(q)|$ and $x[\rho'(t)] = \mathrm{val}(q)[t']$. Let $\ell$ be the least length of the values of the children of $q$ and $\hat{\jmath}$ the least index of a child with this length, so that $|\mathrm{val}(q)| = d_q \ell + \hat{\jmath}$ (Definition 3.3). As $j, \hat{\jmath} < d_q$, $d_q t + j < d_q \ell + \hat{\jmath}$ gives $t < \ell$, or $t = \ell$ and $j < \hat{\jmath}$; in both cases $t < |\mathrm{val}(q')|$, as $|\mathrm{val}(q')| \ge \ell$, with equality only if $j \ge \hat{\jmath}$. So Definition 3.3 gives $\mathrm{val}(q)[t'] = \mathrm{val}(q')[t]$, and $x[\rho'(t)] = \mathrm{val}(q')[t]$.

(b) Let $\pi \neq \pi'$ be root paths to $q$. Neither extends the other: otherwise the extra edges would form a path from $q$ to $q$ with at least one edge, a cycle in the acyclic graph of $P$. So they agree up to a vertex $w$, with a common occurrence $\eta$ of $w$, and then leave $w$ through edges $j \neq j'$. By induction on the number of further edges, every occurrence obtained by extending this common prefix through edge $j$ of $w$ has the form

$$t \mapsto \eta(d_w f(t) + j) \quad \text{for a map } f : \mathbb{N} \to \mathbb{N};$$

after edge $j$, $f(t) = t$, and following an edge $j''$ of a vertex $v$ replaces $f$ by $t \mapsto f(d_v t + j'')$ (Definition 3.6). So its image lies in $\eta(j + d_w\mathbb{N})$, and these sets are disjoint for $j \neq j'$, as $j, j' < d_w$ and $\eta$ is injective. ☐

## A.3. The Conversion

By Lemma 3.7 (b), the minimum $F_q(t)$ is attained by exactly one occurrence. So $I_\rho = \{t : \rho(t) < n \text{ and } \rho(t) < \eta(t) \text{ for all } \eta \in \mathrm{occ}_q \setminus \{\rho\}\}$. Each condition defines an interval of integers, as $\rho$ is increasing and $\eta - \rho$ is affine; hence $I_\rho$ is an interval. As a minimum of finitely many increasing maps, $F_q$ is increasing, so the non-empty lowest intervals of $q$ partition the initial interval $[n_q]$.

**Lemma 4.4** (Conversion, restated)**.** *Partition every lowest interval into intervals, its* pieces*. There is an ARS for $x$ consisting of the following, for each piece $J$ of the lowest interval of an occurrence $\rho$ of $q$ and each edge $j$ of $q$ that $J$ meets:*

- *if $j$ leads to a letter $a$ and $j \in J$: the literal $(a, F_q(j))$ (as $\mathrm{val}(q)[j] = a$);*
- *if $j$ leads to a vertex, with $U$ and the comparison sets of Definition 4.3 for $\rho$, $J$ and $j$:*
  - *no equation if all comparison sets are empty;*
  - *at most one if all non-empty comparison sets are initial segments of $U$ (a set equal to $U$ counts as initial), in particular if $U$ has one element;*
  - *at most two in any case.*

*Proof.* Step 1 shows that the literals hold, Step 2 that each piece and edge adds at most two equations that hold, and Step 3, by induction on the height, that every position is covered or given by a literal.

1. *For every edge $j$ of a vertex $q$ with $\delta(q)[j] = a \in \Sigma$ and $F_q(j) < n$, the literal $(a, F_q(j))$ holds, that is, $x[F_q(j)] = a$.*
   Let $\rho \in \mathrm{occ}_q$ be lowest at $j$, so $\rho(j) = F_q(j) < n$. By Lemma 3.7 (a), $j < |\mathrm{val}(q)|$ and $x[F_q(j)] = \mathrm{val}(q)[j]$. By Definition 3.3, $\mathrm{val}(q)[j] = \mathrm{val}(a)[0] = a$, as $j = d_q \cdot 0 + j$ with $j < d_q$.
2. *Let $J$ be a piece of $\rho \in \mathrm{occ}_q$ and $(q, j)$ an edge with $\delta(q)[j] = q'$ and $U \neq \varnothing$. The equations $(\eta_i, \chi)$ restricted to $U_{\eta_i}$, for the longest non-empty initial comparison set $U_{\eta_1}$ and the longest non-empty final one $U_{\eta_2} \neq U$ (if they exist), hold in $x$ and cover $\chi(t)$ for every $t$ in a comparison set; there is none if all comparison sets are empty, at most one if all non-empty ones are initial segments of $U$, and at most two in any case.*
   As $J$ is an interval and $t \mapsto d_q t + j$ is increasing, $U$ is an interval, and as $J \subseteq I_\rho$,
   $$\chi(t) = \rho\big(d_q t + j\big) = F_q\big(d_q t + j\big) < n \quad \text{for all } t \in U.$$
   For $\eta \in \mathrm{occ}_{q'}$, $U_\eta = \{t \in U : (\chi - \eta)(t) > 0\}$ with $\chi - \eta$ affine, so $U_\eta$ is empty, an initial segment of $U$ or a final segment of $U$. The initial ones are nested, and so are the final ones; a final one equal to $U$ is also initial. So, with an absent $U_{\eta_i}$ read as $\varnothing$,
   $$\bigcup_{\eta \in \mathrm{occ}_{q'}} U_\eta = U_{\eta_1} \cup U_{\eta_2}.$$

For each $i \in \{1,2\}$ for which $U_{\eta_i}$ exists, let $t_i := \min U_{\eta_i}$; the equation $(\eta_i, \chi)$ restricted to $U_{\eta_i}$ and reindexed from 0 is the pair $(\alpha_i, \alpha_i')$ with

$$\alpha_i(s) := \eta_i(t_i + s), \quad \alpha_i'(s) := \chi(t_i + s), \quad s < |U_{\eta_i}|.$$

For $t \in U_{\eta_i}$, $\eta_i(t) < \chi(t) < n$, so by Lemma 3.7 (a)

$$x[\eta_i(t)] = \mathrm{val}(q')[t] = x[\chi(t)].$$

So $|\alpha_i| = |\alpha_i'|$ and $x[\alpha_i] = x[\alpha_i']$, as all these positions lie below $n$: the equation holds, and it covers $\alpha_i'(s) = \chi(t_i + s)$, as $\alpha_i(s) < \alpha_i'(s)$. So no equation is added if all comparison sets are empty, and at most one if all non-empty ones are initial. If $|U| = 1$, every non-empty comparison set is $U$, and one equation suffices.

3. *For every vertex $q$ and every $t'$ with $F_q(t') < n$, the position $F_q(t')$ is covered or given by a literal; in particular every position of $[n]$ is.*

   By induction on the height of $q$ (Definition 3.5); a child of $q$ has smaller height. As $F_{q_0}(t') = t'$, the case $q = q_0$ covers $[n]$. Let $\rho \in \mathrm{occ}_q$ be lowest at $t'$, so $t' \in I_\rho$ and $\rho(t') = F_q(t') < n$; let $J$ be its piece containing $t'$, and write $t' = d_q t + j$ with $j < d_q$.

   *Case $\delta(q)[j] = a \in \Sigma$.* As $\rho(t') < n$, Lemma 3.7 (a) and Definition 3.3 give $t' < |\mathrm{val}(q)|$ and $\mathrm{val}(q)[t'] = \mathrm{val}(a)[t]$. As $|\mathrm{val}(a)| = 1$, $t = 0$, so $t' = j \in J$, and the literal $\big(a, F_q(j)\big)$ of Step 1, which the statement lists for $J$ and $j$, gives $F_q(t') = \rho(t')$.

   *Case $\delta(q)[j] = q' \in Q$.* Then $t \in U$, so $U \neq \varnothing$, and $\chi \in \mathrm{occ}_{q'}$ gives

$$F_{q'}(t) \leq \chi(t) = \rho(t') = F_q(t') < n.$$

   If $F_{q'}(t) < \chi(t)$, the occurrence $\eta$ lowest at $t$ has $t \in U_\eta$, and Step 2 covers $\chi(t)$. Otherwise $\chi(t) = F_{q'}(t) < n$ is covered or given by a literal by induction.

So the literals of Step 1 and the equations of Step 2 form the claimed ARS for $x$: each listed literal $\big(a, F_q(j)\big)$ has $j \in J \subseteq I_\rho$, so $F_q(j) < n$, and it holds (Step 1, Definition 3.2 (a)); each equation has $|\alpha_i| = |\alpha_i'|$ and holds (Step 2, Definition 3.2 (b)); and every $i < n$ is covered or given by a literal $(a, i)$, which has $a = x[i]$ (Step 3, Definition 3.2 (c)). □

## A.4. Least Occurrences and Exit Edges

**Lemma 4.6** (Least paths, restated)**.**

*(a) Let $\rho \in \mathrm{occ}_q$, let $\zeta \in \mathrm{occ}_z$ be the occurrence along a prefix of its root path, ending at $z$, and let $\pi$ be the rest of the path (from $z$ to $q$). Then there is an integer $b$ with $0 \leq b < p(\pi)$ such that $\rho(t) = \zeta(p(\pi)t + b)$ for all $t$, and $\zeta$ is lowest at $p(\pi)t + b$ whenever $\rho$ is lowest at $t$.*

*(b) Every prefix of a least path is a least path. Hence $\mathcal{T}$ is a spanning tree of $P$ (Figure 1), so $m_{\mathcal{T}} = |Q| - 1$, and a root path all of whose edges lie in $\mathcal{T}$ is the least path of its end.*

*Proof.* (a) Let $\pi = (j_1, \ldots, j_k)$, with vertices $w_0 := z, \ldots, w_k := q$ (Definition 3.5), and put $p_i := d_{w_0} \cdots d_{w_{i-1}}$, so $p_0 = 1$ and $p_k = p(\pi)$. For $\eta \in \mathrm{occ}_z$, following $\pi$ from $\eta$ gives the occurrences $\eta^{(0)} := \eta$ and $\eta^{(i)}(t) := \eta^{(i-1)}\big(d_{w_{i-1}} t + j_i\big)$ of $w_i$ (Definition 3.6). By induction on $i$,

$$\eta^{(i)}(t) = \eta(p_i t + b_i), \quad b_i := \sum_{i'=1}^{i} p_{i'-1} j_{i'}, \quad 0 \leq b_i < p_i :$$

the step is $\eta^{(i)}(t) = \eta\Big(p_{i-1}\big(d_{w_{i-1}} t + j_i\big) + b_{i-1}\Big) = \eta(p_i t + b_i)$, and $b_i = b_{i-1} + p_{i-1} j_i \leq (p_{i-1} - 1) + p_{i-1}\big(d_{w_{i-1}} - 1\big) = p_i - 1$, as $j_i < d_{w_{i-1}}$. So with $b := b_k$, which depends on $\pi$ only, following $\pi$ from any $\eta \in \mathrm{occ}_z$ gives the occurrence $t \mapsto \eta(p(\pi)t + b)$ of $q$; for $\eta = \zeta$ it is $\rho$. Let $\rho$ be lowest at $t$ and put $\tilde{t} := p(\pi)t + b$. For every $\eta \in \mathrm{occ}_z$, the occurrence $t \mapsto \eta(p(\pi)t + b)$ of $q$ is not below $\rho$ at $t$, so

$$\eta(\tilde{t}) \geq \rho(t) = \zeta(\tilde{t});$$

hence $\zeta$ is lowest at $\tilde{t}$.

(b) We prove the three claims of (b) in turn.

1. *For every vertex $q$, every prefix of the least path of $q$ is the least path of its end.*
   Split the least path of $q$ into a prefix ending at $z$ and a path $\pi$ from $z$, let $\zeta(t) = St + \iota$ be the occurrence of $z$ along the prefix, and let $p := p(\pi)$ and $b$ be as in (a), so that $\rho_q^*(t) = \zeta(pt + b) = Spt + (Sb + \iota)$. Suppose that some $\eta \in \mathrm{occ}_z$, $\eta(t) = S't + \iota'$, precedes $\zeta$: $S' < S$, or $S' = S$ and $\iota' < \iota$. By (a), following $\pi$ from $\eta$ gives the occurrence of $q$
   $$t \mapsto \eta(pt + b) = S'pt + (S'b + \iota'), \quad \text{of step } S'p \text{ and offset } S'b + \iota' < S'p,$$
   as $b < p$ and $\iota' < S'$. If $S' < S$, then $S'p < Sp$; if $S' = S$ and $\iota' < \iota$, then $S'b + \iota' < Sb + \iota$. Either way it precedes $\rho_q^*$, which contradicts the choice of $\rho_q^*$. So no occurrence of $z$ precedes $\zeta$, that is, $\zeta = \rho_z^*$, and as the root path of an occurrence is unique (Lemma 3.7 (b)), the prefix is the least path of $z$.
2. *$\mathcal{T}$ is a spanning tree of $P$, and $m_{\mathcal{T}} = |Q| - 1$.*
   Every vertex $q \neq q_0$ is the end of a root path (Definition 3.5), so $\mathrm{occ}_q \neq \varnothing$ and $q$ has a least path; it has an edge, as $q \neq q_0$, and its last edge is a tree edge (Definition 4.5). This edge leads to $q$, so distinct vertices have distinct tree edges, and $m_{\mathcal{T}} = |Q| - 1$. By Step 1, the least path of $q \neq q_0$ is the least path of the vertex where its last edge starts, followed by that edge; by induction on the length, every edge of a least path is a tree edge, so $\mathcal{T}$ connects every vertex to $q_0$. Being connected on $Q$ with $|Q| - 1$ edges, the $m_{\mathcal{T}}$ tree edges form a spanning tree. Of the $m - m_\Sigma$ edges that lead to a vertex, $m_{\mathcal{T}}$ are tree edges.
3. *A root path all of whose edges lie in $\mathcal{T}$ is the least path of its end.*
   By induction on the number of edges. The empty path ends at $q_0$, and it is the root path of the only occurrence of $q_0$. Let $\pi = \pi_0 e$ with $e = (w, j) \in \mathcal{T}$ leading to $q$; by induction, $\pi_0$ is the least path of $w$. As $e$ is a tree edge into $q$, it is the last edge of the least path of $q$, and by Step 1 the part before it is the least path of $w$, that is, $\pi_0$. So $\pi$ is the least path of $q$. □

**Lemma A.1** (Exit edges)**.**

*(a) For a vertex $z$, the occurrences $\zeta_e$ of the edges $e \notin \mathcal{T}$ into $z$ differ from each other and from $\rho_z^*$, and $\zeta_e$ has exit edge $e$.*

*(b) Every occurrence of a vertex $q$ other than $\rho_q^*$ is a descendant of $\zeta_e$ for exactly one $e \notin \mathcal{T}$, its exit edge. No least occurrence is a descendant of any $\zeta_e$.*

*Proof.* For every root path $\pi$,

$$\pi \text{ is the least path of its end} \quad \Leftrightarrow \quad \text{every edge of } \pi \text{ lies in } \mathcal{T}. \quad (*)$$

If $\pi$ is a least path, each of its edges is the last edge of a prefix of $\pi$, which is the least path of a vertex other than $q_0$ (Lemma 4.6 (b)), so it is a tree edge; the converse is Lemma 4.6 (b).

(a) Let $e = (w, j) \notin \mathcal{T}$ lead to $z$. The root path of $\zeta_e$ is the least path of $w$ followed by $e$ (Definition 4.7, Definition 3.6), and that of $\rho_z^*$ ends with the tree edge of $z$ ($z \neq q_0$, as $z$ occurs in $\delta(w)$). So for distinct edges $e \notin \mathcal{T}$ into $z$, these root paths end with pairwise different edges, and the occurrences are distinct (Lemma 3.7 (b)). By $(*)$, all edges of the least path of $w$ lie in $\mathcal{T}$, and $e \notin \mathcal{T}$, so $e$ is the first edge of the root path of $\zeta_e \neq \rho_z^*$ outside $\mathcal{T}$, its exit edge.

(b) Let $\rho \in \mathrm{occ}_q$, $\rho \neq \rho_q^*$, have root path $\pi$. As an occurrence is determined by its root path, $\pi$ is not the least path of $q$, so by $(*)$ some edge of $\pi$ lies outside $\mathcal{T}$; let $e = (w, j)$ be the first one, the exit edge of $\rho$; it leads to a vertex, as every edge of a root path does (Definition 3.5). The part of $\pi$ before $e$ has all its edges in $\mathcal{T}$, so by $(*)$ it is the least path of $w$; hence $\pi$ extends the root path of $\zeta_e$, and $\rho$ is a descendant of $\zeta_e$. Conversely, the root path of a descendant of $\zeta_{e'}$, $e' = (w', j') \notin \mathcal{T}$, begins with the least path of $w'$, all of whose edges lie in $\mathcal{T}$ by $(*)$, followed by $e'$; so $e'$ is its first edge outside $\mathcal{T}$. Hence $e' = e$ for $\rho$, and, by $(*)$, no least occurrence is a descendant of any $\zeta_{e'}$. □

## A.5. Early and Late Pieces

**Lemma 4.10** (Early pieces, restated). *The early pieces cost at most $\mu(m - m_{\mathcal{T}})$.*

*Proof.* For each pair $(q, t)$ with $q \neq q_0$ and $t < \min(n_q, \mu)$, we find a free early singleton (one that costs nothing) at a vertex with an edge that leads to $q$, and subtract these from the count of all early singletons.

1. *Every early piece costs at most 1. For every vertex $q$, the early pieces of the occurrences of $q$ are the $\{t'\}$ with $t' < \min(n_q, \mu d_q)$, one for each such $t'$.*
   A singleton $\{t'\}$ at $q$ meets only edge $j := t' \bmod d_q$. If $j$ leads to a letter, Lemma 4.4 gives at most one literal for it; if $j$ leads to a vertex, $U = \{\lfloor t'/d_q \rfloor\}$ has one element, so at most one equation. As the non-empty lowest intervals of $q$ partition $[n_q]$ (Appendix A.3), every $t' < n_q$ lies in $I_\rho$ for exactly one $\rho \in \mathrm{occ}_q$, and no other $t'$ lies in any $I_\rho$.
2. *Let $q \neq q_0$ and $t < \min(n_q, \mu)$, let $\rho \in \mathrm{occ}_q$ be lowest at $t$, let $(w, j)$ be the last edge of its root path and $\eta \in \mathrm{occ}_w$ the occurrence along this path without that edge, and put $t' := d_w t + j$. Then $\{t'\}$ is an early piece of $\eta$ that costs nothing, and the map $(q, t) \mapsto (w, t')$ is injective.*
   As $t < n_q$, $\rho(t) = F_q(t) < n$. By Definition 3.6, $\rho(s) = \eta(d_w s + j)$ for all $s$, so by Lemma 4.6 (a) (with the path $(j)$ and $b = j$), $\eta$ is lowest at $t'$; and $\eta(t') = \rho(t) < n$, so $t' \in I_\eta$. As $t \leq \mu - 1$ and $j \leq d_w - 1$, $t' \leq \mu d_w - 1$, so $\{t'\}$ is an early piece of $\eta$. It meets only edge $j$, which leads to $q$, with $U = \{t\}$ and $\chi = \rho$ (Definition 4.3). For every $\zeta \in \mathrm{occ}_q$, $\zeta(t) \geq F_q(t) = \rho(t)$, so its comparison set $U_\zeta = \{s \in U : \zeta(s) < \rho(s)\}$ is empty, and Lemma 4.4 gives no literal and no equation. Finally $t'$ determines $j = t' \bmod d_w$ and $t = \lfloor t'/d_w \rfloor$, and $q = \delta(w)[j]$.

By Step 2 and the uniqueness in Step 1, the free singletons of distinct pairs $(q, t)$ are distinct early pieces. So, by Step 1, the early pieces cost at most

$$\sum_q \min(n_q, \mu d_q) - \sum_{q \neq q_0} \min(n_q, \mu) \leq \mu d_0 + \sum_{q \neq q_0} \mu(d_q - 1)$$
$$= \mu \sum_q d_q - \mu(|Q| - 1) = \mu(m - m_{\mathcal{T}}),$$

as $\min(a, \mu d) - \min(a, \mu) \leq \mu(d - 1)$, $m = \sum_q d_q$ and $|Q| - 1 = m_{\mathcal{T}}$ (Lemma 4.6 (b)). □

**Lemma 4.12** (Late pieces of descendants, restated). *Let $\rho$ be an occurrence of a vertex $q$. Then $C_{\mathrm{late}(\rho)} = 0$ if $R_\rho < \mu d_q$, and $C_{\mathrm{late}(\rho)} \leq 4(R_\rho/\mu - 1)$ otherwise.*

*Proof.* By induction on the height of $q$ (Definition 3.5): a child $q' = \delta(q)[j] \in Q$ has smaller height than $q$, as every path from $q'$ to a letter, preceded by the edge $(q, j)$, is one from $q$. Let $c$ be the cost of the late piece of $\rho$ ($c := 0$ if there is none); sums over $j$ run over the edges $j$ of $q$ that lead to a vertex.

1. *$C_{\mathrm{late}(\rho)} = c + \sum_j C_{\mathrm{late}(\rho_j)}$, and $d_q R_{\rho_j} \leq R_\rho$ for each such $j$.*

A descendant of $\rho$ other than $\rho$ continues the root path of $\rho$ by an edge $j$ of $q$ that leads to a vertex, so it is a descendant of the child occurrence $\rho_j$ (Definition 3.6) for exactly one such $j$; this gives the sum. If $s \in I_{\rho_j}$, then $\rho$ is lowest at $d_q s + j$ (Lemma 4.6 (a), with the path $(j)$ and $b = j$) and $\rho\big(d_q s + j\big) = \rho_j(s) < n$, so $d_q s + j \in I_\rho$ and $d_q s \le R_\rho$. Taking $s := R_{\rho_j}$ gives the bound if $I_{\rho_j} \neq \varnothing$; otherwise $R_{\rho_j} = 0$.

2. *If $R_\rho < \mu d_q$, then $C_{\mathrm{late}(\rho)} = 0$.*
   All $t \in I_\rho$ have $t \le R_\rho < \mu d_q$, so $\rho$ has no late piece and $c = 0$. For each $j$, Step 1 gives $R_{\rho_j} \le R_\rho / d_q < \mu \le \mu d_{q'}$, where $q' := \delta(q)[j]$; so $C_{\mathrm{late}(\rho_j)} = 0$ by induction.
3. *If $R_\rho \ge \mu d_q$, then $C_{\mathrm{late}(\rho)} \le 4\big(R_\rho/\mu - 1\big)$.*
   The late piece of $\rho$ carries no literal, as an edge index $j < d_q \le \mu d_q$ lies in no late piece, and it gets at most two equations for each of the at most $d_q$ edges $j$ (Lemma 4.4); so $c \le 2d_q$. By induction and Step 1, each $C_{\mathrm{late}(\rho_j)}$ is 0 or at most $4\big(R_{\rho_j}/\mu - 1\big) \le 4\big(R_\rho/\big(d_q\mu\big) - 1\big)$, which is $\ge 0$ as $R_\rho \ge \mu d_q$. So, by Step 1 and as $d_q \ge 2$,
$$C_{\mathrm{late}(\rho)} \le 2d_q + d_q \cdot 4\left(\frac{R_\rho}{d_q \mu} - 1\right) = \frac{4R_\rho}{\mu} - 2d_q \le 4\left(\frac{R_\rho}{\mu} - 1\right).$$
□

**Lemma 4.13** (Adding up, restated)**.** *Let $\mu \ge 1$ be an integer and $\lambda \ge \mu + 2$ real. If $N \le \mu^2 \lambda(\lambda - \mu)/4$, then $\mathrm{ARC}(x) \le \lambda m$.*

*Proof.* We show $\mathrm{ARC}(x) \le \lambda m$; by Lemma 4.4, it is at most the total cost of the pieces. Let $\mathcal{E}$ be the set of edges $e \notin \mathcal{T}$ that lead to a vertex, and for a vertex $z$ let $\mathcal{E}_z := \{e \in \mathcal{E} : e \text{ leads to } z, R_e \ge 2\mu\}$. Of the $m - m_\Sigma$ edges that lead to a vertex, $m_\mathcal{T}$ lie in $\mathcal{T}$ (Definition 4.5), so $|\mathcal{E}| = m - m_\Sigma - m_\mathcal{T}$.

1. *The pieces cost at most $\mu m_\Sigma + 2m_\mathcal{T} + \sum_{e\in\mathcal{E}}\big(\mu + 2 + C_{\mathrm{late}(\zeta_e)}\big)$, and for $e \in \mathcal{E}$, $C_{\mathrm{late}(\zeta_e)} = 0$ if $R_e < 2\mu$, and $C_{\mathrm{late}(\zeta_e)} \le 4(R_e/\mu - 1)$ otherwise.*
   By Lemma 4.10 and Lemma 4.11, the early pieces and the late pieces of the least occurrences cost at most
$$\mu(m - m_\mathcal{T}) + 2(m - m_\Sigma) = \mu m_\Sigma + 2m_\mathcal{T} + (\mu + 2)(m - m_\Sigma - m_\mathcal{T}),$$
   and $m - m_\Sigma - m_\mathcal{T} = |\mathcal{E}|$. By Lemma A.1, the other occurrences are the descendants of the $\zeta_e$, $e \in \mathcal{E}$, each a descendant of exactly one; so their late pieces cost $\sum_{e\in\mathcal{E}} C_{\mathrm{late}(\zeta_e)}$ (Definition 4.9). Let $e \in \mathcal{E}$ lead to $z$. By Lemma 4.12 with $\rho = \zeta_e \in \mathrm{occ}_z$, $C_{\mathrm{late}(\zeta_e)} = 0$ if $R_e < \mu d_z$, in particular if $R_e < 2\mu$, as $d_z \ge 2$; and otherwise $C_{\mathrm{late}(\zeta_e)} = 0$ or $C_{\mathrm{late}(\zeta_e)} \le 4(R_e/\mu - 1)$, which is $> 0$.
2. *Let $z$ be a vertex with $k := |\mathcal{E}_z| \ge 1$. Then $2 + \sum_{e\in\mathcal{E}_z}\big(\mu + 2 + C_{\mathrm{late}(\zeta_e)}\big) \le \lambda(1 + k)$.*
   As the edges of $\mathcal{E}_z$ lead to $z$, $z \neq q_0$ (Definition 3.4). Their $\zeta_e$ are distinct members of $\big\{\rho \in \mathrm{occ}_z : I_\rho \neq \varnothing, \rho \neq \rho^*_z\big\}$ (Lemma A.1; $R_e > 0$). So by Lemma 4.8, $r := \sum_{e\in\mathcal{E}_z} R_e$ has $r(r + 2) < kN$, hence $r < \sqrt{kN}$. By Step 1,
$$\begin{aligned} 2 + \sum_{e\in\mathcal{E}_z}\big(\mu + 2 + C_{\mathrm{late}(\zeta_e)}\big) &\le 2 + (\mu + 2)k + \frac{4r}{\mu} - 4k \le \mu k + \frac{4\sqrt{kN}}{\mu} \\ &\le \mu k + (\lambda - \mu)k + \frac{4N}{\mu^2(\lambda - \mu)} \le \lambda(1 + k), \end{aligned}$$
   the second as $2 - 2k \le 0$ and $r < \sqrt{kN}$, the third by the inequality of arithmetic and geometric means (as $\lambda - \mu \ge 2$), the fourth by the hypothesis $N \le \mu^2\lambda(\lambda - \mu)/4$.

Every $e \in \mathcal{E}$ leads to a vertex $z \neq q_0$, and exactly one tree edge leads to each $z \neq q_0$ (Lemma 4.6 (b)). Grouping the tree edge into $z$ with $\mathcal{E}_z$, Step 1 gives

$$\mathrm{ARC}(x) \le \mu m_\Sigma + \sum_{z \ne q_0} \left( 2 + \sum_{e \in \mathcal{E}_z} \left( \mu + 2 + C_{\mathrm{late}(\zeta_e)} \right) \right)$$
$$+ \sum_{e \in \mathcal{E},\, R_e < 2\mu} (\mu + 2) \le \lambda(m_\Sigma + m_\mathcal{T} + |\mathcal{E}|) = \lambda m,$$

as each term of the sum over $z$ is at most $\lambda(1 + |\mathcal{E}_z|)$ (Step 2 if $\mathcal{E}_z \ne \varnothing$, and $2 \le \lambda$ otherwise), and $\mu < \mu + 2 \le \lambda$. ☐

**Lemma A.2** (Small $N$). *Let $\mu \ge 1$ be an integer. If $N \le 4\mu(\mu + 1)$, then* $\mathrm{ARC}(x) \le (\mu + 2)m$*.*

*Proof.* Take $\mathcal{E}$ and $\mathcal{E}_z$ as in the proof of Lemma 4.13; its Step 1 does not use the hypothesis of Lemma 4.13. Every $\mathcal{E}_z$ is empty: if $k := |\mathcal{E}_z| \ge 1$, then, as in Step 2, $r := \sum_{e \in \mathcal{E}_z} R_e$ has $r(r + 2) < kN$, and $r \ge 2\mu k$, so

$$4\mu k(\mu k + 1) \le r(r + 2) < kN, \quad \text{hence} \quad N > 4\mu(\mu k + 1) \ge 4\mu(\mu + 1).$$

So $R_e < 2\mu$, hence $C_{\mathrm{late}(\zeta_e)} = 0$, for all $e \in \mathcal{E}$, and by Step 1

$$\mathrm{ARC}(x) \le \mu m_\Sigma + 2m_\mathcal{T} + (\mu + 2)(m - m_\Sigma - m_\mathcal{T}) \le (\mu + 2)m.$$ ☐

**Theorem 4.1** (restated). $\mathrm{ARC}(x) \le 2^{5/4} N^{1/4} m$*.*

*Proof.* Put $v := (2N)^{1/4}$, so that $2^{5/4} N^{1/4} = 2v$.

*Case* $N \le 24$. Let $\mu \in \{1, 2\}$ be least with $N \le 4\mu(\mu + 1)$. Lemma A.2 gives $\mathrm{ARC}(x) \le (\mu + 2)m$, and $\mathrm{ARC}(x) \le n = Nd_0 \le Nm$ (the $n$ literals form an ARS), as $d_0 \le m$. If $\mu = 2$, then $N > 8$, so $v > 2$ and $\mu + 2 < 2v$. If $\mu = 1$, then $\min(N, 3) \le 3^{3/4} N^{1/4} < 2v$, as $\min(a, b) \le a^{1/4} b^{3/4}$ and $27 < 32$. Either way, $\mathrm{ARC}(x) \le \min(N, \mu + 2)m \le 2vm$.

*Case* $N > 24$. Then $v^4 > 48 > 39.0625 = 2.5^4$, so $v > 2.5$. Put $\mu := \lceil v \rceil \ge 3$, $\lambda := 2v$ and $b := \mu/v$; then $1 \le b < 1 + 1/v < 1.4$. If $\mu = 3$, then $\mu + 2 = 5 < 2v$; otherwise $v > 3$ and $\mu + 2 < v + 3 < 2v$. The hypothesis $N \le \mu^2 \lambda(\lambda - \mu)/4$ of Lemma 4.13 holds, as

$$\mu^2 \lambda(\lambda - \mu) = 2b^2(2 - b)v^4 \ge 2v^4 = 4N,$$

since $b^2(2 - b) - 1 = (b - 1)(1 + b - b^2) \ge 0$ for $1 \le b \le 1.4$ ($1 + b - b^2$ decreases for $b \ge 1/2$ and is 0.44 at $b = 1.4$). So $\mathrm{ARC}(x) \le \lambda m = 2vm$. ☐

## A.6. Remarks

Every vertex is the end of a root path: as $q_0$ is the only vertex in no list, we can move from any other vertex to one in whose list it occurs, and as the graph is finite and acyclic, repeating this reaches $q_0$; reversed, the walk is a root path. By induction along root paths, a map $\varphi$ as in Definition 3.4 has $\varphi(w) = p(\pi)$ for every root path $\pi$ with end $w$; conversely, if all root paths with the same end have the same product, this product is such a $\varphi$. So a ZLP is synchronous iff all root paths with the same end have the same product, equivalently, all occurrences of each vertex have the same step (Definition 3.6).

**Remark A.3** (Occurrences of two steps). Let $\delta(q) = (a, b, b)$, $\delta(q_1) = (q, q)$ and $\delta(q_0) = (q_1, q)$ (Figure 1). Then $\mathrm{val}(q) = abb$, $\mathrm{val}(q_1) = \mathrm{zip}(abb, abb) = aabbbb$, and $\mathrm{val}(q_0) = \mathrm{zip}(aabbbb, abb) = aaabbbb$, of length $2 \cdot 3 + 1 = 7$, as the shortest child $q$ is on edge 1. Here $q$ has three occurrences: $2t + 1$ (edge 1 of the root), and $4t$ and $4t + 2$ (through $q_1$). ◁

**Remark A.4** (The example, continued). In the program of Remark A.3, $x = \mathrm{val}(P) = aaabbbb$, $n = 7$, $d_0 = 2$, $N = 7/2$, $m = 7$, $m_\Sigma = 3$ and $m_\mathcal{T} = 2$. As $q$ has occurrences of steps 2 and 4, this program is not synchronous. For a synchronous $P$, every $x \sqsubseteq \mathrm{val}(P)$ has an ARS of size at most $|P|$ [6, Prop. 4.9]. Thus, if $\mathrm{ARC}(x) > \mathrm{ZLC}(x)$, no ZLP of size

ZLC($x$) for $x$ is synchronous, so each has a vertex with occurrences of different steps. The least occurrence of $q$ is $2t+1$, and $\mathcal{T}$ consists of the two edges of the root, so the two edges of $q_1$ are the edges outside $\mathcal{T}$ that lead to a vertex. The occurrence $4t$ has the exit edge $(q_1, 0)$ and is $\zeta_e$ itself, for $e = (q_1, 0)$; its reach is $R_e = 0$, as $I_{4t} = \{0\}$. With every lowest interval as one piece, the conversion gives the literals $(a, 0)$, $(b, 3)$ and $(b, 5)$ at the first positions of $(q, 0)$, $(q, 1)$ and $(q, 2)$, and three equations (Figure 1, arcs): none at edge 0 of the root, where $q_1$ has one occurrence; at edge 1 of the root ($U = \{0, 1, 2\}$, $\chi = 2t+1$, comparison sets $\{0\}$ of $4t$ and $\varnothing$ of $4t+2$) $x[0] = x[1]$, covering position 1; at edge 0 of $q_1$ ($U = \{0, 1\}$, $\chi = 4t$, the only non-empty comparison set the final segment $\{1\}$ of $2t+1$) $x[3] = x[4]$, covering 4; and at edge 1 of $q_1$ ($U = \{0, 1\}$, $\chi = 4t+2$, the comparison sets of $2t+1$ and $4t$ both $U$) one equation, say $x[4t] = x[4t+2]$ for $t < 2$, covering 2 and 6. This ARS has size 6. ◁

**Remark A.5** (The constant)**.** A constant $C$ in an inequality $\mathrm{ARC}(x) \le Cn^{1/4}\,\mathrm{ZLC}(x)$ for all words is at least $2^{-1/4}$, already because of the word $aa$: ARC($aa$) = 2 (position 0 needs a literal, position 1 a literal or an equation; two literals suffice) and ZLC($aa$) = 2 (the ZLP with $\delta(q_0) = (a, a)$ has two edges, the least possible), so $2 \le C2^{1/4} \cdot 2$. ◁

**Remark A.6** (Arities that are powers of one integer)**.** If the arities of all non-root vertices of $P$ that have a vertex child are powers of one integer $b \ge 2$, then $\mathrm{ARC}(x) \le 2\,|P| - m_\Sigma \le 2\,|P|$. Indeed, every step of an occurrence of a vertex $q \ne q_0$ is then $d_0$ times a power of $b$, so an occurrence of $q$ of step $s$ and offset $i$ other than $\rho^*_q = s^*t + i^*$ has $s = s^*$ and $i > i^*$, and is never lowest, or $s \ge bs^* \ge 2s^*$, and then $st + i \ge s^*t + s^* > s^*t + i^*$ for $t \ge 1$. So $F_q(t) = \rho^*_q(t)$ for $t \ge 1$, and the non-empty lowest intervals of $q$ are at most that of $\rho^*_q$ and, if the occurrence of least offset is not $\rho^*_q$ and $F_q(0) < n$, the singleton $\{0\}$ of that occurrence. Apply Lemma 4.4 with each non-empty lowest interval as one piece. For an edge $(q, j)$ of the tree $\mathcal{T}$ leading to $q'$, $\rho^*_q(d_q t + j) = \rho^*_{q'}(t)$ by Lemma 4.6 (b), so every comparison set is contained in $\{0\}$ and a non-empty one is an initial segment: at most one equation. Each of the $|P| - m_\Sigma - m_\mathcal{T}$ other edges to vertices costs at most two equations for the piece of $\rho^*_q$. Each additional singleton $\{0\}$ of a non-least occurrence meets only edge 0 of its vertex: at most one equation if that edge leads to a vertex, none if it leads to a letter (its literal is among the at most $m_\Sigma$ literals). There is at most one such singleton for each vertex, and the root has only the occurrence $t \mapsto t$, so there are at most $|Q| - 1 = m_\mathcal{T}$ of them. In all $m_\Sigma + m_\mathcal{T} + 2(|P| - m_\Sigma - m_\mathcal{T}) + m_\mathcal{T} = 2\,|P| - m_\Sigma$. ◁

## B. Forward Equations and Copying New Bits

We prove the forward restriction used in the proof of Lemma 5.5 (Lemma B.1) and Lemma 5.2.

**Lemma B.1** (Forward restriction)**.** *Let $x \in \Sigma^n$ and let $\alpha$, $\alpha'$ be arithmetic progressions with $|\alpha| = |\alpha'|$ and $x[\alpha] = x[\alpha']$. Then $T := \{t \in [|\alpha|] : \alpha(t) < \alpha'(t) < n\}$ is an interval of integers; if it is not empty, $\alpha$ and $\alpha'$ restricted to $T$ and reindexed from 0 form a forward equation in $x$, whose targets $\alpha'(T)$ are the positions $i < n$ covered by $(\alpha, \alpha')$.*

*Proof.* Let $\alpha(t) = \sigma + \tau t$ and $\alpha'(t) = \sigma' + \tau' t$ for $t \in [\ell]$, $\ell := |\alpha| = |\alpha'|$, and $k_\alpha := |\{t \in [\ell] : \alpha(t) < n\}|$, $k_{\alpha'} := |\{t \in [\ell] : \alpha'(t) < n\}|$. As $\alpha$ and $\alpha'$ are increasing, these sets are $[k_\alpha]$ and $[k_{\alpha'}]$.

1. *$T = \{t \in [k_{\alpha'}] : (\sigma' - \sigma) + (\tau' - \tau)t > 0\}$ is an interval.*

The equality holds as $\alpha'(t) - \alpha(t) = (\sigma' - \sigma) + (\tau' - \tau)t$. The $t \in \mathbb{Z}$ where this affine function is positive form a half-line, $\mathbb{Z}$ or $\varnothing$, and its intersection with $[k_{\alpha'}]$ is an interval.

2. *$k_\alpha = k_{\alpha'}$ and $\forall s < k_{\alpha'} : x[\alpha(s)] = x[\alpha'(s)]$.*
   By Definition 3.1, $x[\alpha] = x[\alpha(0)]...x[\alpha(k_\alpha - 1)]$ is a word of length $k_\alpha$, and likewise for $\alpha'$. Equal words have equal lengths and equal letters.
3. *If $T \neq \varnothing$, let $t_0 := \min T$ and $\alpha_T(s) := \alpha(t_0 + s)$, $\alpha'_T(s) := \alpha'(t_0 + s)$ for $s \in [|T|]$. Then $(\alpha_T, \alpha'_T)$ is a forward equation in $x$ with targets $\alpha'(T)$.*
   By Step 1, $T = \{t_0 + s : s \in [|T|]\}$, so $\alpha_T, \alpha'_T$ are arithmetic progressions of length $|T|$ (steps $\tau, \tau'$) and $\alpha'_T([|T|]) = \alpha'(T)$. For $s \in [|T|]$, $t := t_0 + s \in T$ gives $\alpha_T(s) < \alpha'_T(s) < n$; so Definition 3.1 reads all $|T|$ positions of each, and $t \in [k_{\alpha'}]$ (Step 1) with Step 2 gives $x[\alpha_T(s)] = x[\alpha'_T(s)]$. Hence $x[\alpha_T] = x[\alpha'_T]$, and Definition 5.4 applies.
4. *$\{i \in [n] : (\alpha, \alpha')$ covers $i\} = \alpha'(T)$.*
   By Definition 3.2, $(\alpha, \alpha')$ covers $i$ iff $\exists t \in [\ell] : \alpha(t) < \alpha'(t) = i$. For $i < n$ such a $t$ lies in $T$; conversely, each $t \in T$ gives $i := \alpha'(t) < n$. □

**Lemma 5.2** (Copying new bits, restated)**.** *Let $u \geq 2$ be an integer, $x$ a slot word with pool size $u^2$, and $j_0 < i_0$, ..., $j_{u-1} < i_{u-1}$ positions of $x$ with $|\Omega_{\text{new}}(\{i_0, ..., i_{u-1}\})| = u$. Then* $\Pr[\forall l \in [u] : x[j_l] = x[i_l]] \leq \exp(6)2^{-u}$.

*Proof.* By hypothesis, $\psi(i_l) = (\omega_l, r_l)$ with distinct slots $\omega_l$, and $\psi(j) \neq \psi(i_l)$ for all $j < i_l$. List the at most $2u$ slots of $\Omega' := \{\omega \in \Omega : \psi(i_l)$ or $\psi(j_l)$ lies in $\{\omega\} \times \mathbb{N}$ for some $l \in [u]\}$ in a fixed order as $\omega'_0, \omega'_1, ...$; let $z_c := \mathbb{1}[\beta(\omega'_c) = \beta(\omega'_{c'})$ for some $c' < c]$ and $\nu := \sum_c z_c$. Call the indices in $I := \{l : \beta(\omega) \neq \beta(\omega_l)$ for all $\omega \in \Omega' \setminus \{\omega_l\}\}$ *isolated.* We show

$$\Pr[\forall l : x[j_l] = x[i_l]] = \mathbb{E}[\Pr[\forall l : x[j_l] = x[i_l] \mid \beta]] \leq \mathbb{E}\big[2^{-|I|}\big] \leq 2^{-u}\mathbb{E}[4^\nu] \leq \exp(6)2^{-u}.$$

1. $\Pr[\forall l : x[j_l] = x[i_l] \mid \beta] \leq 2^{-|I|}$.
   Condition on a realization of $\beta$; the pool bits stay independent and uniform. For $l \in I$ let $d_l := x[i_l] \oplus x[j_l]$; these bits are independent and uniform: renumber so that $I = [k]$ and $i_0 < ... < i_{k-1}$, and let $b_l := Y_{\beta(\omega_l), r_l}$; these are distinct pool bits, as isolated slots have distinct pointers. A position $p \in \{i_m, j_m : m \in [u]\}$ with $\psi(p) = (\omega, r)$ has $\omega \in \Omega'$ and $x[p] = Y_{\pi(p)}$, where $\pi(p) := (\beta(\omega), r)$; by isolation, $\pi(p) = \pi(i_l)$ iff $\psi(p) = \psi(i_l)$. So $x[i_l] = b_l$. If $\psi(j_l)$ is undefined, $x[j_l] = 0$; otherwise, for $m \in [k]$, $\pi(j_l) = \pi(i_m)$ implies $\psi(j_l) = \psi(i_m)$ by isolation, hence $i_m \leq j_l < i_l$ and $m < l$. Hence
   $$d_l = b_l \oplus f_l(b_0, ..., b_{l-1}, \text{pool bits other than } b_0, ..., b_{k-1})$$
   for functions $f_l$, so for fixed other pool bits $(b_l)_{l \in [k]} \mapsto (d_l)_{l \in [k]}$ is a bijection of $\{0,1\}^k$. The claim's probability is at most $\Pr[\forall l \in I : d_l = 0 \mid \beta] = 2^{-|I|}$.
2. $u - |I| \leq 2\nu$.
   Let $\Omega'_y := \{\omega \in \Omega' : \beta(\omega) = y\}$. Then $l$ is isolated iff $|\Omega'_{\beta(\omega_l)}| = 1$, and $\nu = \sum_y \max\big(|\Omega'_y| - 1, 0\big)$, since in each nonempty $\Omega'_y$ exactly the slot $\omega'_c$ with the least $c$ has $z_c = 0$. As the slots $\omega_l$ are distinct,
   $$u - |I| \leq \sum_{y:\, |\Omega'_y| \geq 2} |\Omega'_y| \leq \sum_{y:\, |\Omega'_y| \geq 2} 2\big(|\Omega'_y| - 1\big) = 2\nu.$$
3. $\mathbb{E}[4^\nu] \leq \exp(6)$.
   Let $\mathcal{F}_c$ be the pointers $\beta(\omega'_0), ..., \beta(\omega'_{c-1})$. Given $\mathcal{F}_c$, the pointer $\beta(\omega'_c)$ is uniform on $[u^2]$, and $z_c = 1$ iff it is one of these at most $c$ values; as $4^{z_c} = 1 + 3z_c$,
   $$\mathbb{E}[4^{z_c} \mid \mathcal{F}_c] = 1 + 3\Pr[z_c = 1 \mid \mathcal{F}_c] \leq 1 + 3c/u^2.$$

As $z_0, ..., z_{c-1}$ are functions of $\mathcal{F}_c$, induction on the number of factors, $|\Omega'| \le 2u$ and $1 + z \le \exp(z)$ give

$$\mathbb{E}[4^{\nu}] = \mathbb{E}\left[\prod_{c < |\Omega'|} 4^{z_c}\right] \le \prod_{c<2u}\left(1 + \frac{3c}{u^2}\right) \le \exp\left(\sum_{c<2u} \frac{3c}{u^2}\right) = \exp\left(\frac{3(2u-1)}{u}\right) \le \exp(6).$$

□

## C. Construction of the Relay Words

In Appendix C and Appendix D, ln is the natural logarithm. Appendix C.1 builds the random ZLP $\mathcal{P}_k$ and its word $x_{k,L}$ (Definition C.8): the runners give each column $H$ occurrences that take turns being lowest, and the slopes, free levels and shift gadgets fix where they lie; Appendix C.2 defines the critical set, and Appendix C.3 shows that it is spread with high probability.

### C.1. The Program

**Definition C.1** (Parameters, good integers, types). For integers $k \ge 2$ and $L \ge L_k := \lceil 512k \log(k+2) \rceil$ put

$$X := 2^L, \qquad B := X^{k+1}, \qquad W := B/8,$$
$$u := 10(4k+6)L, \qquad E := 128 \cdot 4^k u, \qquad M := 8EW^2,$$
$$g := \log B = (k+1)L, \qquad \Lambda := 16k!\binom{2k+2}{k}(4L)^k, \qquad H := \lceil B/\Lambda \rceil.$$

Let $\mathbb{P}$ be the set of primes in $[X, 4X)$. An integer $s \in [M, 2M)$ is called *good* if $s = \alpha_1 \cdots \alpha_k a$ with primes $\alpha_1 < \alpha_2 < ... < \alpha_k$ in $\mathbb{P}$ and an integer $a \ge 2$; we call the tuple $(\alpha_1, ..., \alpha_k)$ a *type* of $s$.

Table 1: The parameters for the specialization $L = L_k$, fixed by $k$ (Definition C.1, Definition C.3; bounds: Lemma D.1, Lemma C.15, Lemma C.13); $\Lambda := 16k!\binom{2k+2}{k}(4L)^k$ is the loss in the density of slopes (Lemma C.2).

| | | | |
|---|---|---|---|
| $L$ | $L = L_k := \lceil 512k \log(k+2) \rceil$, $X := 2^L$ | $B$ | $X^{k+1}$, the length of a gadget's list |
| $u$ | $u := 10(4k+6)L$, $D := u^2$ words | $E$ | $128 \cdot 4^k u$, the arity of a column |
| $H$ | $\lceil B/\Lambda \rceil \ge BX^{-1/32}$ runners | $R$ | $H/X^k \le R \le X/2$, root arity |
| $s_h$ | $\Psi_h a_h \in [M, M + B/8)$, $M := EB^2/8$ | $n$ | $RBs_{H-1}E(H+1) \le X^{4k+6}$ |
| | $10 \log n \le u \le 16 \log n$; $12u^2, \Lambda < X^{1/32}$ | | $n = \Theta(RE^2H^4\Lambda^3) = H^{4+O(1/k)}$ |

**Lemma C.2** (Dense slopes; proof in Appendix D). *Every window $[y, y+W) \subseteq [M, 2M)$ contains at least $B/\Lambda$ good integers.*

**Definition C.3** (Slopes and their types). Let the *slopes* $s_0 > s_1 > ... > s_{H-1}$ be the $H$ largest good integers in the window $[M, M+W)$ (there are at least $H$, by Lemma C.2, as their number is an integer), and fix for each $s_h$ its lexicographically least type $(\alpha_1(h), ..., \alpha_k(h))$, which we call the *type* of $h$; put $\Psi_h := \alpha_1(h) \cdots \alpha_k(h)$, $a_h := s_h/\Psi_h \ge 2$, $\Psi_* := \max_h \Psi_h$ and $\Delta_h := s_h - s_{h+1}$ for $h < H-1$.

As $s_0 > ... > s_{H-1}$ are integers, $\Delta_h \ge 1$ and $s_h \ge s_{H-1}$. Moreover $X^k < \Psi_h < (4X)^k$, as every $\alpha_i(h)$ is a prime in $(X, 4X)$, and so $\Psi_* < (4X)^k$.

**Definition C.4** (Runners, free levels, placement). Given vertices $w_0, ..., w_{H-1}$, the *runners*, with their types, we define the *free levels* as the following vertices. For $1 \le i \le k$ and each prefix $(\alpha_1, ..., \alpha_i)$ of length $i$ of the type of some runner, there is one vertex of level $i$, which carries this prefix and has arity $\alpha_i$. Its *real children* are, for $i < k$, the vertices of level $i+1$ whose prefix extends its own, and, for $i = k$, the runners of that type. Let $R$ be the number of level-1 vertices; a further vertex, the *root*, has them as its $R$ children.

We call the following random choices, all independent, the *placement*: the root assigns the level-1 vertices to its $R$ edges by a uniform bijection, and every free vertex assigns its real children to distinct random edges, by a uniform injection into its edge indices. An edge that gets no real child will lead to a zero word (Definition C.8). For a runner $w_h$, let $\pi_h$ be the index of the edge of the root that leads to its level-1 ancestor, and $j_i(h)$ the index of the edge of its level-$i$ ancestor that leads to its level-$(i+1)$ ancestor (to $w_h$, for $i = k$), and put $\zeta_h := \sum_{i=1}^{k} \alpha_1(h)\cdots\alpha_{i-1}(h)\, j_i(h)$.

**Definition C.5** (Shift gadget)**.** For a list $U_0, ..., U_{B-1}$ of vertices (not necessarily distinct) and $g = \log B$, let $V_{g,z} := U_z$ and, for $i = g-1, ..., 0$, let $V_{i,z}$ be a new vertex, distinct for distinct $(i, z)$ and distinct from the $U_\kappa$, with the two children $V_{i+1,z}$ and $V_{i+1,z+2^i}$, indices modulo $B$; put $G_z := V_{0,-z}$. These are $gB$ vertices of arity 2.

**Definition C.6** (Intercepts and edge indices)**.** Put

$$c_0 := 0, \quad c_{h+1} := c_h + B\Delta_h(E(h+1) - 1/2)\ (h < H-1), \quad \Gamma := \Psi_* \mathbb{Z} \cap [0, s_{H-1}/4).$$

For a placement, $h < H$ and $\gamma \in \Gamma$, let $\eta_h(\gamma) \in [1, \Psi_h]$ be the integer with $c_h + \gamma + \eta_h(\gamma) \equiv \zeta_h \pmod{\Psi_h}$, and $q_h(\gamma) := (c_h + \gamma + \eta_h(\gamma) - \zeta_h)/\Psi_h$.

**Lemma C.7** (Edge indices; proof in Appendix C.3)**.**

*(a) Every $c_h$ is an integer, and $c_h < Bs_{H-1}/8$ for all $h < H$.*

*(b) For every placement and every $h < H$: $q_h(\gamma) \in \mathbb{N}$ for all $\gamma \in \Gamma$, the residues $q_h(\gamma) \bmod a_h$, $\gamma \in \Gamma$, are pairwise distinct, and $\lfloor q_h(\gamma)/a_h \rfloor + \kappa < B$ for all $\kappa < B/2$.*

**Definition C.8** (The relay words $x_{k,L}$)**.** Let $k \geq 2$ and $L \geq L_k$, with the objects of Definition C.1, Definition C.3, Definition C.4 and Definition C.6. Put

$$n := RBs_{H-1}E(H+1), \quad J := \lceil \log n \rceil.$$

The program $\mathcal{P}_k$ over $\{0, 1\}$ has the following vertices; vertices not reachable from the root are deleted. (A remaining vertex loses no child, so its value is unchanged.)

- *Pool words* $A_a$, $a < D := u^2$, with the $H+1$ children $Y_{a,0}, ..., Y_{a,H} \in \{0, 1\}$.
- *Columns* $v_{\gamma,\kappa}$, $\gamma \in \Gamma$, $\kappa < B/2$, of arity $E$; we call their edges the *slots*, and edge $\ell$ of $v_{\gamma,\kappa}$ leads to $A_{\beta(\gamma,\kappa,\ell)}$.
- *Zero words*: $Z_A$ with $H+1$ children 0; $Z_v$ with $E$ children $Z_A$; $Z_G$ with $B$ children $Z_v$; and $Z^0 := (0, 0)$, $Z^{i+1} := (Z^i, Z^i)$ for $i + 1 < J$, so that $\mathrm{val}(Z^{J-1})$ has length $2^J \geq n$.
- *Shift gadgets*: for each $\gamma \in \Gamma$, the $gB$ vertices of Definition C.5, with the outputs $G_z^\gamma$, $z < B$, for the list $U_\kappa := v_{\gamma,\kappa}$ $(\kappa < B/2)$, $U_\kappa := Z_v$ $(B/2 \leq \kappa < B)$.
- *Runners* $w_h$, $h < H$, of arity $a_h$. For $\gamma \in \Gamma$, edge $q_h(\gamma) \bmod a_h$ of $w_h$ leads to $G_z^\gamma$ with $z := \lfloor q_h(\gamma)/a_h \rfloor$; all other edges lead to $Z_G$. (These edges are pairwise distinct and $z < B$, by Lemma C.7 (b).)
- The *free levels* (Definition C.4) and the root above the runners; edges without a real child lead to $Z^{J-1}$.

The randomness: the placement; independently, the *pointers* $\beta(\gamma, \kappa, \ell)$, uniform in $[D]$; and the *bits* $Y_{a,t}$, uniform in $\{0, 1\}$; all independent. The word is $x_{k,L} := \mathrm{val}(\mathcal{P}_k)[: n]$. Throughout Appendix C and Appendix D, $L \geq L_k$ is fixed and $x_k := x_{k,L}$, except in Theorem 5.7 and Corollary 5.8, where $L = L_k$ and the placement is the one fixed in the proof of Theorem 5.7.

## C.2. Raw Coordinates and Critical Positions

A column's occurrence through $w_h$ lies at the positions below, where the corrections $\eta_h(\gamma)$ and $\pi_h$ are fixed by the placement (Definition C.6, Definition C.4):

**Definition C.9** (Raw coordinates, slots, rounds, carrying and first positions)**.** For a placement, a runner $h$, a column $(\gamma, \kappa)$ and $t \in \mathbb{N}$ put

$$p_h(\gamma, \kappa, t) := R(s_h(Bt + \kappa) + c_h + \gamma + \eta_h(\gamma)) + \pi_h.$$

We call the triples $(\gamma, \kappa, t)$ with $t < E(H+1)$ the *raw coordinates*, and $(\gamma, \kappa, t \bmod E)$ and $\lfloor t/E \rfloor$ the *slot* and the *round* of $(\gamma, \kappa, t)$.
A position $p < n$ *carries* the raw coordinate $v$ if $p = p_h(v)$ for a runner $h$. It carries at most one, as it has a unique root path to a letter, through at most one runner and column (Lemma C.11 (b)). A position is the *first position* of the raw coordinate it carries if no smaller position carries it.

Runner $h$ shows the bits of round $h$ first for most slots (Lemma C.12), the core slots:

**Definition C.10** (Core slots, points, inner slots, $S'$, critical set $S$)**.** A slot $\omega = (\gamma, \kappa, \ell)$ with $2 \le \ell < E - 2$ is called a *core* slot; for $h < H$ we call $y_h(\omega) := p_h(\gamma, \kappa, Eh + \ell)$ its *point* of round $h$. A core slot is called *inner* if $B/8 \le \kappa < 3B/8$ and $E/4 + 1 \le \ell < 3E/4 - 1$. Put $S' := \{y_h(\omega) : \omega \text{ core}, h < H\}$ and call $S := \{y_h(\omega) : \omega \text{ inner}, h < H\}$ the *critical set.* We say that $S$ is *spread in* $S'$ if $|\{\omega \text{ core} : y_h(\omega) \in A \text{ for some } h < H\}| \ge u$ for every finite arithmetic progression $A$ with $|A \cap S| \ge u^2$.

Inner slots are core slots, so $S \subseteq S'$; $y_h(\omega) < n$ (Lemma C.12 (a)) carries $(\gamma, \kappa, Eh + \ell)$, of round $h$. By Definition C.9, which position carries which raw coordinate, and so $S$ and $S'$, depend on the placement only.

**Lemma C.11** (Structure; proof in Appendix C.3)**.** *For every outcome of the random choices:*

*(a)* $\mathcal{P}_k$ *is a ZLP.*

*(b)* $|\mathrm{val}(\mathcal{P}_k)| \ge n$*, and every position* $p < n$ *either equals* $p_h(\gamma, \kappa, t)$ *for exactly one tuple* $(h, \gamma, \kappa, t)$ *with* $h < H$*,* $\gamma \in \Gamma$*,* $\kappa < B/2$ *and* $t < E(H+1)$*, and then* $x_k[p] = Y_{\beta(\gamma,\kappa,t \bmod E),\lfloor t/E \rfloor}$*, or equals no such* $p_h(\gamma, \kappa, t)$*, and then* $x_k[p] = 0$*.*

*So, for every placement,* $x_k$ *is a slot word with pool size* $D = u^2$ *whose slots are the edges of the columns, whose layout maps a position that carries a raw coordinate* $v$ *to* $(\mathrm{slot}(v), \mathrm{round}(v))$*; as these determine* $v$*, a position* $p$ *is the first position of* $v$ *iff no* $p' < p$ *has the same layout value.*

**Lemma C.12** (Cores and round order; proof below)**.** *For every placement:*

*(a) for every column* $(\gamma, \kappa)$ *and all* $h < H$ *and* $t$ *with* $Eh + 2 \le t \le E(h+1) - 3$*:* $p_h(\gamma, \kappa, t) < p_{h'}(\gamma, \kappa, t)$ *for every* $h' \ne h$*, and* $p_h(\gamma, \kappa, t) < n$*;*

*(b) every point of* $S'$ *is the first position of its raw coordinate;*

*(c) every point of* $S'$ *of round* $h$ *is smaller than every point of* $S'$ *of round* $h+1$*.*

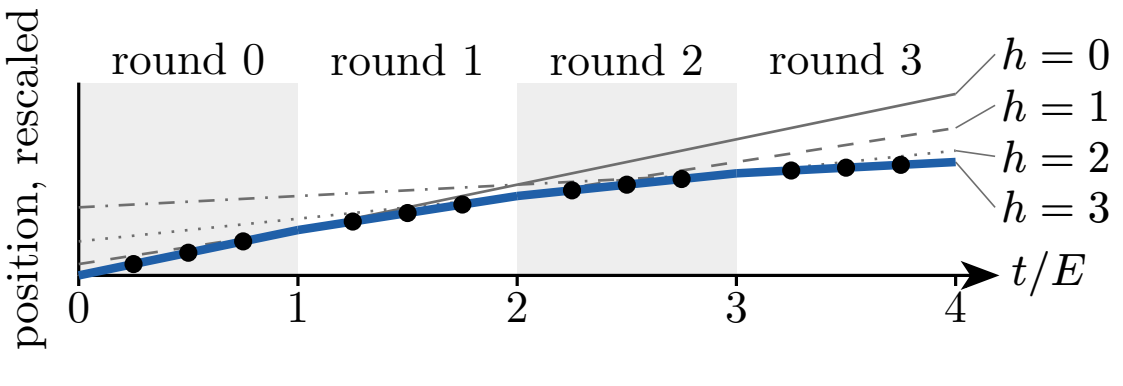

Figure 2: The relay, schematic ($H = 4$): the occurrences of one column through the runners against its index, in units of the round length $E$ (a common drift subtracted, step differences exaggerated); stripes: the rounds; bold: each runner is lowest in turn; dots: runner $h$ first shows round $h$ (for most slots).

*Proof.* The comparison of neighbouring runners below drives the proof; (a) chains it, and (c) compares two rounds directly. Fix a column $(\gamma, \kappa)$ and $t \in \mathbb{N}$, and write $p_h := p_h(\gamma, \kappa, t)$. We show

$$\forall h < H-1: \quad t \le E(h+1) - 3 \Rightarrow p_h < p_{h+1}, \quad t \ge E(h+1) + 2 \Rightarrow p_h > p_{h+1}.$$

Write $g_h := s_h(Bt + \kappa) + c_h$, so that $p_h(\gamma, \kappa, t) = R(g_h + \gamma + \eta_h(\gamma)) + \pi_h$. As $s_h - s_{h+1} = \Delta_h \ge 1$, $c_{h+1} - c_h = B\Delta_h(E(h+1) - 1/2)$ (Definition C.6) and $0 \le \kappa < B/2$, for $h < H - 1$

$$g_h - g_{h+1} = \Delta_h(B(t - E(h+1)) + \kappa + B/2) \begin{cases} \le -2B & \text{if } t \le E(h+1) - 3, \\ \ge 5B/2 & \text{if } t \ge E(h+1) + 2. \end{cases}$$

The terms $\gamma$ are common to all runners, and $|(\eta_h(\gamma) + \pi_h/R) - (\eta_{h'}(\gamma) + \pi_{h'}/R)| < \Psi_* + 1 \le B/2$, as $1 \le \eta_h(\gamma) \le \Psi_* < (4X)^k \le B/4$ (Lemma D.1 (b)) and $0 \le \pi_h < R$. As $(p_h - p_{h+1})/R = (g_h - g_{h+1}) + (\eta_h(\gamma) + \pi_h/R) - (\eta_{h+1}(\gamma) + \pi_{h+1}/R)$, this quotient is $< -2B + B/2 < 0$ if $t \le E(h+1) - 3$, and $> 5B/2 - B/2 > 0$ if $t \ge E(h+1) + 2$, which is the comparison.

(a) Let $h < H$ and $Eh + 2 \le t \le E(h+1) - 3$. For $h \le h' \le H - 2$, $t \le E(h'+1) - 3$, so $p_{h'} < p_{h'+1}$. For $h' < h$, $t \ge E(h'+1) + 2$, so $p_{h'} > p_{h'+1}$. Hence

$$p_j > p_{j+1} \quad (0 \le j < h), \quad p_j < p_{j+1} \quad (h \le j < H-1),$$

so $p_h < p_{h'}$ for all $h' \ne h$: $p_h$ is the least. Moreover $p_h \le p_{H-1}$ by the case $h' \ge h$, and by Lemma C.7 (a) and $t < EH$,

$$\begin{aligned} p_{H-1}(\gamma, \kappa, t) &< R(s_{H-1}(Bt + B/2) + Bs_{H-1}/8 + s_{H-1}/4 + \Psi_* + 1) \\ &\le Rs_{H-1}B(t+1) \le Rs_{H-1}BEH < n. \end{aligned}$$

(b) Let $y \in S'$. By Definition C.10, $y = y_h(\omega) = p_h(\gamma, \kappa, t)$ for some $h < H$ and core slot $\omega = (\gamma, \kappa, \ell)$, with $t := Eh + \ell$, so $Eh + 2 \le t \le E(h+1) - 3$. By (a), $y < n$, so $y$ carries $v := (\gamma, \kappa, t)$, and every position that carries $v$ is some $p_{h'}(v) \ge y$. By Lemma C.11 (b), $v$ is the only raw coordinate that $y$ carries, and $p_h(\gamma, \kappa, t)$ is thus the first position of $(\gamma, \kappa, t)$.

(c) Let $y, y' \in S'$ be of rounds $h$ and $h+1$. As in (b), and as a position carries at most one raw coordinate (Definition C.9), $y = p_h(\gamma, \kappa, t)$ and $y' = p_{h+1}(\gamma', \kappa', t')$ for core slots $(\gamma, \kappa, t - Eh)$ and $(\gamma', \kappa', t' - E(h+1))$; so $t \le E(h+1) - 3$ and $t' \ge E(h+1) + 2$. So

$$\begin{aligned} (y' - y)/R &> s_{h+1}Bt' + c_{h+1} - s_h(Bt + B/2) - c_h - s_{H-1}/4 - \Psi_* - 1 \\ &\ge 2Bs_h + (5/2)Bs_{h+1} - s_{H-1}/4 - \Psi_* - 1 > 0 : \end{aligned}$$

the first inequality holds as $0 \le \kappa < B/2$, $0 \le \gamma < s_{H-1}/4$, $1 \le \eta_h(\gamma) \le \Psi_*$ and $0 \le \pi_h < R$; the second by these bounds on $t$ and $t'$, $c_{h+1} - c_h = B\Delta_h(E(h+1) - 1/2)$ and $s_h = s_{h+1} + \Delta_h$; and the last as $s_{H-1} \le s_h$ and $\Psi_* + 1 \le B$. □

**Lemma C.13** (Size and length; proof in Appendix D)**.**

*(a) The columns have* $C_{\text{col}} := |\Gamma|\,(B/2)E$ *edges and* $|\mathcal{P}_k| \le 2C_{\text{col}}$*; there are* $I := |\Gamma|\,(B/4)(E/2 - 2) \ge C_{\text{col}}/6 \ge |\mathcal{P}_k|\,/12$ *inner slots, and* $|S| = IH \ge H\,|\mathcal{P}_k|\,/12$.

*(b)* $X^{3(k+1)} \le n \le X^{4k+6}$*; so* $10 \log n \le u \le 16 \log n$*; and* $\log n = \Theta(k^2 \log k)$ *if* $L = L_k$.

## C.3. Structure and Spread

**Definition C.14** (Descent)**.** In a ZLP $P$, Definition 3.3 gives $\text{val}(w)[p] = \text{val}(c)[\lfloor p/d_w \rfloor]$ for a vertex $w$ and $p < |\,\text{val}(w)|$, with $c$ the child on edge $p \bmod d_w$ of $w$. Applied repeatedly from the root, this rule gives each $p < |\,\text{val}(P)|$ a unique root path to a letter, its *descent*, and $\text{val}(P)[p]$ is that letter.

**Lemma C.15** (Free levels)**.**

(a) *A vertex of level $i < k$ has at most $\alpha_i/2$, a vertex of level $k$ at most $\alpha_k/8$ real children, so the injections of the placement exist; $H/X^k \le R \le X/2$, and the free levels and the root have at most $5B$ edges.*
(b) *The map $(j_1(h), ..., j_k(h)) \mapsto \zeta_h$ is a bijection from $\prod_{i=1}^k [\alpha_i(h)]$ onto $[\Psi_h]$.*
(c) *Let $q_0$ be the root and $p < |\operatorname{val}(q_0)|$. The rule of Definition C.14, applied from the root through the free levels, reaches either an edge without a real child, and then $\operatorname{val}(q_0)[p]$ is a letter of the word that this edge leads to, or a unique runner $w_h$ at an index $t < |\operatorname{val}(w_h)|$, and then $\operatorname{val}(q_0)[p] = \operatorname{val}(w_h)[t]$ and $p = R(\Psi_h t + \zeta_h) + \pi_h$.*
(d) *If every edge without a real child leads to a word of length at least $n$, then $|\operatorname{val}(q_0)| \ge \min(n, \min_h R\Psi_h\ |\operatorname{val}(w_h)|)$.*

*Proof.* (a) The real children of a vertex of level $i < k$ with prefix $(\alpha_1, ..., \alpha_i)$ have prefixes $(\alpha_1, ..., \alpha_i, p)$ for distinct primes $p \in \mathbb{P}$; so there are at most $|\mathbb{P}| \le X/2 \le \alpha_i/2$ of them (Lemma D.1 (c)). A vertex of level $k$ has at most $X/8 \le \alpha_k/8$ real children, as a type is the type of at most $X/8$ runners: the slopes of the runners of one type are distinct multiples of its product $\Psi > X^k$ in $[M, M+W)$, which contains at most $\lceil W/\Psi \rceil \le W/X^k = X/8$ of them, as $W/X^k$ is an integer. A vertex of level $i$ has $\alpha_i$ edges, and the root $R$ edges for its $R$ children; so each injection of Definition C.4 exists. The level-1 vertices carry distinct primes of $\mathbb{P}$, so $R \le |\mathbb{P}| \le X/2$. Level $i$ has at most $|\mathbb{P}|^i \le (X/2)^i$ vertices (one per prefix), each of arity below $4X$; as $X \ge 4$ and $k \ge 2$, the edges number at most

$$R + \sum_{i=1}^{k} 4X\left(\frac{X}{2}\right)^i \le \frac{X}{2} + \frac{4X(X/2)^k}{1 - 2/X} \le \frac{X}{2} + 8 \cdot 2^{-k} X^{k+1} \le \frac{X}{2} + 2B \le 5B.$$

Below a level-1 vertex lie at most $|\mathbb{P}|^{k-1} \le (X/2)^{k-1}$ types, each of at most $X/8$ runners, and each runner lies below exactly one level-1 vertex; so

$$H \le R \cdot (X/2)^{k-1} \cdot X/8 < RX^k, \quad \text{so} \quad R \ge H/X^k.$$

(b) $\zeta_h = j_1 + \alpha_1(j_2 + \alpha_2(...))$ with digits $j_i < \alpha_i$ is the mixed-radix representation of $[\Psi_h]$.
(c) Put $t_0 := p$ and let $q_i$ be the vertex of level $i$ that the descent reaches. By Definition 3.3, $\operatorname{val}(q_i)[t_i] = \operatorname{val}(q_{i+1})[t_{i+1}]$ for the child $q_{i+1}$ on edge $e_i$ of $q_i$, where

$$t_i = d_{q_i} t_{i+1} + e_i, \quad e_i < d_{q_i} \quad (i = 0, 1, ...),$$

as long as $q_i$ is the root or a free vertex. Each decomposition is unique. If this descent reaches an edge without a real child, $\operatorname{val}(q_0)[p]$ is a letter of the word it leads to. Otherwise it reaches some $q_{k+1} = w_h$ at the index $t := t_{k+1} < |\operatorname{val}(w_h)|$, with $e_0 = \pi_h$, $e_i = j_i(h)$ and $d_{q_i} = \alpha_i(h)$ for $1 \le i \le k$; composing gives, by Definition C.4,

$$\begin{aligned} p &= R\left(\sum_{r=1}^{k} \alpha_1(h) \cdots \alpha_{r-1}(h)\, j_r(h) + \alpha_1(h) \cdots \alpha_k(h)\, t\right) + \pi_h \\ &= R(\Psi_h t + \zeta_h) + \pi_h. \end{aligned}$$

(d) By Definition 3.3, $|\operatorname{zip}(u_0, ..., u_{a-1})| = \min_j \big(a\ |u_j| + j\big)$, the smallest position that would read past the end of an argument. Along the tree of the free levels put $\rho_{q_0}(t) := t$ and $\rho_c(t) := \rho_q\big(d_q t + j\big)$ for the real child $c$ on edge $j$ of $q$; each $\rho_q$ is increasing with $\rho_q(t) \ge t$, and $\rho_{w_h}(t) = R(\Psi_h t + \zeta_h) + \pi_h$ by the composition in (c). As $\rho_q(|\operatorname{val}(q)|) = \min_j \rho_q\big(d_q\ |\operatorname{val}(\delta(q)[j])| + j\big)$ and $\rho_q\big(d_q\ |\operatorname{val}(c)| + j\big) = \rho_c(|\operatorname{val}(c)|)$ for a real child $c$, induction from level $k$ up gives

$$|\operatorname{val}(q_0)| = \min_{(q,j)} \rho_q\big(d_q\ |\operatorname{val}(\delta(q)[j])| + j\big)$$

over the edges $(q, j)$ of the root and the free vertices that do not lead to a free vertex. The term of the edge of a runner $w_h$ is $\rho_{w_h}(|\operatorname{val}(w_h)|) \geq R\Psi_h\ |\operatorname{val}(w_h)|$; that of an edge without a real child is at least $|\operatorname{val}(\delta(q)[j])| \geq n$, by the hypothesis. ☐

For $B = 4$ and values $\operatorname{val}(U_i)$ of a common length: $V_{1,z} = (U_z, U_{z+2})$ and $V_{0,z} = (V_{1,z}, V_{1,z+1})$, so, with $u_i := \operatorname{val}(U_i)$, $\operatorname{val}(V_{0,z}) = \operatorname{zip}(\operatorname{zip}(u_z, u_{z+2}), \operatorname{zip}(u_{z+1}, u_{z+3}))$, which is $\operatorname{zip}(u_z, u_{z+1}, u_{z+2}, u_{z+3})$ by the identity in the proof of Lemma C.16; and $\operatorname{val}(G_z) = \operatorname{val}(V_{0,-z})$ zips the values of $U_{-z}, U_{1-z}, \ldots$.

**Lemma C.16** (All cyclic shifts). *If the values of the $U_\kappa$ have a common length, then* $\operatorname{val}(G_z) = \operatorname{zip}(\operatorname{val}(U_{0-z}), \operatorname{val}(U_{1-z}), \ldots, \operatorname{val}(U_{B-1-z}))$*, indices modulo $B$. Moreover, the paths of $g$ edges from $G_z$ through the gadget are indexed by $i \in [B]$: writing $i = \sum_{j=0}^{g-1} 2^j b_j$ with $b_j \in \{0, 1\}$, path $i$ takes edge $b_j$ at level $j$ and ends at the list entry $U_{(i-z) \bmod B}$; so the endpoint indices give a bijection between these paths and the list indices. Applying the zip rule repeatedly to an existing position $Bt + i$ of* $\operatorname{val}(G_z)$ *follows path $i$ and reaches index $t$ of that list entry.*

*Proof.*

1. *For $(b_0, \ldots, b_{g-1}) \in \{0, 1\}^g$ and $i := \sum_{j<g} 2^j b_j$, the path from $G_z$ with edges $b_0, \ldots, b_{g-1}$ ends at $U_{(i-z) \bmod B}$, and Definition 3.3 applied to a position $Bt + i < |\operatorname{val}(G_z)|$ follows this path and reaches index $t$.*
   By Definition C.5, edge $b$ of $V_{m,w}$ leads to $V_{m+1,w+2^m b}$; so, by induction on $m$, the edges $b_0, \ldots, b_{m-1}$ from $G_z = V_{0,-z}$ end at $V_{m,-z+\sum_{j<m} 2^j b_j}$, and for $m = g$ at $U_{(i-z) \bmod B}$. Since $B = 2^g$, binary expansion is a bijection between $[B]$ and $\{0, 1\}^g$, and $i \mapsto (i - z) \bmod B$ is a bijection of $[B]$; this gives the stated bijection. At the binary vertex $V_{m,w}$, Definition 3.3 reads index $2^{g-m} t + \sum_{m \leq j < g} 2^{j-m} b_j$ on edge $b_m$, at index $2^{g-m-1} t + \sum_{m<j<g} 2^{j-m-1} b_j$ of the child; at $m = 0$ the index is $Bt + i$, and after $g$ edges it is $t$.
2. *For words $a_j, a'_j$ ($j < m$) of a common length,* $\operatorname{zip}(\operatorname{zip}(a_0, \ldots, a_{m-1}), \operatorname{zip}(a'_0, \ldots, a'_{m-1}))$ $= \operatorname{zip}(a_0, a'_0, a_1, a'_1, \ldots, a_{m-1}, a'_{m-1})$.
   Both sides have $2m$ times this length, and the letter at index $2(mt + j) + c$ ($j < m$, $c < 2$) of the left side is the letter at index $t$ of $a_j$ ($c = 0$) or $a'_j$ ($c = 1$), as on the right.
3. *If the values of the $U_\kappa$ have a common length, then* $\operatorname{val}(V_{i,z}) = \operatorname{zip}(\operatorname{val}(U_{z+2^i j}) : j < 2^{g-i})$ *for all $i \leq g$ and $z$.*
   We show this by induction from $i = g$ down; for $i = g$ it holds as a zip of one word is that word. For $i < g$, $V_{i,z}$ has the children $V_{i+1,z}$ and $V_{i+1,z+2^i}$; by the induction hypothesis their values are the zips of $a_j := \operatorname{val}(U_{z+2^{i+1}j})$ and $a'_j := \operatorname{val}(U_{z+2^i+2^{i+1}j})$, $j < 2^{g-i-1}$, words of a common length, and Step 2 with $(a_0, a'_0, a_1, a'_1, \ldots) = (\operatorname{val}(U_{z+2^i j}) : j < 2^{g-i})$ is the induction step.

For $i = 0$, Step 3 gives $\operatorname{val}(V_{0,z}) = \operatorname{zip}(\operatorname{val}(U_z), \operatorname{val}(U_{z+1}), \ldots)$, and $G_z = V_{0,-z}$. With Step 1 this is the lemma. ☐

*Proof of Lemma C.7.* (a) Each $c_{h+1}$ is an integer, as $B$ is even. As $E(h' + 1) - 1/2 < EH$ for $h' < H - 1$, $\sum_{h'<h} \Delta_{h'} = s_0 - s_h < W$ (one window, Definition C.3), $H \leq W$ (as $\Lambda \geq 16$) and $M = 8EW^2$,

$$c_h = \sum_{h'<h} B\Delta_{h'}(E(h' + 1) - 1/2) \leq BEH(s_0 - s_h) < BEW^2 = BM/8 \leq Bs_{H-1}/8.$$

(b) Fix a placement and $h < H$. By Definition C.6, $\Psi_h q_h(\gamma) = c_h + \gamma + \eta_h(\gamma) - \zeta_h$ with $1 \leq \eta_h(\gamma) \leq \Psi_h$; by Lemma C.15 (b), $0 \leq \zeta_h < \Psi_h$.

1. $\forall \gamma \in \Gamma : q_h(\gamma) \in \mathbb{N}$.

By the congruence of Definition C.6, $q_h(\gamma)$ is an integer, and $\Psi_h q_h(\gamma) \geq 1 - \zeta_h > -\Psi_h$, as $c_h, \gamma \geq 0$; so $q_h(\gamma) > -1$, hence $q_h(\gamma) \in \mathbb{N}$.

2. $\forall \gamma \in \Gamma : \gamma + \Psi_* \in \Gamma \Rightarrow q_h(\gamma + \Psi_*) \geq q_h(\gamma) + 1$.
$\Psi_h(q_h(\gamma + \Psi_*) - q_h(\gamma)) = \Psi_* + \eta_h(\gamma + \Psi_*) - \eta_h(\gamma) \geq \Psi_* + 1 - \Psi_h > 0$, as $\Psi_* \geq \Psi_h$; so the integer $q_h(\gamma + \Psi_*) - q_h(\gamma)$ is positive.
3. $\forall \gamma < \gamma'$ in $\Gamma : 0 < q_h(\gamma') - q_h(\gamma) < a_h$; *so the residues* $q_h(\gamma) \bmod a_h$, $\gamma \in \Gamma$, *are pairwise distinct.*
The first inequality is Step 2, as $\Gamma = \Psi_*\mathbb{Z} \cap [0, s_{H-1}/4)$. For the second, as $\gamma' - \gamma < s_{H-1}/4$, $\eta_h(\gamma') - \eta_h(\gamma) < \Psi_h$, $s_{H-1} \leq s_h = \Psi_h a_h$ and $a_h \geq 2$ (Definition C.3),
$$q_h(\gamma') - q_h(\gamma) < \frac{s_{H-1}/4 + \Psi_h}{\Psi_h} \leq \frac{a_h}{4} + 1 < a_h.$$
Two distinct integers whose difference lies in $(0, a_h)$ are not congruent modulo $a_h$.
4. $\forall \gamma \in \Gamma, \kappa < B/2 : \lfloor q_h(\gamma)/a_h \rfloor + \kappa < B$.
First $a_h = s_h/\Psi_h > M/(4X)^k \geq 4$, as $s_h \geq M$, $\Psi_h < (4X)^k$ and $M \geq B^2 \geq 4(4X)^k$ (Definition C.1, Lemma D.1 (b)). By $q_h(\gamma) \leq (c_h + \gamma + \Psi_h)/\Psi_h$, (a) and $\gamma < s_{H-1}/4 \leq s_h/4$,
$$\frac{q_h(\gamma)}{a_h} \leq \frac{c_h + \gamma + \Psi_h}{s_h} < \frac{B}{8} + \frac{1}{4} + \frac{1}{a_h} \leq \frac{B}{8} + \frac{1}{2},$$
so $\lfloor q_h(\gamma)/a_h \rfloor \leq B/8$, as $B/8$ is an integer, and $\lfloor q_h(\gamma)/a_h \rfloor + \kappa < B/8 + B/2 < B$. □

*Proof of Lemma C.11.* (a) Every edge goes downwards in the order of layers: the root, the free levels, the runners or $Z^{J-1}$, the gadget vertices or $Z_G$, the columns or $Z_v$, the pool words or $Z_A$, the letters; inside the free levels and the gadgets it goes one level down, and the zero words $Z^i$ only lead to smaller ones. So the graph is acyclic; all arities are at least 2 ($R \geq 2$: Lemma C.15 (a), Lemma D.1); after the deletion only the root occurs in no list.
(b) Each $p < |\operatorname{val}(\mathcal{P}_k)|$ has a descent, a unique root path to a letter, and $\operatorname{val}(\mathcal{P}_k)[p]$ is that letter (Definition C.14). We say that the descent of $p$ *passes through* a vertex $q$ if $p = \rho(t)$ for an occurrence $\rho \in \operatorname{occ}_q$ and some $t \in \mathbb{N}$, that is, if the root path of $\rho$ is an initial segment of the descent; then $\operatorname{val}(\mathcal{P}_k)[p] = \operatorname{val}(q)[t]$ (Lemma 3.7 (a)). The steps below give (b): Step 1 gives $|\operatorname{val}(\mathcal{P}_k)| \geq n$; by Step 2 and Step 3, a position $p < n$ whose descent passes through a zero word has $x_k[p] = 0$ and equals no $p_h(\gamma, \kappa, t)$ with $t < E(H+1)$, and every other $p < n$ equals $p_h(\gamma, \kappa, t)$ for exactly one tuple, with $x_k[p] = Y_{\beta(\gamma,\kappa,t \bmod E),\lfloor t/E \rfloor}$.

1. $|\operatorname{val}(\mathcal{P}_k)| \geq n$.
The pool words and $Z_A$ have length $H+1$, the columns and $Z_v$ length $E(H+1)$, the gadget outputs $G_z^\gamma$ and $Z_G$ length $BE(H+1)$ (Lemma C.16), and $w_h$ length $a_h BE(H+1)$. By Lemma C.15 (d), as the edges without a real child lead to $Z^{J-1}$, of length $2^J \geq n$, $|\operatorname{val}(\mathcal{P}_k)| \geq \min(n, \min_h R\Psi_h a_h BE(H+1)) = n$, as $\Psi_h a_h = s_h \geq s_{H-1}$.
2. *For every $p < n$, either the descent of $p$ passes through a zero word, so that $x_k[p] = 0$, or $p = p_h(\gamma, \kappa, t)$ for a tuple $(h, \gamma, \kappa, t)$ with $t < E(H+1)$, the descent of $p$ passes through $w_h$ and $v_{\gamma,\kappa}$, and $x_k[p] = Y_{\beta(\gamma,\kappa,t \bmod E),\lfloor t/E \rfloor}$.*
By Lemma C.15 (c), as $p < n \leq |\operatorname{val}(\mathcal{P}_k)|$ (Step 1), the descent of $p$ reaches either an edge without a real child, which leads to $Z^{J-1}$, so that $x_k[p] = 0$, or a unique runner $w_h$ at an index $t'$, with $p = R(\Psi_h t' + \zeta_h) + \pi_h$. In the second case, Definition 3.3 at $w_h$ and, at a gadget output, Lemma C.16 decompose the index as
$$t' = a_h t'' + \sigma, \quad \sigma < a_h, \qquad t'' = Bt + i, \quad i < B,$$

and the descent continues to the child $c$ on edge $\sigma$ of $w_h$, at index $t''$. If $c = Z_G$, the letter is 0. Otherwise $\sigma = q_h(\gamma) \bmod a_h$ for exactly one $\gamma$ (Lemma C.7 (b)), and $c = G_z^\gamma$ with $z = \lfloor q_h(\gamma)/a_h \rfloor$ (Definition C.8); the descent continues along path $i$ to $U_{(i-z) \bmod B}$, at index $t$. If this is $Z_v$, the letter is 0. Otherwise $U_{(i-z) \bmod B} = v_{\gamma,\kappa}$ with $\kappa = (i - z) \bmod B < B/2$, so $i = \kappa + z$, as $i < B$ and $\kappa + z < B$ (Lemma C.7 (b)). By Definition 3.3 at the column and at the pool word, $x_k[p] = Y_{\beta(\gamma,\kappa,t \bmod E),\lfloor t/E \rfloor}$, with $t < E(H+1)$. In this case, as $a_h z + \sigma = q_h(\gamma)$ and $\Psi_h q_h(\gamma) + \zeta_h = c_h + \gamma + \eta_h(\gamma)$,

$$\begin{aligned} p &= R(\Psi_h(a_h(Bt + \kappa + z) + \sigma) + \zeta_h) + \pi_h \\ &= R(s_h(Bt + \kappa) + \Psi_h q_h(\gamma) + \zeta_h) + \pi_h = p_h(\gamma, \kappa, t). \end{aligned}$$

3. *The tuple of Step 2 is unique, and no* $p_h(\gamma,\kappa,t) < n$ *with* $t < E(H+1)$ *has a descent through a zero word.*

   Each of these decompositions is unique. Conversely, let $p := p_h(\gamma, \kappa, t) < n$ with $t < E(H+1)$, $z := \lfloor q_h(\gamma)/a_h \rfloor$ and $\sigma := q_h(\gamma) \bmod a_h$. The display of Step 2, read backwards, gives

$$p = R(\Psi_h t' + \zeta_h) + \pi_h, \quad t' = a_h t'' + \sigma, \quad t'' = Bt + \kappa + z,$$

   with $\pi_h < R$, $\zeta_h < \Psi_h$, $\sigma < a_h$ and $\kappa + z < B$ (Lemma C.15 (b), Lemma C.7 (b)). So the descent of $p$ takes edge $\pi_h$ of the root and the edges $j_i(h)$ of the free levels to $w_h$, at index $t'$ (Lemma C.15 (c)); edge $\sigma$ of $w_h$ to $G_z^\gamma$, at index $t''$ (Definition C.8); and path $\kappa + z$ of the gadget to $U_\kappa = v_{\gamma,\kappa}$, at index $t$ (Lemma C.16). It passes through $w_h$ and $v_{\gamma,\kappa}$ and through no zero word. As the descent of $p$ is unique, it determines $h$ (its runner), $\gamma$ (the edge $\sigma$ of $w_h$, by Lemma C.7 (b)), $\kappa$ (the list index) and $t$ (the index at the column); so the tuple is unique. □

**Remark C.17** (Long lowest intervals)**.** Take $P = \mathcal{P}_k$ in the terms of Section 4. The column $v_{\gamma,\kappa}$ has exactly $H$ occurrences, one through each runner: the free levels form a tree, runner $h$ has exactly one edge that leads to a gadget vertex of $\gamma$ (Definition C.8, Lemma C.7 (b)), and below this vertex the column occurs once (Lemma C.16). The occurrence through runner $h$ is $t \mapsto p_h(\gamma, \kappa, t)$ (Step 2 in the proof of Lemma C.11), and by Lemma C.12 (a) it is lowest at every $t$ with $Eh + 2 \le t \le E(h+1) - 3$. So each of the $H$ occurrences of every column has a lowest interval of at least $E - 4$ indices. In a synchronous program, only the least occurrence of a vertex can have a non-empty lowest interval: all occurrences of a vertex have the same step, and the one of least offset lies below the others. ◁

**Lemma C.18** (Few runners take prescribed digits)**.** *Fix for every runner $h$ either a value $\xi_h$ that the tuple $(\pi_h, j_1(h), ..., j_k(h))$ can take, or no value; we say that runner $h$* hits *if a value was fixed for it and its tuple equals that value. For every integer $u$ with $1 \le u \le \min(R, X)/4$, at least $u$ runners hit with probability at most* $\exp(4/3)2^{-u}$ *over the placement.*

*Proof.* Let $N_{\text{hit}}$ be the number of runners that hit. Each vertex other than the root is a real child of exactly one vertex; for a set $T$ of runners let $V(T)$ be the set of the vertices other than the root on the root paths of the members of $T$, $\hat{v}$ the vertex directly above $v$, and $w(v) := 1/(d_{\hat{v}} - u)$, with the arity $d_y$ of a vertex $y$ (Definition 3.4; $d_{\text{root}} = R$). For a vertex $v$, let $f_v(\lambda) := \sum_T \lambda^{|T|} \prod_y w(y)$: the sum is over all sets $T$ of runners of the subtree of $v$ ($v$ itself if it is a runner; $\varnothing$ included), and the product over the vertices $y \ne v$ on the paths from $v$ to the members of $T$. We show

$$\Pr[N_{\text{hit}} \ge u] \le \mathbb{E}\binom{N_{\text{hit}}}{u} \le \sum_{|T|=u} \prod_{v \in V(T)} w(v) = [\lambda^u] f_{\text{root}}(\lambda) \le 2^{-u} f_{\text{root}}(2) \le \exp(4/3)2^{-u}.$$

The first inequality holds as $\mathbb{1}_{N_{\rm hit}\ge u} \le \binom{N_{\rm hit}}{u}$, and $\mathbb{E}\binom{N_{\rm hit}}{u}$ is the sum over the $u$-sets $T$ of runners of the probability that all of them hit.

1. *The second inequality:* $\Pr[\text{all members of } T \text{ hit}] \le \prod_{v\in V(T)} w(v)$ *for a $u$-set $T$.*
   The event says that every $v \in V(T)$ is the child on the edge of $\hat{v}$ that the tuple of each member below $v$ prescribes. If these prescriptions conflict, the probability is 0. Otherwise, if $j_y$ vertices of $V(T)$ lie directly below a vertex $y$ (so $j_y \le u$), the probability that they are the children on $j_y$ given distinct edges of $y$ is $1/\big(d_y\big(d_y-1\big)\cdots\big(d_y-j_y+1\big)\big) \le \big(d_y-u\big)^{-j_y}$, and the placements at distinct vertices are independent.
2. *The equality and the next inequality.*
   The equality is the definition of $f_{\rm root}$, whose product for a set $T$ runs over $V(T)$; the inequality holds as all coefficients are $\ge 0$. For a runner, $f_v(\lambda) = 1+\lambda$; for the root or a free vertex, $f_v(\lambda) = \prod_c(1+w(c)(f_c(\lambda)-1))$ over its real children $c$, as a set below $v$ is a family of sets below its children, and $c$ is on a path exactly when its set is non-empty.
3. *The last inequality:* $f_{\rm root}(2) \le \exp(4/3)$.
   By the recursion of Step 2 and $1+r \le \exp(r)$, by induction from level $k$ up,
$$f_v(2) \le \exp\left(\sum_c w(c)(f_c(2)-1)\right) \le \begin{cases} \exp(1/6\cdot 2) < 2 & \text{if } v \text{ has level } k,\\ \exp(2/3\cdot 1) < 2 & \text{if } v \text{ has level } i<k,\\ \exp(4/3\cdot 1) & \text{if } v \text{ is the root.}\end{cases}$$
   Indeed $d_v - u \ge 3d_v/4$ at a free vertex $v$, as $u \le X/4 \le d_v/4$; so, by Lemma C.15 (a), at level $k$, $\sum_c w(c) \le (d_v/8)/(3d_v/4) = 1/6$ and $f_c(2)-1 = 2$ (the $c$ are runners); at level $i<k$, $\sum_c w(c) \le (d_v/2)/(3d_v/4) = 2/3$; at the root, $\sum_c w(c) = R/(R-u) \le 4/3$; in these two cases $f_c(2)-1<1$ by the case of the level below. ☐

**Lemma C.19** (Spread critical positions; proof below)**.**

*(a) With probability at least $1-4n^{-7}$ over the placement, $S$ is spread in $S'$.*

*(b) For every outcome, $|S| \ge H\ |\mathcal{P}_k|\ /12$.*

*Proof.* With high probability, no progression of large step has $u$ points of one inner slot (Step 1); for small steps, the round order of Lemma C.12 (c) and a pigeonhole count give points of $u$ slots (Step 2).

1. *With probability at least $1-4n^{-7}$, no progression of step greater than $\tau_0 := R\Psi_*$ contains $u$ points of one inner slot.*
   The points of $S'$ are below $n$ (Lemma C.12 (a)), and a progression of step at least $n$ has at most one element below $n$, fewer than $u$; so only steps $\tau < n$ matter. A progression of step $\tau$ lies in $\nu+\tau\mathbb{Z}$ for some $\nu \in [\tau]$, and its points of a slot $\omega$ are points $y_h(\omega)$ of distinct runners $h$. So it suffices that, with probability at least $1-4n^{-7}$, for no inner slot $\omega$ and integers $\tau, \nu$ with $\tau_0 < \tau < n$ and $0 \le \nu < \tau$ do at least $u$ runners $h$ have $y_h(\omega) \in \nu+\tau\mathbb{Z}$.
   Fix such $\omega = (\gamma, \kappa, \ell)$, $\tau$ and $\nu$. With $\overline{y}_h := s_h(B(Eh+\ell)+\kappa)+c_h+\gamma$, independent of the placement,
$$y_h(\omega) = R(\overline{y}_h + \eta_h(\gamma)) + \pi_h \in [R(\overline{y}_h+1), R(\overline{y}_h+\Psi_h+1)).$$
   By Lemma C.15 (b), and as $\eta_h(\gamma) \in [1, \Psi_h]$ is determined by $\zeta_h$ bijectively, the tuple $(\pi_h, j_1(h), ..., j_k(h))$ determines $y_h(\omega)$ bijectively among these $R\Psi_h \le \tau_0 < \tau$ consecutive integers; so at most one tuple $\xi_h$ puts $y_h(\omega)$ into $\nu+\tau\mathbb{Z}$. Fix these $\xi_h$ in Lemma C.18, applicable as $u \le X^{31/32}/4 \le R/4$ (Lemma C.15 (a), Lemma D.1); as a runner with $y_h(\omega) \in \nu+\tau\mathbb{Z}$ hits,
$$\Pr[|\{h<H : y_h(\omega) \in \nu+\tau\mathbb{Z}\}| \ge u] \le \Pr[|\{h<H : h \text{ hits}\}| \ge u] \le \exp(4/3)2^{-u}.$$

There are fewer than $n^3$ triples $(\omega, \tau, \nu)$: fewer than $n$ inner slots, as their points are distinct and below $n$, and both $\tau$ and $\nu$ below $n$; and $u \geq 10 \log n$ (Lemma C.13 (b)); so a union bound gives the claim, as $\exp(4/3) n^3 2^{-u} \leq 4n^{-7}$.

2. *On the event of Step 1, every finite arithmetic progression with $u^2$ points of $S$ contains points of $S'$ of at least $u$ slots.*

   Suppose, for a contradiction, that a finite arithmetic progression $\Pi$ of step $\tau$ has $|\Pi \cap S| \geq u^2$ but $|\{\omega \text{ core} : \exists h < H : y_h(\omega) \in \Pi\}| < u$; then $\Pi \subseteq \nu + \tau\mathbb{Z}$ for some $\nu$. Each point of $S$ is $y_h(\omega)$ for exactly one inner slot $\omega$ and round $h$ (it carries one raw coordinate, Definition C.9), and inner slots are core slots; so

   $$u^2 \leq |\Pi \cap S| = \sum_{\omega \text{ inner}} |\{h < H : y_h(\omega) \in \Pi\}|,$$

   a sum of fewer than $u$ non-zero terms. As $(u-1)u < u^2$, some inner slot $\omega = (\gamma, \kappa_0, \ell_0)$ has at least $u+1$ points in $\Pi$, of rounds $h_1 < h_2 < ... < h_{u+1}$; so $\tau \leq \tau_0$ (Step 1). Let $h_* := h_2$, so that $y_{h_*}(\omega)$ lies in $\Pi$ and, as $u \geq 2$, $h_1 < h_* < h_{u+1}$. By Lemma C.12 (c), every point of $S'$ of round $h_*$ lies strictly between $y_{h_1}(\omega)$ and $y_{h_{u+1}}(\omega)$, so $\Pi$, of step $\tau$, contains each of them that lies in $\nu + \tau\mathbb{Z}$. For a core slot $(\gamma, \kappa, \ell)$ of the same $\gamma$, by Definition C.9,

   $$y_{h_*}(\gamma, \kappa, \ell) - y_{h_*}(\omega) = Rs_{h_*}((B\ell + \kappa) - (B\ell_0 + \kappa_0)),$$

   so, with $\tau' := \tau / \gcd\big(\tau, Rs_{h_*}\big)$,

   $$\tau' \mid (B\ell + \kappa) - (B\ell_0 + \kappa_0) \quad \Rightarrow \quad y_{h_*}(\gamma, \kappa, \ell) \in \nu + \tau\mathbb{Z} \quad \Rightarrow \quad y_{h_*}(\gamma, \kappa, \ell) \in \Pi.$$

   Let $\delta_\kappa := B/8 - 1$ and $\delta_\ell := E/4 - 2$. As $\omega$ is inner, every $(\kappa, \ell)$ with $|\kappa - \kappa_0| \leq \delta_\kappa$ and $|\ell - \ell_0| \leq \delta_\ell$ has $1 \leq \kappa < B/2$ and $3 \leq \ell < E - 3$, so $(\gamma, \kappa, \ell)$ is a core slot. The $(\delta_\ell + 1)(\delta_\kappa + 1)$ integers $Bi + j$, $0 \leq i \leq \delta_\ell$, $0 \leq j \leq \delta_\kappa$, are distinct, as $\delta_\kappa < B$; and

   $$(\delta_\ell + 1)(\delta_\kappa + 1) \geq (E/5)(B/8) = BE/40 \geq u 4^k B/2 \geq uR\Psi_* \geq u\tau',$$

   using $E \geq 400$ (Lemma D.1 (b)), $E = 128 \cdot 4^k u$ (Definition C.1), $R\Psi_* \leq (X/2)(4X)^k = 4^k B/2$ and $\tau' \leq \tau \leq \tau_0$. So $u$ of them, $Bi_r + j_r$ $(r < u)$, lie in one residue class modulo $\tau'$. Put $\ell_r := \ell_0 + i_r - i_0$ and $\kappa_r := \kappa_0 + j_r - j_0$: then $|\ell_r - \ell_0| \leq \delta_\ell$ and $|\kappa_r - \kappa_0| \leq \delta_\kappa$, so $(\gamma, \kappa_r, \ell_r)$ is a core slot, and $(B\ell_r + \kappa_r) - (B\ell_0 + \kappa_0) = (Bi_r + j_r) - (Bi_0 + j_0)$ is divisible by $\tau'$. These $u$ distinct slots have points of round $h_*$ in $\Pi$, a contradiction.

The steps give (a), as Lemma C.12 (a) and Lemma C.11 (b), with Definition C.10, show $y_h(\omega)$ carries a raw coordinate of slot $\omega$. Part (b) is the bound $|S| \geq H\ |\mathcal{P}_k|\ /12$ of Lemma C.13 (a). □

## C.4. Proof of the Lower Bound

**Theorem 5.7** (Large spread sets, restated). *For every $k \geq 2$, the relay word $x_k$ with suitable runner offsets (random ones work with high probability) is a slot word of length $n$ with pool size $u^2$ and a set $\Omega$ of slots, such that*

*(a)* $5 \log n \leq u = O(\log n)$ *and* $\log n = \Theta(k^2 \log k)$*;*

*(b)* $x_k$ *has a spread set* $S$ *with* $|S| \geq n^{1/4 - O(1/k)}\ |\Omega|$*;*

*(c)* $\mathrm{ZLC}(x_k) \leq 2\ |\Omega|$ *for every outcome of the pointers and the bits.*

*Proof.* Take $L = L_k$, and $\mathcal{P}_k$, $n$, $H$, $u$ and $x_k$ of Definition C.8 and $S \subseteq S'$ of Definition C.10. By Lemma C.13 (b), $10 \log n \leq u \leq 16 \log n$ and $\log n = \Theta(k^2 \log k)$: this is (a). As $4n^{-7} < 1$, Lemma C.19 (a) gives a placement for which $S$ is spread in $S'$; we fix it (these are the runner offsets of the relay word in Theorem 5.7). By Lemma C.11, $x_k$ is then a slot word with pool size $u^2$ whose slots are the edges of the columns, so $|\Omega| = C_{\text{col}}$ (Lemma C.13 (a)); its layout $\psi$, like $S$ and $S'$, depends on the placement only (Definition C.9).

1. *$S$ is spread: every arithmetic progression $A \subseteq [n]$ with $|A \cap S| \geq u^2$ has $|\Omega_{\text{new}}(A)| \geq u$.*
   Fix such an $A$ and let $\Omega_A := \{\omega \text{ core} : \exists h < H : y_h(\omega) \in A\}$; as $S$ is spread in $S'$, $|\Omega_A| \geq u$ (Definition C.10). We show $\Omega_A \subseteq \Omega_{\text{new}}(A)$. Let $\omega = (\gamma, \kappa, \ell) \in \Omega_A$ and $h < H$ with $i := y_h(\omega) \in A$. The position $i = p_h(\gamma, \kappa, Eh + \ell) < n$ (Lemma C.12 (a)) carries $v := (\gamma, \kappa, Eh + \ell)$, whose slot is $\omega$ and whose round is $h$ (as $\ell < E$), so $\psi(i) = (\omega, h)$; and $i \in S'$ is the first position of $v$ (Lemma C.12 (b)), so $\psi(j) \neq (\omega, h)$ for all $j < i$ (Lemma C.11). Hence $\omega \in \Omega_{\text{new}}(A)$ (Definition 5.1).
2. $|S| \geq (H/6)\ |\Omega| \geq n^{1/4-9/(64k+96)}\ |\Omega|$.
   By Lemma C.13 (a), $|S| = IH \geq HC_{\text{col}}/6 = (H/6)\ |\Omega|$. By Lemma D.1 (a), $H \geq BX^{-1/32}$ and $6 < 12u^2 < X^{1/32}$; with $B = X^{k+1}$ and $n \leq X^{4k+6}$ (Lemma C.13 (b)),
   $$H/6 > X^{k+15/16} \geq n^{(k+15/16)/(4k+6)} = n^{1/4-9/(64k+96)}.$$
   This is (b), as $9/(64k + 96) = O(1/k)$.
3. *For every outcome of the pointers and the bits,* $\text{ZLC}(x_k) \leq |\mathcal{P}_k| \leq 2\ |\Omega|$.
   $\mathcal{P}_k$ is a ZLP (Lemma C.11 (a)) whose value has $x_k$ as a prefix (Definition C.8), so $\text{ZLC}(x_k) \leq |\mathcal{P}_k|$ (Definition 3.4); and $|\mathcal{P}_k| \leq 2C_{\text{col}} = 2\ |\Omega|$ (Lemma C.13 (a)). This is (c). □

**Theorem C.20** (Relay words, explicit constants). *For every $k \geq 2$, the relay word $x_k$ of length $n$ satisfies*
$$\text{ARC}(x_k) \geq |S|\ /(u^2 - 1) \geq H\ |\mathcal{P}_k|\ /(12u^2) \geq n^{1/4-9/(64k+96)}\, \text{ZLC}(x_k)$$
*with probability at least $1 - 408n^{-6}$ over the placement, the pointers and the bits. The last two hold for every outcome; $H$, $u$, $\mathcal{P}_k$ and $S$ are those of Definition C.8 and Definition C.10.*

*Proof.* Let $\mathcal{E}$ be the event that $S$ is spread in $S'$ and every forward equation in $x_k$ has fewer than $u^2$ targets in $S$. On $\mathcal{E}$, Lemma 5.5 ($h = u^2 - 1$) gives $\text{ARC}(x_k) \geq |S|\ /(u^2 - 1)$; so the first inequality follows from Step 2, and the other two from Step 3.

1. *For every placement for which $S$ is spread in $S'$:* $\Pr[\mathcal{E} \mid \text{placement}] \geq 1 - \exp(6)n^{-6}$.
   Condition on such a placement; the pointers and the bits stay independent and uniform. By Lemma C.11 and the proof of Step 1 in the proof of Theorem 5.7 (which uses neither $L = L_k$ nor more of the placement than that $S$ is spread in $S'$), $x_k$ is a slot word with pool size $u^2$, and $S$ is spread. Here $\exp(6)2^{-u} \leq \exp(6)n^{-10}$, as $u \geq 10 \log n$ (Lemma C.13 (b)). So, given the placement, every forward equation in $x_k$ has fewer than $u^2$ targets in $S$, and $\text{ARC}(x_k) \geq |S|\ /(u^2 - 1)$, with probability at least $1 - \exp(6)n^4 n^{-10} = 1 - \exp(6)n^{-6}$ (Proposition 5.6, Lemma 5.5).
2. $\Pr[\mathcal{E}] \geq 1 - 408n^{-6}$.
   As $\exp(6) < 404$ and $n \geq 1$, Lemma C.19 (a) and Step 1 bound the failure probability:
   $$\Pr[\neg\mathcal{E}] \leq \Pr[S \text{ is not spread in } S'] + \exp(6)n^{-6} \leq 4n^{-7} + \exp(6)n^{-6} \leq 408n^{-6}.$$
3. *For every outcome,* $|S|\ /(u^2 - 1) \geq H\ |\mathcal{P}_k|\ /(12u^2) \geq n^{1/4-9/(64k+96)}\, \text{ZLC}(x_k)$.
   By Lemma C.13 (a), $|S|\ /(u^2 - 1) \geq H\ |\mathcal{P}_k|\ /(12u^2)$, and by Lemma D.1 (a), $H/(12u^2) \geq BX^{-1/32}X^{-1/32} = X^{k+15/16}$, as $\log(12u^2) < L/32$ and $B = X^{k+1}$. As $n \leq X^{4k+6}$ (Lemma C.13 (b)), $X^{k+15/16} \geq n^{(k+15/16)/(4k+6)} = n^{1/4-9/(64k+96)}$. Finally $\text{ZLC}(x_k) \leq |\mathcal{P}_k|$ (Lemma C.11). □

**Theorem 1.2** (Lower bound, restated). *There are constants $C > 0$ and $n_0$ such that for every integer $n \geq n_0$ some binary word $x$ of length $n$ satisfies*

$$\mathrm{ARC}(x) \geq n^{1/4} \exp\left(-C\sqrt{\log n \log\log n}\right) \mathrm{ZLC}(x).$$

*In particular, for every $a < 1/4$ and every large enough $n$, some binary word $x$ of length $n$ has* $\mathrm{ARC}(x) \geq n^a\, \mathrm{ZLC}(x)$.

*Proof.* We reach every large length with prefixes of the words $x_{k,L}$, $L \geq L_k$, of Definition C.8. $N_{k,L}$ is the length of $x_{k,L}$, $N$ that of the word at hand, $m$ that of a prefix; $H, \Lambda, \mathcal{P}_k, \ldots$ are the objects of $(k, L)$, primed for $(k, L-1)$ where $L > L_k$.

1. *Let $k \geq 2$ and $L \geq L_k$. Some outcome of the random choices gives a relay word $x_{k,L}$ such that every prefix $z$ of $x_{k,L}$ of length $m \geq 2N/H$ satisfies* $\mathrm{ARC}(z)/\mathrm{ZLC}(z) \geq mH/(24Nu^2) \geq m/(2N) \cdot N^{1/4-9/(64k+96)}$.

   By Step 2 in the proof of Theorem C.20 (for $(k, L)$, with $N$ for $n$), the event $\mathcal{E}$ there, that $S$ is spread in $S'$ and every forward equation in $x_{k,L}$ has fewer than $u^2$ targets in $S$, fails with probability at most $408N^{-6} < 1$. Fix an outcome in $\mathcal{E}$, let $z := x_{k,L}[:m]$ and $q := \min(H, \lfloor m(H+1)/N \rfloor)$, and let $I$ be the number of inner slots. We show

   $$\frac{\mathrm{ARC}(z)}{\mathrm{ZLC}(z)} \geq \frac{|S \cap [m]|}{(u^2-1)\,|\mathcal{P}_k|} \geq \frac{qI}{(u^2-1)\,|\mathcal{P}_k|} \geq \frac{q}{12u^2} \geq \frac{mH}{24Nu^2}.$$

   For the first inequality, a forward equation in $z$ has all its positions below $m$ (Definition 5.4), so it is a forward equation in $x_{k,L}$, with fewer than $u^2$ targets in $S$; so Lemma 5.5, for $z$, $S \cap [m]$ and the bound $u^2 - 1$, gives $\mathrm{ARC}(z) \geq |S \cap [m]|\,/(u^2-1)$; and $\mathrm{ZLC}(z) \leq |\mathcal{P}_k|$, as $z$ is a prefix of $\mathrm{val}(\mathcal{P}_k)$ (Definition 3.4). For the second, we show

   $$\forall \omega \text{ inner}, \forall h < q: \quad y_h(\omega) \in S \cap [m].$$

   Let $\omega = (\gamma, \kappa, \ell)$ be inner and $h < q$ (so $h < H$ and $y_h(\omega) \in S$), and put $t := Eh + \ell$, so $Eh + 2 \leq t \leq E(h+1) - 3$. By the chain in the proof of Lemma C.12 (a) and Definition C.8,

   $$y_h(\omega) \leq p_{H-1}(\gamma, \kappa, t) < Rs_{H-1}B(t+1) \leq Rs_{H-1}BE(h+1) = \frac{N(h+1)}{H+1} \leq m,$$

   as $h + 1 \leq q \leq m(H+1)/N$. These $qI$ points are distinct positions, as distinct pairs $(\omega, h)$ give distinct tuples $(h, \gamma, \kappa, Eh + \ell)$ and the points lie below $N$, the $n$ of Definition C.8 (Lemma C.11 (b)). The third inequality holds as $I \geq |\mathcal{P}_k|\,/12$ (Lemma C.13 (a)), the last as $q \geq \lfloor mH/N \rfloor \geq mH/(2N)$: $\lfloor mH/N \rfloor \leq H$ as $m \leq N$, and $\lfloor y \rfloor > y - 1 \geq y/2$ for $y := mH/N \geq 2$. Finally, $H/(12u^2) \geq N^{1/4-9/(64k+96)}$ by Step 3 of Theorem C.20.

2. $(4k+4)L \leq \log N_{k,L} \leq (4k+6)L$*;* $N_{k,L} < 2^{4k+5}\Lambda' N_{k,L-1}$ *if* $L > L_k$.

   With $R \geq H/X^k$ (Lemma C.15 (a)), $s_{H-1} \geq M$ and $H \geq BX^{-1/32}$ (Lemma D.1 (a)):

   $$N = RBs_{H-1}E(H+1) \geq \frac{H}{X^k} \cdot B \cdot \frac{EB^2}{8} \cdot EH \geq \frac{E^2}{8} X^{4k+79/16} \geq X^{4k+4};$$

   the upper bound is Lemma C.13 (b). Next, $R/R' \leq (X/2)X'^k\Lambda'/B' = \Lambda'$, $B/B' = 2^{k+1}$, $s_{H-1}/s'_{H'-1} < (M+W)/M' = (1 + 1/(EB))2^{2k+2}E/E'$, $E/E' = L/(L-1)$ and $(H+1)/(H'+1) \leq 2^{k+1}$ $(\Lambda \geq \Lambda')$; so $N/N' < 2^{4k+4}\Lambda'(1 + 1/(EB))(L/(L-1))^2 < 2^{4k+5}\Lambda'$, as $L \geq L_2 = 2048$.

3. *Fix $n \geq N_{2,L_2}$. Let $t := \log n$, $k$ the largest index with $N_{k,L_k} \leq n$, and $L \geq L_k$ the least with $N := N_{k,L} \geq n$ (both exist by Step 2); put $\Delta := \log N - t$. Then $0 \leq \Delta < 4k + 5 + \log \Lambda < 4k + 5 + L/32$, and Step 1 gives, for the prefix $z$ of length $n$ of $x_{k,L}$, $r :=$* $\log(\mathrm{ARC}(z)/\mathrm{ZLC}(z)) \geq \log H - \Delta - \log(24u^2)$.

If $L = L_k$, then $N = n$ and $\Delta = 0$; otherwise $N_{k,L-1} < n$, and Step 2 gives $\Delta < 4k + 5 + \log \Lambda' \le 4k + 5 + \log \Lambda$. Thus in both cases $\Delta < 4k + 5 + \log \Lambda < 4k + 5 + L/32$ by Lemma D.1 (a). As $\log H \ge (k + 31/32)L$ (Lemma D.1 (a)) and $L \ge 1024k$,

$$\log(nH/N) = \log H - \Delta > (k + 31/32)L - 4k - 5 - L/32 \ge 1,$$

that is, $n \ge 2N/H$; with $n \le N$, Step 1 applies to $m = n$, and $\log(nH/(24Nu^2)) = \log H - \Delta - \log(24u^2)$.

4. $t/4 - r < 11L/32$.
   As $\log N = t + \Delta$, $\log H = (t + \Delta - \log(N/H^4))/4$, so
   $$t/4 - r \le 3\Delta/4 + \log(N/H^4)/4 + \log(24u^2)$$
   $$< \frac{L}{4} + \frac{\log E}{2} + \frac{3}{2}\log \Lambda + \log(12u^2) + 3k + 4 < \frac{43L}{128} + 4k + \frac{15}{2} \le \frac{11L}{32} :$$
   the first by Step 3. The second adds $3\Delta/4 < 3k + 15/4 + (3/4)\log \Lambda$ (Step 3), $\log(24u^2) = 1 + \log(12u^2)$ and $\log(N/H^4)/4 < L/4 + (\log E)/2 + (3/4)\log \Lambda - 3/4$, as $N = RBs_{H-1}E(H+1)$ with $R \le X/2$ (Lemma C.15 (a)), $s_{H-1} < M + W = (1 + 1/(EB))EB^2/8$ and $B \le \Lambda H$ give
   $$N/H^4 < (X/16)E^2\Lambda^3(1 + 1/(EB))(1 + 1/H) < XE^2\Lambda^3/8.$$
   The third uses $\log E = 7 + 2k + \log u$ and, by Lemma D.1 (a), $\log \Lambda < L/32$ and $2\log u < \log(12u^2) < L/32$; the last $4k + 15/2 \le 8k \le L/128$, as $k \ge 2$ and $L \ge 1024k$.
5. $L^2 < 64.2t \log t$.
   By the choice of $k$, $L$ and Step 2, $(4k+4)(L-1) < t < (4k+10)L_{k+1}$ and $t \ge (4k+4)L_k$, $L_k \ge \max(2048, 1024k)$; so $t \ge (4k+4)\max(2048, 1024k)$, hence $t \ge 2048(4k+4)$, $t \ge 4096k^2$ and $\log t > 12 + 2\log k$. With $b_{k+1} := \log(k+3)$, $L_{k+1} \le 512(k+1)b_{k+1} + 1 \le 513(k+1)b_{k+1}$ and $(1 + 3/(2k+2))b_{k+1} < \log k + 2.5$ (as $b_{k+1} - \log k \le \log(5/2)$, $3b_{k+1}/(2k+2) \le 1.17$), so
   $$\frac{L^2}{t} < \left(1 + \frac{1}{2048}\right)^2 \frac{t}{(4k+4)^2} < 128.4\left(1 + \frac{3}{2k+2}\right)b_{k+1} < 128.4(\log k + 2.5) < 64.2 \log t :$$
   the first as $L < t/(4k+4) + 1 \le (1 + 1/2048)t/(4k+4)$; the second as $t < (4k+10) \cdot 513(k+1)b_{k+1}$ and $(1 + 1/2048)^2 \cdot 513(4k+10)(k+1)/(4k+4)^2 = (1 + 1/2048)^2(513/4)(1 + 3/(2k+2))$ with $(1 + 1/2048)^2 \cdot 513/4 < 128.4$; the last as $\log t > 2\log k + 5$.
6. *Every* $n \ge N_{2,L_2}$ *is the length of a prefix* $z$ *of some* $x_{k,L}$ *with* $\mathrm{ARC}(z) \ge n^{1/4}2^{-2.76\sqrt{\log n \log\log n}}\,\mathrm{ZLC}(z)$.
   By Step 4 and Step 5, $r > t/4 - (11/32)\sqrt{64.2t \log t} > t/4 - 2.76\sqrt{t \log t}$. Here $(11/32)\sqrt{64.2} < 2.755$, and $2^r = \mathrm{ARC}(z)/\mathrm{ZLC}(z)$, $2^{t/4} = n^{1/4}$, $\log t = \log\log n$.

The binary word $z$ of length $n$ of Step 6 gives the first statement, with $C := 2.76 \ln 2 < 1.92$ and $n_0 := N_{2,L_2}$. For the second, as $\sqrt{\log n \log\log n} = o(\log n)$, for $a < 1/4$ and all large $n$, $n^{1/4}\exp(-C\sqrt{\log n \log\log n}) \ge n^a$, so they have $\mathrm{ARC}(z) \ge n^a\,\mathrm{ZLC}(z)$. □

**Remark C.21** (The regime of the words)**.** The words $z$ of length $n$ in the proof of Theorem 1.2 have $\mathrm{ZLC}(z) = n^{1/2+o(1)}$. There $k \to \infty$ as $n \to \infty$; $N$ and $n$ lie in $[X^{4k+3}, X^{4k+6}]$ (Step 2 of Theorem 1.2, $\Delta < L$); and $C_{\mathrm{col}} = |\Gamma|\,(B/2)E = X^{2k+O(1)}$ (Definition C.6, Lemma D.1 (a)), so $C_{\mathrm{col}} = n^{1/2+o(1)}$. Now $\mathrm{ZLC}(z) \le |\mathcal{P}_k| \le 2C_{\mathrm{col}}$ (Lemma C.13 (a)), and by Theorem 1.1 and Step 1 of Theorem 1.2, $\mathrm{ZLC}(z) \ge \mathrm{ARC}(z)/(2n^{1/4}) \ge (nH/(2N))(C_{\mathrm{col}}/(6u^2))/(2n^{1/4}) = C_{\mathrm{col}}n^{-o(1)}$. ◁

**Corollary C.22** (Synchronous versus general programs)**.** *For every* $a < 1/4$*, at every large enough length* $n$ *some binary word* $x$ *has* $\mathrm{sZLC}(x) \ge n^a\,\mathrm{ZLC}(x)$.

*Proof.* $\mathrm{sZLC}(x) \geq \mathrm{ARC}(x)$ [6, Prop. 4.9], and Theorem 1.2. □

For $L = L_k$, the words are very long: $n \geq 2^{18432}$ for every $k$ (Lemma C.13 (b)), and for $k = 2$ (where $L = 2048$), $n \leq 2^{28672}$ and the exponent is $1/4 - 9/224 \approx 0.21$.

## D. Arithmetic of the Relay Words

**Lemma D.1** (Size of the parameters). *For all integers $k \geq 2$ and $L \geq L_k$:*

*(a) $\log \Lambda < L/32$ and $\log(12u^2) < L/32$; so $H \geq BX^{-1/32}$.*

*(b) $2(4X)^k \leq M$, $(4X)^k \leq B/4$, $2^{5k+9}u^2 \leq X$, $u \leq X^{31/32}/4$, $E \geq 400$ and $L \geq 20$.*

*(c) $|\mathbb{P}| \leq X/2$ and $S_{\mathbb{P}} := \sum_{p\in\mathbb{P}} 1/p \geq 1/(4L)$.*

*Proof.* Put $b := \log(k+2)$; as $k \geq 2$, $b \geq 2$, so $L \geq 512kb \geq 1024k$. We first show

$$\log L \leq \frac{L}{64k}.$$

Let $f(\ell) := \ell/(64k) - \log \ell$ for real $\ell > 0$, and $\ell_0 := 512kb$. As $\log k < b$ and $\log b \leq b - 1$ (true at $b = 2$; the left side grows more slowly),

$$\log \ell_0 = 9 + \log k + \log b < 8 + 2b \leq 8b = \frac{\ell_0}{64k},$$

so $f(\ell_0) > 0$; and $f'(\ell) = 1/(64k) - 1/(\ell \ln 2) \geq 0$ for $\ell \geq 93k$, as $64/\ln 2 < 93$. As $L \geq \ell_0 \geq 93k$, $f(L) \geq f(\ell_0) > 0$.

(a) As $k!\binom{2k+2}{k} = \prod_{j=k+3}^{2k+2} j \leq (2k+2)^k$, $\log(2k+2) \leq 1 + b$ and $k \log L \leq L/64$,

$$\log \Lambda = 4 + \log\left(k!\binom{2k+2}{k}\right) + k(2 + \log L)$$
$$\leq 4 + 3k + kb + L/64 < 8kb + L/64 \leq L/32,$$

as $4 + 3k + kb < 8kb$ $(b \geq 2)$ and $8kb \leq L/64$. Next, $u = (40k + 60)L$ and $40k + 60 \leq 64(k + 2)$, so $\log u \leq 6 + b + \log L$; with $\log 12 < 4$ and $2 \log L \leq L/64$,

$$\log(12u^2) < 4 + 2(6 + b + \log L) \leq 16 + 2b + L/64 < 16b + L/64 \leq L/32,$$

as $16 < 14b$ and $16b \leq 8kb \leq L/64$. Hence $\Lambda < 2^{L/32} = X^{1/32}$, and $H = \lceil B/\Lambda \rceil \geq B/\Lambda \geq BX^{-1/32}$.

(b) With $\log u \leq 6 + b + \log L$ from (a) and $2 \log L \leq L/64$:

$$M = EB^2/8 \geq B^2 = X^{2k+2} \geq 2 \cdot 4^k X^k, \quad (4X)^k \leq X^{k+1}/4 = B/4,$$
$$\log(2^{5k+9}u^2) \leq 5k + 21 + 2b + L/64 \leq L, \quad \log(4u) \leq 8 + b + L/64 \leq 31L/32,$$

the first as $E \geq 8$ and $X^{k+2} \geq 2^{2k+1}$, the second as $X \geq 4^{k+1}$, i.e. $L \geq 2k + 2$, the last two as $5k + 21 + 2b \leq 63L/64$ and $8 + b \leq 29L/32$ ($L \geq 1024k$, $b \leq k + 1$). So $2^{5k+9}u^2 \leq X$ and $u \leq X^{31/32}/4$. Finally $E = 128 \cdot 4^k u > 400$ and $L \geq 1024k \geq 20$.

(c) In this part $p$ denotes a prime, and $\theta(x) := \sum_{p\leq x} \ln p$.

1. *$\theta(x) \leq x \ln 4$ for all real $x \geq 0$, and $\theta(x) \geq x \ln 2 - \ln(x+1) - 2\sqrt{x} \ln x$ for all integers $x \geq 1$.*

   The upper bound is [23, Thm. 415] for integers, and extends to real $x$ by $\theta(x) = \theta(\lfloor x \rfloor)$. For the lower bound,

   $$x \ln 2 - \ln(x+1) \leq \ln\binom{x}{\lfloor x/2 \rfloor} \leq \theta(x) + \sqrt{x} \ln x :$$

   the central binomial coefficient is the largest of the $x + 1$ coefficients summing to $2^x$; by Legendre's formula [23, Thm. 416] each prime power $p^e$ dividing it exactly has $p^e \leq$ $x$, so its prime factors exceeding $\sqrt{x}$ occur at most once, and each of the at most $\sqrt{x}$ remaining primes contributes at most $\ln x$ to its logarithm.

2. $|\mathbb{P}| \le X/2$.
The powers of two $X$ and $4X$ are not prime, so $\sum_{p\in\mathbb{P}} \ln p = \theta(4X) - \theta(X) \le 4X \ln 4 = 8X \ln 2$ (Step 1); as $\ln p \ge \ln X = L \ln 2$ for $p \in \mathbb{P}$, $|\mathbb{P}| \le 8X/L \le X/2$ (as $L \ge 16$).
3. $\theta(4X) - \theta(X) \ge X$.
By Step 1, the lower bound at $x = 4X$ and the upper one at $X$,
$$\theta(4X) - \theta(X) \ge 4X \ln 2 - \ln(4X+1) - 4\sqrt{X}\ln(4X) - 2X \ln 2$$
$$\ge 2X \ln 2 - 5\sqrt{X}(L+3) \ge X,$$
using $\ln(4X+1) \le \ln(8X) \le L+3$ and $\ln(4X) \le L+2$, then $5\sqrt{X}(L+3) \le X/4$ (that is, $2^{L/2} \ge 20(L+3)$, true for $L \ge 20$, as at $L = 20$ and since $2^{L/2}/(L+3)$ increases for $L \ge 3$) and $2\ln 2 - 1/4 > 1$.
4. $S_{\mathbb{P}} \ge 1/(4L)$.
Each $p \in \mathbb{P}$ has $1/p > \ln p/(4X\ln(4X))$, as $p < 4X$; hence, by Step 3 and $\ln(4X) = (L+2)\ln 2$,
$$S_{\mathbb{P}} \ge \frac{\theta(4X) - \theta(X)}{4X\ln(4X)} \ge \frac{1}{4(L+2)\ln 2} \ge \frac{1}{4L},$$
as $(L+2)\ln 2 \le L$ for $L \ge 5$. □

**Lemma C.2** (Dense slopes, restated). *Every window $[y, y+W) \subseteq [M, 2M)$ contains at least $B/\Lambda$ good integers.*

*Proof.* Fix a window $\mathcal{W} := [y, y+W) \subseteq [M, 2M)$. Let $\mathcal{A}$ be the set of all types (the $k$-element subsets of $\mathbb{P}$, written increasingly). For a type $\alpha$ write $\Psi(\alpha) := \alpha_1 \cdots \alpha_k$ (so $X^k < \Psi(\alpha) < (4X)^k$), and for an integer $s$ put $A(s) := \#\{\alpha \in \mathcal{A} : \Psi(\alpha) \mid s\}$.

1. $\sum_{s\in\mathcal{W}} A(s) \ge WS_{\mathbb{P}}^k/(2k!)$, *and every $s \in \mathcal{W}$ with $A(s) \ge 1$ is good.*
In this proof $p$, $p_1, \ldots, p_k$ denote primes. Every multiple $s \in \mathcal{W}$ of $\Psi(\alpha)$ has quotient $s/\Psi(\alpha) \ge M/(4X)^k \ge 2$, as $M \ge 2(4X)^k$ (Lemma D.1 (b)); so it is good, with type $\alpha$. Put $S := S_{\mathbb{P}}$. By Lemma D.1 (b), $X \ge 2^{5k+9}u^2 \ge 8k^2L$ and $X \ge 32 \cdot 4^k$, as $k^2L \le u^2$ (since $u \ge L$ and $u \ge k^2$). We show
$$\sum_{s\in\mathcal{W}} A(s) \ge (3W/4)\sum_{\alpha\in\mathcal{A}} 1/\Psi(\alpha) \ge (3W/(4k!)) \cdot \big(3S^k/4\big) \ge WS^k/(2k!).$$
For the first inequality, $\sum_{s\in\mathcal{W}} A(s) = \sum_{\alpha\in\mathcal{A}} \#\{s \in \mathcal{W} : \Psi(\alpha) \mid s\}$, and the window $\mathcal{W}$, of $W$ consecutive integers, contains at least $\lfloor W/\Psi(\alpha)\rfloor \ge W/\Psi(\alpha) - 1 \ge 3W/(4\Psi(\alpha))$ multiples of $\Psi(\alpha)$, as $W/\Psi(\alpha) > W/(4X)^k = X/\big(8 \cdot 4^k\big) \ge 4$. For the second, let $\mathcal{Q}$ be the set of sequences $(p_1, \ldots, p_k)$ of distinct primes of $\mathbb{P}$; each type is ordered in $k!$ ways into one of them, so $k!\sum_{\alpha\in\mathcal{A}} 1/\Psi(\alpha) = \sum_{\mathcal{Q}} 1/(p_1 \cdots p_k)$. Put $t_j := j/(XS)$; then $\sum_{j<k} t_j = \binom{k}{2}/(XS) \le 2k^2L/X \le 1/4$ by Lemma D.1 (c) and $X \ge 8k^2L$, so $t_j \in [0,1]$. For $j < k$ and distinct $p_1, \ldots, p_j \in \mathbb{P}$, as every $p \in \mathbb{P}$ has $p \ge X$,
$$\sum_{p\in\mathbb{P}\setminus\{p_1,\ldots,p_j\}} 1/p \ge S - j/X = S\big(1 - t_j\big) \ge 0;$$
evaluating the sum over $\mathcal{Q}$ from the innermost index $p_k$ outwards, and as $\prod_j\big(1 - t_j\big) \ge 1 - \sum_j t_j$ for $t_j \in [0,1]$,
$$k!\sum_{\alpha\in\mathcal{A}} \frac{1}{\Psi(\alpha)} \ge \prod_{j<k}\left(S - \frac{j}{X}\right) \ge S^k - \binom{k}{2}\frac{S^{k-1}}{X} \ge \frac{3}{4}S^k.$$
2. $A(s) \le \binom{2k+2}{k}$ *for every integer* $1 \le s < 2M$.

The primes of a type of $s$ are distinct prime divisors of $s$ in $[X, 4X)$. If $s \geq 1$ had $2k+3$ distinct prime divisors $\geq X$, then $s \geq X^{2k+3}$; but $2M = 32 \cdot 4^k u X^{2k+2} \leq X^{2k+3}$, as $2^{2k+5} u \leq 2^{5k+9} u^2 \leq X$ (Lemma D.1 (b)). So $s$ has at most $2k+2$ distinct prime divisors in $[X, 4X)$, and its types are $k$-subsets of them: $A(s) \leq \binom{2k+2}{k}$.

3. $\#\{s \in \mathcal{W} : s \text{ is good}\} \geq B/\Lambda$.
By Step 1, every $s \in \mathcal{W}$ with $A(s) \geq 1$ is good, and by Step 2 (applicable, as $s \geq M \geq 1$) $A(s) \leq \binom{2k+2}{k}$ for each of them. Hence, by the bound of Step 1 on $\sum_{s \in \mathcal{W}} A(s)$, $S_{\mathbb{P}} \geq 1/(4L)$ (Lemma D.1 (c)), $\Lambda = 16k!\binom{2k+2}{k}(4L)^k$ and $W = B/8$ (Definition C.1),

$$\#\{s \in \mathcal{W} : s \text{ is good}\} \geq \#\{s \in \mathcal{W} : A(s) \geq 1\} \geq \sum_{s \in \mathcal{W}} \frac{A(s)}{\binom{2k+2}{k}}$$

$$\geq \frac{W S_{\mathbb{P}}^k}{2k!\binom{2k+2}{k}} \geq \frac{W}{2k!\binom{2k+2}{k}(4L)^k} = \frac{8W}{\Lambda} = \frac{B}{\Lambda}.$$

□

**Lemma C.13** (Size and length, restated)**.**

*(a) The columns have* $C_{\text{col}} := |\Gamma|\ (B/2)E$ *edges and* $|\mathcal{P}_k| \leq 2C_{\text{col}}$*; there are* $I := |\Gamma|\ (B/4)(E/2-2) \geq C_{\text{col}}/6 \geq |\mathcal{P}_k|\ /12$ *inner slots, and* $|S| = IH \geq H\ |\mathcal{P}_k|\ /12$.

*(b)* $X^{3(k+1)} \leq n \leq X^{4k+6}$*; so* $10 \log n \leq u \leq 16 \log n$*; and* $\log n = \Theta(k^2 \log k)$ *if* $L = L_k$.

*Proof.* (a) By Definition C.10, the inner slots are the $(\gamma, \kappa, \ell)$ with $\gamma \in \Gamma$, $B/8 \leq \kappa < 3B/8$ and $E/4 + 1 \leq \ell < 3E/4 - 1$ (8 divides $B$, 4 divides $E$), so, as $E \geq 12$ (Lemma D.1 (b)),

$$I = |\Gamma| \cdot \frac{B}{4} \cdot \left(\frac{E}{2} - 2\right) = |\Gamma|\ \frac{BE}{2}\left(\frac{1}{4} - \frac{1}{E}\right) = C_{\text{col}}\left(\frac{1}{4} - \frac{1}{E}\right) \geq \frac{C_{\text{col}}}{6}.$$

An inner slot $\omega$ has the $H$ points $y_h(\omega) = p_h(\gamma, \kappa, Eh + \ell)$, $h < H$; as $2 \leq \ell \leq E - 3$, they lie below $n$ (Lemma C.12 (a)), and so, being $p_h$ of distinct tuples $(h, \gamma, \kappa, Eh + \ell)$, all these points are pairwise distinct (Lemma C.11 (b)). So $|S| = IH$. As $\Gamma = \Psi_* \mathbb{Z} \cap [0, s_{H-1}/4)$ (Definition C.6), $s_{H-1} \geq M = EB^2/8$ and $\Psi_* < (4X)^k \leq B/4$ (Lemma D.1 (b)),

$$|\Gamma| \geq \frac{s_{H-1}}{4\Psi_*} \geq \frac{M}{4(4X)^k}, \quad C_{\text{col}} = |\Gamma|\ \frac{BE}{2} \geq \frac{MBE}{8(4X)^k} \geq \frac{E^2 B^2}{16}.$$

Counting the edges before the deletion, which only removes edges,

$$|\mathcal{P}_k| \leq C_{\text{col}} + 2gB\ |\Gamma| + \sum_h a_h + 5B + u^2(H+1) + (H + 1 + E + B + 2J)$$
$$\leq \big(1 + 4g/E + 2^{2k+1}/E\big)C_{\text{col}} + (u^2 + 9)B \leq (1 + 3/4)C_{\text{col}} \leq 2C_{\text{col}},$$

where the terms count, in this order:

- the columns, $C_{\text{col}}$;
- the gadgets, $|\Gamma|$ times $gB$ vertices of arity 2 (Definition C.5): $2gB\ |\Gamma| = (4g/E)C_{\text{col}}$;
- the runners: at most $H \leq W$ arities $a_h = s_h/\Psi_h$, with $s_h < 2M$ and $\Psi_h > X^k$, so

$$\sum_h a_h \leq \frac{2MW}{X^k} = \frac{MB}{4X^k} = \frac{2^{2k+1}}{E} \cdot \frac{MBE}{8(4X)^k} \leq \frac{2^{2k+1}}{E} C_{\text{col}};$$

- the free levels and the root, at most $5B$ (Lemma C.15 (a));
- the pool, $u^2(H+1) \leq u^2 B$;
- the zero words, $H + 1 + E + B + 2J \leq 4B$, as $H + 1 \leq B/16 + 2 \leq B/4$ ($\Lambda \geq 16$), $E = 2^{2k+7}u \leq 2^{2k+5}X \leq B$ (Lemma D.1 (b)) and $2J \leq 2\log n + 2 \leq B$ by (b), whose proof does not use (a).

By Definition C.1, $4g/E = (k+1)L/\big(32 \cdot 4^k u\big) < 1/4$ and $2^{2k+1}/E = 1/(64u) < 1/4$; and $(u^2 + 9)B \leq 2u^2 B \leq E^2B^2/64 \leq C_{\text{col}}/4$, as $E \geq u$ and $B \geq 128$.

(b) As $R, E, H+1 \geq 1$ and $M = EB^2/8 \geq B^2$,

$$n = RBs_{H-1}E(H+1) \geq Bs_{H-1} \geq BM \geq B^3 = X^{3(k+1)}.$$

With $R \leq X/2$ (Lemma C.15 (a)), $s_{H-1} < 2M = 32 \cdot 4^k uB^2$, $E = 128 \cdot 4^k u$, $H+1 \leq B/4$ and $2^{5k+9}u^2 \leq X$ (Lemma D.1 (b)),

$$n \leq \frac{X}{2} \cdot B \cdot 32 \cdot 4^k uB^2 \cdot 128 \cdot 4^k u \cdot \frac{B}{4} = 2^{4k+9}u^2XB^4 \leq 2^{5k+9}u^2XB^4 \leq X^{4k+6}.$$

Hence $3(k+1)L \leq \log n \leq (4k+6)L$, so $10\log n \leq 10(4k+6)L = u$, and, as $10(4k+6)/(3k+3)$ decreases in $k$, for $k \geq 2$

$$\frac{u}{\log n} \leq \frac{10(4k+6)}{3k+3} \leq \frac{140}{9} < 16.$$

If $L = L_k$, then $L = \Theta(k \log k)$, so $\log n = \Theta(kL) = \Theta(k^2 \log k)$. □

## D.1. The Loss in the Exponent

As $B/\Lambda \leq H \leq 2B/\Lambda$, $s_{H-1} \in [M, 2M)$, $M = EB^2/8$ and $R, E, \Lambda \leq X^2 \leq H^{2/k}$ (Definition C.1, Lemma D.1, Lemma C.15 (a), Lemma C.13), with absolute constants in $\Theta$,

$$n = RBs_{H-1}E(H+1) = \Theta(RE^2H^4\Lambda^3) = H^{4+O(1/k)}.$$

## E. The Reverse Comparison and Other Measures

By Theorem 1.2 and Proposition E.1, ARC is no polylogarithmic approximation of ZLC. No polynomial-time approximation of ZLC within a polylogarithmic factor is known. We prove Proposition E.1 and Remark E.8; $\mathrm{bit}_j(i)$ is bit $j$ of $i$ (bit 0 least significant).

**Proposition E.1** (The reverse comparison; proof below)**.**

*(a) There is an absolute constant $C$ such that $\mathrm{ZLC}(x) \leq Cn\log(\mathrm{ARC}(x)+1)/\log(n+1)$ for every finite alphabet $\Sigma$ and every word $x \in \Sigma^n$, $n \geq 2$.*

*(b) For every $n \geq 2$ there is a binary word $x$ of length $n$ with $\mathrm{ZLC}(x) \geq n\,\mathrm{ARC}(x)/(64\log^2 n)$ (log to base 2).*

For $n \geq 56$, the words of (b) have $\mathrm{ARC}(x) < 2\log n + 1$ and $\mathrm{ZLC}(x) > n/(32\log n)$.

**Definition E.2** (Sections and comparability)**.** For a word $u \in \Sigma^*$, $s \geq 1$ and $0 \leq i < s$, the *section* $u[i + s\mathbb{N}] := u[i]u[i+s]u[i+2s]...$, up to the end of $u$. Two words $u, v$ are called *comparable* if $u \sqsubseteq v$ or $v \sqsubseteq u$.

**Lemma E.3** (Programs from few letters)**.** *There is an absolute constant $C_0$ such that, for every finite alphabet $\Sigma$, every word $x \in \Sigma^n$ with $n \geq 2$ and $d$ distinct letters has $\mathrm{ZLC}(x) \leq C_0 n\log(d+1)/\log(n+1)$. Moreover $d \leq \mathrm{ARC}(x)$; so $\mathrm{ZLC}(x) \leq C_0 n\log(\mathrm{ARC}(x)+1)/\log(n+1)$.*

*Proof.*

1. $d \leq \mathrm{ARC}(x)$.
   Let $(L, \Phi)$ be an ARS for $x$ of size $\mathrm{ARC}(x)$, and for each letter $a$ of $x$ let $i_a := \min\{i < n : x[i] = a\}$. No equation $(\alpha, \alpha') \in \Phi$ covers $i_a$: $\alpha(t) < \alpha'(t) = i_a$ would give $x[\alpha(t)] = x[i_a] = a$ by Lemma B.1, an $a$ before $i_a$. So $(a, i_a) \in L$ for each of the $d$ letters $a$ (Definition 3.2 (c)), and $d \leq |L| \leq \mathrm{ARC}(x)$.
2. *If $d = 1$, then $\mathrm{ZLC}(x) \leq 2\lceil \log n \rceil$. For every integer $\ell \geq 1$, $\mathrm{ZLC}(x) \leq \lceil n/\ell \rceil + \ell d^\ell$.*
   For the first claim: if $d = 1$, then $x = a^n$, and the vertices $Z^0 := (a, a)$, $Z^{i+1} := (Z^i, Z^i)$, $i + 1 < \lceil \log n \rceil$, have $\mathrm{val}(Z^i) = a^{2^{i+1}}$; with root $Z^{\lceil \log n \rceil - 1}$ they give $\mathrm{ZLC}(x) \leq 2\lceil \log n \rceil$.

For the second claim, let $\ell \ge 1$, $s := \lceil n/\ell \rceil$, $x' := x\, x[0]^{s\ell - n}$ and $u_i := x'[i + s\mathbb{N}]$ for $i < s$. As $|x'| = s\ell$, each $u_i = x'[i]x'[i+s]...x'[i+(\ell-1)s]$ is a word of length $\ell$ over the $d$ letters of $x$, so $U := \{u_i : i < s\}$ has $|U| \le d^\ell$; and by Definition 3.3, as $p = (p \bmod s) + s\lfloor p/s \rfloor$,

$$x' = \mathrm{zip}(u_0, ..., u_{s-1}).$$

Let $P$ have a vertex $q_u$ with $\delta(q_u) := (u[0], ..., u[\ell-1])$ for each $u \in U$ if $\ell \ge 2$, and the root $q_0$ with $\delta(q_0) := (c_0, ..., c_{s-1})$, where $c_i := q_{u_i}$ if $\ell \ge 2$ and $c_i := u_i$ (a letter) if $\ell = 1$; if $s = 1$, then $\ell \ge n \ge 2$ and the root is $q_{u_0}$ instead. Then $\mathrm{val}(q_u) = u$, $\mathrm{val}(P) = x'$ and

$$\mathrm{ZLC}(x) \le |P| \le s + \ell\ |U| \le \lceil n/\ell \rceil + \ell d^\ell.$$

3. $\mathrm{ZLC}(x) \le C_0 n \log(d+1)/\log(n+1)$.
   Let $d \ge 2$. If $\log_d n < 4$, take $\ell = 1$: $\mathrm{ZLC}(x) \le n + d \le 2n < 8n \log d/\log n$. Otherwise take $\ell := \lfloor 1/2 \log_d n \rfloor$. Then $\ell \ge 1/4 \log_d n$, $\ell \le \log n$ and $d^\ell \le \sqrt{n}$, so by Step 2
   $$\mathrm{ZLC}(x) \le \frac{n}{\ell} + 1 + \ell d^\ell \le \frac{4n \log d}{\log n} + 1 + \sqrt{n} \log n = O\!\left(\frac{n \log d}{\log n}\right),$$
   as $1 + \sqrt{n}\log n = O(n/\log n)$ and $\log d \ge 1$. As $\log(n+1) \le 2\log n$ and $2\lceil \log n \rceil = O(n/\log(n+1))$, in all cases $\mathrm{ZLC}(x) \le C_0 n \log(d+1)/\log(n+1)$.

Finally $\log(d+1) \le \log(\mathrm{ARC}(x)+1)$ by Step 1. □

**Definition E.4** (The words $z_m$ and $x_n$, blocks)**.** For $m \ge 3$ let $B := 2^m$, $K := mB$, $T := B/8$ and $N_m := (2m+1)B$. Let $y_m \in \{0,1\}^{mB}$ be given by $y_m[jB + i] := \mathrm{bit}_j(i)$ for $j < m$ and $i < B$; we call its factor $j$, of length $B$, its *block* $j$. Let $z_m := 0^K 1^B y_m$, a binary word of length $N_m$; block $j$ of $y_m$ occupies the positions $a_j, ..., a_j + B - 1$ of $z_m$, where $a_j := (m + 1 + j)B$. As $N_m$ increases with $m$ and $N_3 = 56$, every $n \ge 56$ has exactly one $m \ge 3$ with $N_m \le n < N_{m+1}$; for it, put $x_n := z_m 1^{n - N_m}$. For $2 \le n < 56$ put $x_n := 0^n$.

So $y_m = \prod_{j<m} \left(0^{2^j} 1^{2^j}\right)^{B/2^{j+1}}$; for example, $y_3 = 01010101\,00110011\,00001111$: block $j$ lists bit $j$ of $0, ..., 7$, and block $j+1$ doubles each letter of the first half of block $j$. As $2^{j+1}$ divides $a_j$, $z_m[p] = \mathrm{bit}_j(p)$ for every position $p$ of block $j$ in $z_m$.

**Lemma E.5** (ARC of $x_n$)**.** *If $m \ge 3$ and $N_m \le n < N_{m+1}$, then* $\mathrm{ARC}(x_n) \le 2m + 5$.

*Proof.* Write $x := x_n$, and write an equation $(\alpha, \alpha')$ of length $\ell$ as $x[\alpha(t)] = x[\alpha'(t)]$ $(t < \ell)$. Let $L := \{(0,0), (1,K)\}$, and let $\Phi$ consist of the equations

$$x[t] = x[t+1] \quad (t < K-1), \quad x[K+t] = x[K+1+t] \quad (t < B-1),$$
$$x[t] = x[a_0 + 2t], \quad x[K+t] = x[a_0 + 2t + 1] \quad (t < B/2),$$
$$x\left[a_{j-1} + t\right] = x\left[a_j + 2t + \varepsilon\right] \quad (t < B/2) \quad \text{for } 1 \le j < m \text{ and } \varepsilon \in \{0,1\},$$
$$x[N_m - 1 + t] = x[N_m + t] \quad (t < n - N_m) \quad \text{if } n > N_m.$$

We show that $(L, \Phi)$ is an ARS for $x$ (Definition 3.2) of size at most $2m+5$.

1. $|L| + |\Phi| \le 2m + 5$.
   $|L| = 2$, and the four lines of $\Phi$ hold at most $2 + 2 + 2(m-1) + 1$ equations; in all at most $2m+5$.
2. *Every equation $(\alpha, \alpha') \in \Phi$, of length $\ell$, has $\alpha(t) < \alpha'(t) < n$ and $x[\alpha(t)] = x[\alpha'(t)]$ for all $t < \ell$.*
   By Definition E.4, $x = 0^K 1^B$ on $[a_0]$ $(a_0 = K + B)$, $x\left[a_j + i\right] = y_m[jB + i] = \mathrm{bit}_j(i)$ for $j < m$ and $i < B$, and $x = 1$ on $[n] \setminus [N_m]$ $(N_m = a_m)$. The largest targets are $K - 1$, $a_0 - 1$, $a_j + B - 1 \le N_m - 1$ and $n - 1$, all below $n$, and each source is below its target, as $K < a_0$ and $a_{j-1} < a_j$. The letters agree: the runs repeat one letter, and for $t < B/2 < K$,

$$x[t] = 0 = \mathrm{bit}_0(2t) = x[a_0 + 2t], \quad x[K + t] = 1 = \mathrm{bit}_0(2t + 1) = x[a_0 + 2t + 1],$$

$$x\big[a_{j-1} + t\big] = \mathrm{bit}_{j-1}(t) = \mathrm{bit}_j(2t + \varepsilon) = x\big[a_j + 2t + \varepsilon\big],$$

and $x[N_m - 1] = \mathrm{bit}_{m-1}(B - 1) = 1$.

3. *Every $i < n$ is 0, $K$ or a target $\alpha'(t)$ of an equation $(\alpha, \alpha') \in \Phi$.*
   As $a_0 = K + B$, $a_j + B = a_{j+1}$ and $a_m = N_m$,
   $$[n] = [K] \cup (K + [B]) \cup \bigcup_{j<m} \big(a_j + [B]\big) \cup (N_m + [n - N_m]).$$
   The targets of the four lines of $\Phi$ are $([K] \setminus \{0\}) \cup ((K + [B]) \setminus \{K\})$, then $\big\{a_j + 2t + \varepsilon : t < B/2, \varepsilon \in \{0, 1\}\big\} = a_j + [B]$ for $j = 0$ and for each $1 \le j < m$, and $N_m + [n - N_m]$. So the literals and the targets cover $[n]$.

By Step 2 and Step 3, $(L, \Phi)$ satisfies (b) and (c) of Definition 3.2, and (a) as $x[0] = 0$ and $x[K] = 1$; so $\mathrm{ARC}(x) \le 2m + 5$ by Step 1. □

**Lemma E.6** (Sections)**.** *Let $m \ge 3$, $1 \le s, s' \le T$, $0 \le i < s$ and $0 \le i' < s'$. If the sections $z_m[i + s\mathbb{N}]$ and $z_m[i' + s'\mathbb{N}]$ are comparable, then $(s, i) = (s', i')$.*

*Proof.* By symmetry let $s \le s'$.

1. $0 \le 1/s - 1/s' < 2/K$.
   As $s \le T < B$, the section $z_m[i + s\mathbb{N}]$ has a letter in the run $1^B$ on $K + [B]$, so its first 1 is at the index $k := \min\{k' \in \mathbb{N} : i + sk' \ge K\}$ ($z_m$ begins with $0^K$). So $i + s(k - 1) < K \le i + sk$, and as $0 \le i < s$,
   $$\frac{K}{s} - 1 < k < \frac{K}{s} + 1.$$
   The two sections are comparable and both contain a 1, so their first 1 is at the same index $k$, which satisfies these bounds for $s$ and for $s'$. Hence $0 \le 1/s - 1/s' < 2/K$, as $s \le s'$.
2. *For every $j < m$ there is $t \in \mathbb{N}$ such that, for both $t' \in \{t, t+1\}$, $i + st', i' + s't' \in a_j + [B]$ (block $j$) and $\mathrm{bit}_j(i + st') = \mathrm{bit}_j(i' + s't')$.*
   Let $R_j := \big\{t \in \mathbb{Z} : a_j/s \le t \le \big(a_j + B\big)/s' - 1\big\}$. For $t \in R_j$, the bounds on $t$ and $0 \le i < s \le s'$ give
   $$a_j \le st \le i + st < s(t + 1) \le s'(t + 1) \le a_j + B,$$
   and likewise $a_j \le s't \le i' + s't < s'(t + 1) \le a_j + B$; so both positions lie in $a_j + [B]$. As $a_j \le 2K$ and $B/s' \ge 8$, by Step 1 the bounds of $R_j$ differ by
   $$\frac{a_j + B}{s'} - 1 - \frac{a_j}{s} = \frac{B}{s'} - a_j\Big(\frac{1}{s} - \frac{1}{s'}\Big) - 1 > 8 - 2K \cdot \frac{2}{K} - 1 = 3,$$
   so $R_j$ holds integers $t, t + 1$. At both, the positions are below $N_m$, so the comparable sections have equal letters there: $z_m[i + st'] = z_m[i' + s't']$, and these are the bits $j$ of the positions (Definition E.4).
3. *$i \equiv i'$ and $s \equiv s'$ $(\mathrm{mod}\, 2^j)$ for every $j \le m$; hence $(s, i) = (s', i')$.*
   By induction on $j$, trivial for $j = 0$. If it holds for $j < m$, then $i + st \equiv i' + s't \pmod{2^j}$ for all $t$; with $t$ of Step 2, equal bits $j$ at $t$ and $t + 1$ give (a residue modulo $2^{j+1}$ is given by the residue modulo $2^j$ and bit $j$)
   $$i + st' \equiv i' + s't' \pmod{2^{j+1}} \quad \text{for } t' \in \{t, t + 1\}.$$

Subtracting the case $t$ from the case $t+1$ gives $s \equiv s' \pmod{2^{j+1}}$, then $i \equiv i'$. For $j = m$ they hold modulo $B$, so they are equalities, as $i < s \le s' \le T < B$ and $i' < s' < B$. □

**Lemma E.7** (ZLC of $z_m$)**.** *For every $m \ge 3$, $\mathrm{ZLC}(z_m) \ge B/8$.*

*Proof.* Let $P$ be a ZLP with $z_m \sqsubseteq \mathrm{val}(P)$. We unfold $P$ from the root until the step exceeds $T$. A root path $\pi$ with end $q_\pi$ gives the occurrence $\rho_\pi \in \mathrm{occ}_{q_\pi}$, of step $s_\pi := p(\pi)$ (Definition 3.6); followed by edge $j$ of $q_\pi$ it gives $\pi j$, a root path or a root path to a letter, and we put $s_{\pi j} := d_{q_\pi} s_\pi$ in both cases. The *expanded* paths and the *leaves* are

$$\mathcal{E} := \{\pi \text{ root path} : s_{\pi'} \le T \text{ for every prefix } \pi' \text{ of } \pi\},$$

$$\mathcal{F} := \left\{\pi j : \pi \in \mathcal{E}, j < d_{q_\pi}, \pi j \notin \mathcal{E}\right\}$$

(the prefixes include $\pi$ itself). The empty root path, of step $1 \le T$, is in $\mathcal{E}$, and $\mathcal{E}$ is finite, as the graph of $P$ is acyclic.

1. *Every vertex $q$ has at most one occurrence of step at most $T$.*
   An occurrence $\rho(t) = i + st$ of a vertex $q$ $(0 \le i < s)$ has $z_m[i + s\mathbb{N}] \sqsubseteq \mathrm{val}(q)$ by Lemma 3.7 (a); so two occurrences of $q$ give comparable sections, and distinct occurrences have distinct $(s, i)$ by Lemma 3.7 (b). So by Lemma E.6, $q$ has at most one occurrence of step at most $T$.
2. *$s_f > T$ for every $f \in \mathcal{F}$; hence $|\mathcal{F}| > T$.*
   Occurrences of step at most $T$ are expanded: a leaf $\pi j$ that is a root path is not in $\mathcal{E}$, while all prefixes of $\pi$ are, so $s_{\pi j} > T$. A letter $c$ on edge $j$ of an expanded $\rho := \rho_\pi$ of $q := q_\pi$, $\rho(t) = i + st$ with $0 \le i < s$, with $d_q s \le T$ would give
   $$\rho(d_q + j) = i + s(d_q + j) < s + s(2d_q - 1) \le 2T < N_m,$$
   so $\mathrm{val}(q)[d_q + j]$ exists by Lemma 3.7 (a), and by Definition 3.3 it is $\mathrm{val}(c)[1]$, while $\mathrm{val}(c)$ has length 1. Each $\pi \in \mathcal{E}$ has $\sum_{j<d_{q_\pi}} 1/s_{\pi j} = 1/s_\pi$, and as $\mathcal{E}$ is closed under prefixes, the $\pi j$ with $\pi \in \mathcal{E}$ and $j < d_{q_\pi}$ are the elements of $\mathcal{E} \cup \mathcal{F}$ other than the empty path, each once. Summing over $\pi \in \mathcal{E}$ and cancelling the terms of the non-empty paths in $\mathcal{E}$, the leaf weights $1/s_f$, where $s_f > T$, sum to 1:
   $$1 = \sum_{f \in \mathcal{F}} \frac{1}{s_f} < \frac{|\mathcal{F}|}{T}.$$

Distinct $\pi, \pi' \in \mathcal{E}$ with $q_\pi = q_{\pi'}$ would give distinct occurrences $\rho_\pi \ne \rho_{\pi'}$ of step at most $T$ (Lemma 3.7 (b)); so the expanded occurrences are of distinct vertices (Step 1), and $\pi j \mapsto (q_\pi, j)$ maps the leaves injectively to edges of $P$: the last edges of the leaves are distinct edges of $P$, and $|P| \ge |\mathcal{F}| > T = B/8$. □

*Proof of Proposition E.1.* (a) is Lemma E.3. Let $n \ge 2$ and $x := x_n$. If $n < 56$, then $x = 0^n$ has $\mathrm{ARC}(x) \le 2$ (the literal $(0,0)$ and the equation $x[t] = x[t+1]$, $t < n-1$) and $\mathrm{ZLC}(x) \ge 2$ (the root has arity at least 2), and $n/(64 \log^2 n) < 1$; so $n\,\mathrm{ARC}(x)/(64 \log^2 n) < 2 \le \mathrm{ZLC}(x)$. If $n \ge 56$, let $m$ be as in Definition E.4. As $z_m$ is a prefix of $x$, $\mathrm{ZLC}(x) \ge \mathrm{ZLC}(z_m) \ge B/8$ by Lemma E.7. Moreover $n < N_{m+1} = 2(2m+3)B$ and $\log n \ge \log N_m = m + \log(2m+1) > m + 2$, so by Lemma E.5

$$\frac{n\,\mathrm{ARC}(x)}{64 \log^2 n} < \frac{2(2m+3)B(2m+5)}{64(m+2)^2} = \frac{B(4(m+2)^2 - 1)}{32(m+2)^2} < \frac{B}{8} \le \mathrm{ZLC}(x).$$

As $2m + 4 < 2 \log n$, also $\mathrm{ARC}(x) < 2 \log n + 1$ and $\mathrm{ZLC}(x) \ge B/8 > n/(32 \log n)$. □

**Remark E.8** (Interleaving, Lempel–Ziv copying and grammars)**.** Let $g(x)$ be the least number of rules of a grammar in Chomsky normal form for a non-empty word $x$; then $g(x) \geq \mathrm{LZC}(x)$ (by [3, Thm. 1], $g(x)$ is at least the number of factors of the Lempel–Ziv factorisation without self-reference, which is at least $\mathrm{LZC}(x)$).

(a) For every $m \geq 3$, the word $z_m$ of Definition E.4, of length $n = (2m+1)2^m$, has $\mathrm{ZLC}(z_m) \geq ng(z_m)/(16\log^3 n) \geq n\,\mathrm{LZC}(z_m)/(16\log^3 n)$.

(b) For every $a < 1/4$, at every large enough length $n$, some binary word $x$ has $g(x) \geq \mathrm{LZC}(x) \geq n^a\,\mathrm{ZLC}(x)$.

(c) So neither ZLC nor LZC is bounded by the other times $n^{o(1)}$, and neither are ZLC and $g$, nor ZLC and the measures based on substring complexity and on string attractors, which are within polylogarithmic factors of LZC and of $g$ (Section 1). $\lhd$

*Proof.* First, $z_m$ has a grammar in Chomsky normal form with at most $\log^2 n$ rules. The grammar $G$ has the rules

$$\begin{aligned}
&U_0 \to 0, \quad U_i \to U_{i-1}U_{i-1} \quad (1 \leq i \leq m + \lfloor \log m \rfloor),\\
&V_0 \to 1, \quad V_i \to V_{i-1}V_{i-1} \quad (1 \leq i \leq m),\\
&W_{j,0} \to U_jV_j \quad (j < m),\\
&W_{j,k} \to W_{j,k-1}W_{j,k-1} \quad (j < m, 1 \leq k < m-j),
\end{aligned}$$

so $U_i$, $V_i$ and $W_{j,k}$ derive $0^{2^i}$, $1^{2^i}$ and $\left(0^{2^j}1^{2^j}\right)^{2^k}$, and $W_{j,m-j-1}$ derives block $j$ of $y_m$; and $\nu(m) + m$ rules concatenate the $\nu(m) + 1 + m$ pieces

$$z_m = \left(\prod_{b:\mathrm{bit}_b(m)=1} 0^{2^{m+b}}\right) 1^{2^m} \prod_{j<m} \left(0^{2^j}1^{2^j}\right)^{2^{m-j-1}},$$

derived by the $U_{m+b}$, by $V_m$ and by the $W_{j,m-j-1}$ (as $\sum_{b:\mathrm{bit}_b(m)=1} 2^{m+b} = m2^m = K$), where $\nu(m)$ is the binary digit sum of $m$. As $\lfloor \log m \rfloor + \nu(m) \leq 2\log m + 1$, $m^2/2 + m/2 + 1 \geq 2\log m$ and $\log n = m + \log(2m+1) > m+2$, the number of rules is

$$\begin{aligned}
&(m + \lfloor \log m \rfloor + 1) + (m+1) + \frac{m(m+1)}{2} + (\nu(m) + m)\\
&= \frac{m^2 + 7m + 4}{2} + \lfloor \log m \rfloor + \nu(m) \leq \frac{m^2 + 7m + 4}{2} + 2\log m + 1 \leq (m+2)^2 < \log^2 n.
\end{aligned}$$

But $\mathrm{ZLC}(z_m) \geq 2^m/8 \geq n/(16\log n)$ (Lemma E.7; $2m+1 < 2\log n$, as $\log n > m+2$). So, as $\mathrm{LZC}(z_m) \leq g(z_m) \leq \log^2 n$, $\mathrm{ZLC}(z_m) \geq n/(16\log n) \geq ng(z_m)/(16\log^3 n) \geq n\,\mathrm{LZC}(z_m)/(16\log^3 n)$: this is (a). Part (b) follows from $\mathrm{LZC}(x) \geq \mathrm{ARC}(x)$ [6, Prop. 4.3], $g(x) \geq \mathrm{LZC}(x)$ and Theorem 1.2, and (c) from (a) and (b). $\square$

## F. Use of AI Tools in Detail

*Systems and research.* The systems were Anthropic's Claude Opus, Sonnet and Haiku 5.5, run as agents in Claude Code, OpenAI's Codex with gpt-6-astra, and Lean 4 with Mathlib. The question is Kosoy's [6]. The results were found by the AI systems in rounds that the author started, stopped and redirected. Codex found the first polynomial separation (exponent 1/6), the first upper bound of order $n^{1/4}$ (with a logarithmic factor), the constant $2^{5/4}$ and the reverse comparison (Proposition E.1); Claude Opus found the first lower bound with exponent $1/4 - o(1)$, from which the relay words descend. In later rounds aimed at simpler proofs and better constants, Claude Opus found the present proof of the upper bound and the present form of the lower-bound construction, building on earlier constructions of both systems; the extension of the lower bound to every large length was first found by Codex during a validity audit. The author asked for a better exponent, simpler proofs and a different lower-bound construction, chose which proofs the paper uses, and decided to state the lower bound at every sufficiently large length.

*Verification.* The working rules required a blind review by the other model family (Codex for Claude's work, Claude Opus for Codex's), plus an independent audit selected with probability 0.1; the records document such reviews and scoped re-checks of research results, revised proofs and the changed text of each revision round. Agents also ran computational checks of constructions and estimates. Of four full validity audits of the paper by Codex, three found errors, which were corrected and re-checked; the other reported a minor issue that a change in the formalisation made moot. The formalisation (Section 1.3), by Claude agents, uses only the standard axioms of Lean; Codex checked the faithfulness of its statements to the text. Claude agents drafted and revised the text, and the author edited it.

*The author.* The author designed this workflow: its review, audit and formalisation rules, and the writing process, with a style guide drawn from the author's comments and tests with model readers, such as a linear reader with bounded memory, and simulated referees of both families. The author also chose the framing, the scope, the venue, the title and the structure of the paper, and commented on the drafts and the outline. The author read the main text in full and edited it, read the appendices at the level of their arguments, checked particular arguments in detail (finding, for example, a missing converse for synchronous programs), and checked the README's Lean definitions and statements of Theorem 1.1, Theorem 1.2 and Corollary 1.3 against the text.